\documentclass[twocolumn,prb,aps,floatfix,superscriptaddress,nobibnotes]{revtex4-2}

\usepackage{graphicx}
\usepackage{enumerate}
\usepackage[colorlinks=true,citecolor=blue]{hyperref}
\usepackage{textcomp}
\usepackage{amsmath}
\usepackage{amssymb}
\usepackage{pgf}
\usepackage{svg}
\usepackage{subfigure}
\usepackage[english]{babel}
\usepackage[normalem]{ulem}
\usepackage{tabularray}
\usepackage{xcolor}
\usepackage{array}
\usepackage{comment}

\usepackage{dcolumn}
\usepackage{bm}
\usepackage{amsfonts}
\usepackage{siunitx}
\usepackage{dsfont}
\usepackage{placeins}

\DeclareUnicodeCharacter{03BD}{$\nu$}

\definecolor{bluepoli}{RGB}{0,36,179}
\definecolor{redpoli}{RGB}{204,0,51}
\definecolor{greenpoli}{RGB}{45,137,0}
\definecolor{purplepoli}{RGB}{153,102,204}
\definecolor{azzurropoli}{RGB}{51,53,204}
\definecolor{orangepoli}{RGB}{255,124,17}
\usepackage{hyperref}
\hypersetup{
	colorlinks,
	linkcolor={bluepoli}, 
	citecolor={redpoli}, 
	urlcolor={redpoli} %
}

\renewcommand{\rm}[1]{\mathrm{#1}}

\usepackage{soul}

\usepackage[capitalize]{cleveref}

\begin{document}

\title{Fast-driving spectroscopy of current-phase relationships of individual Josephson elements}

\author{Maxim Tjøtta}
\thanks{These authors contributed equally.}
\affiliation{Center for Quantum Spintronics, Department of Physics, NTNU, Norwegian University of Science and Technology, NO-7491 Trondheim, Norway}

\author{Devashish Shah}
\thanks{These authors contributed equally.}
\affiliation{Institute of Science and Technology Austria, Am Campus 1, 3400 Klosterneuburg, Austria}

\author{Kanishk Modi}
\affiliation{Institute of Science and Technology Austria, Am Campus 1, 3400 Klosterneuburg, Austria}
\affiliation{Department of Physics, Indian Institute of Technology Bombay, Mumbai, India}

\author{Marco Valentini}
\affiliation{Institute of Science and Technology Austria, Am Campus 1, 3400 Klosterneuburg, Austria}
\affiliation{Department of Physics, University of California at Santa Barbara, Santa Barbara, CA, USA}

\author{Rub\'en Seoane Souto}
\affiliation{Instituto de Ciencia de Materiales de Madrid (ICMM), Consejo Superior de Investigaciones Cient\'{i}ficas (CSIC), Sor Juana In\'{e}s de la Cruz 3, 28049 Madrid, Spain}

\author{Georgios Katsaros}
\affiliation{Institute of Science and Technology Austria, Am Campus 1, 3400 Klosterneuburg, Austria}

\author{Jeroen Danon}
\affiliation{Center for Quantum Spintronics, Department of Physics, NTNU, Norwegian University of Science and Technology, NO-7491 Trondheim, Norway}

\begin{abstract}
Extracting the current--phase relationship (CPR) of a single superconducting junction is challenging in practice and traditionally involves embedding the junction in a larger superconducting circuit containing SQUIDs and/or resonators.
Applying ac driving to the junction has proven to be a viable and less invasive way to extract information about the few lowest harmonics of the CPR, by locating the integer and fractional Shapiro steps in the $IV$-curve of the driven junction.
Here, we present an alternative driving-based method that allows to extract the full harmonic content of a CPR in a relatively non-invasive way, by fitting the measured critical currents of the driven junction as a function of driving power.
We test our method, both using numerical simulations and in experiments, and we show that it works accurately.
\end{abstract}

\maketitle

\section{Introduction}

Superconducting circuits provide an attractive platform for applications within quantum computing~\cite{clarkeSuperconductingQuantumBits2008,wendinQuantumInformationProcessing2017,krantzQuantumEngineerGuide2019,blaisCircuitQuantumElectrodynamics2020}, sensing~\cite{kgrammSQUIDMagnetometerMangetization1976,fagalySuperconductingQuantumInterference2006,granataNanoSuperconductingQuantum2016,degenQuantumSensing2017} and metrology~\cite{benzApplicationJosephsonEffect2004,schererQuantumMetrologyTriangle2012, tafuriFundamentalsFrontiersJosephson2019}, but could also offer new functionalities for conventional electronics \cite{castellanos-beltranWidelyTunableParametric2007,Braginski_JSNM2019}.
Most of these applications exploit the consequences of the fundamental connection between the macroscopic phase coherence of the superconducting condensate and the dissipationless flow of supercurrent, such as the often unconventional current--phase relationships (CPRs) of individual superconducting elements.

The sinusoidal CPR of the Josephson junction---a junction comprised of two superconductors separated by a tunnel barrier---is the canonical example~\cite{josephsonPossibleNewEffects1962} and it provides the essential non-linear ingredient that creates the anharmonicity of a superconducting $LC$-circuit needed to operate it reliably as a transmon qubit~\cite{krantzQuantumEngineerGuide2019}.
More complex CPRs, containing also higher harmonics and phase offsets, can arise when the tunnel barrier is non-ideal~\cite{Josephson_dynamics_Galaktionov,willschObservationJosephsonHarmonics2024} or when it is replaced by, for instance, a normal metal~\cite{baselmansReversingDirectionSupercurrent1999,golubovCurrentphaseRelationJosephson2004}, a ferromagnet~\cite{buzdinProximityEffectsSuperconductorferromagnet2005,bergeretOddTripletSuperconductivity2005,linderSuperconductingSpintronics2015,ryazanovCouplingTwoSuperconductors2001,sellierHalfIntegerShapiroSteps2004,frolovJosephsonInterferometryShapiro2006,stoutimoreSecondHarmonicCurrentPhaseRelation2018}, graphene~\cite{leeProximityCouplingSuperconductorgraphene2018,heerscheBipolarSupercurrentGraphene2007,englishObservationNonsinusoidalCurrentphase2016,nandaCurrentPhaseRelationBallistic2017,manjarresSkewnessCriticalCurrent2020}, a topological insulator~\cite{fuJosephsonCurrentNoise2009,kayyalhaHighlySkewedCurrent2020}, or a quantum dot~\cite{vandamSupercurrentReversalQuantum2006}.

Especially in superconductor--semiconductor-based junctions, the CPR can be highly non-sinusoidal and tunable, due to the typically high transparency of the junction and the often strong spin--orbit coupling and tunable carrier density in semiconductors~\cite{spantonCurrentPhaseRelations2017,hartCurrentphaseRelationsInAs2019,nicheleRelatingAndreevBound2020,uedaEvidenceHalfintegerShapiro2020,aggarwalEnhancementProximityinducedSuperconductivity2021,valentiniParityconservingCooperpairTransport2024,leblancGateFluxTunable2024,scherublDeterminationCurrentphaseRelation2025,banszerusHybridJosephsonRhombus2025}.
Such more exotic and tunable CPRs allow for the realization of superconducting devices with advanced functionalities:
An asymmetry between the critical current in the two directions of the current gives rise to the superconducting diode effect~\cite{andoObservationSuperconductingDiode2020,bauriedlSupercurrentDiodeEffect2022,baumgartnerSupercurrentRectificationMagnetochiral2022a,nadeemSuperconductingDiodeEffect2023a, fracassiAnomalousSupercurrentDiode2024b}, Josephson junctions with tunable transparency can yield electrically controllable superconducting qubits~\cite{larsenSemiconductorNanowireBasedSuperconductingQubit2015,delangeRealizationMicrowaveQuantum2015,casparisSuperconductingGatemonQubit2018,hertelGatetunableTransmonUsing2022,sagiGateTunableTransmon2024a,kiyookaGatemonQubitGermanium2025}, a junction with a CPR dominated by its second harmonic can be used for realizing protected superconducting qubits~\cite{bellProtectedJosephsonRhombus2014,larsenParityprotectedSuperconductorsemiconductorQubit2020,smithSuperconductingCircuitProtected2020,schradeProtectedHybridSuperconducting2021,danonProtectedSolidstateQubits2021}, and a spin-dependent CPR allows for hosting so-called Andreev spin qubits in the junctions~\cite{chtchelkatchevAndreevQuantumDots2003a,padurariuTheoreticalProposalSuperconducting2010,haysCoherentManipulationAndreev2021,pita-vidalDirectManipulationSuperconducting2023}.

Despite its importance, it has been proven challenging to determine the CPR of a given superconducting element in experiment.
Over the last decades, a few standard approaches have been developed:
The dc-SQUID method determines the CPR of a junction by embedding it in a SQUID that also includes another junction with either a known CPR or a critical current that is much larger than that of the junction of interest.
In both cases, this allows to relate the CPR of the junction directly to the critical current of the SQUID as a function of magnetic flux threading the loop~\cite{dellaroccaMeasurementCurrentPhaseRelation2007, leeUltimatelyShortBallistic2015, englishObservationNonsinusoidalCurrentphase2016, muraniBallisticEdgeStates2017, ginzburgDeterminationCurrentPhase2018,  nandaCurrentPhaseRelationBallistic2017, liZeemanEffectInduced$0textensuremathensuremathpi$Transitions2019, kayyalhaAnomalousLowTemperatureEnhancement2019, kayyalhaHighlySkewedCurrent2020, nicheleRelatingAndreevBound2020, aggarwalEnhancementProximityinducedSuperconductivity2021, endresCurrentPhaseRelation2023, babichLimitationsCurrentPhase2023, messelotDirectMeasurementSin2024, chieppaUnveilingCurrentPhaseRelationship2025a, strambiniJosephsonPhaseBattery2020}.
Alternatively, the junction can be coupled to a resonator with a high quality factor, either embedded in a superconducting loop of small inductance or directly.
By monitoring the effective impedance of the total system under rf driving, the CPR of the junction can be extracted~\cite{silverQuantumStatesTransitions1967, rifkinCurrentphaseRelationPhasedependent1976, frolovMeasurementCurrentphaseRelation2004, troemanTemperatureDependenceMeasurements2008a, hallerPhasedependentMicrowaveResponse2022,uhl2024extracting,scherublDeterminationCurrentphaseRelation2025}.
Finally, with a scanning-SQUID fabricated close to a superconducting loop containing the junction the phase-dependent supercurrent through the loop can be detected directly through the mutual inductance of the SQUID and the loop, thus allowing to map out the CPR of the junction~\cite{sochnikovDirectMeasurementCurrentPhase2013,sochnikovNonsinusoidalCurrentPhaseRelationship2015,spantonCurrentPhaseRelations2017,hartCurrentphaseRelationsInAs2019}.
All these methods have in common that they require the junction to be embedded in a SQUID or other specific superconducting circuits and it is often challenging to study the transport properties of the isolated junction \cite{uhl2024extracting}.

In tandem with transport measurements, Shapiro-step measurements have also been used to gain insight into the CPR of superconducting elements and partially address this issue, as they do not require the junction to be part of a SQUID or other complicating structure.
From the location of the (non-integer) Shapiro steps when subjecting a superconducting junction to ac driving, the relative weight of the first and second harmonic of the CPR can be determined~\cite{sellierHalfIntegerShapiroSteps2004, frolovJosephsonInterferometryShapiro2006, stoutimoreSecondHarmonicCurrentPhaseRelation2018, uedaEvidenceHalfintegerShapiro2020, leblancNonreciprocalCharge4eSupercurrent2024,valentiniParityconservingCooperpairTransport2024,zhaoTimereversalSymmetryBreaking2023, tsarevAllFractionalShapiro2025, iorioHalfintegerShapiroSteps2023a} and the observation of missing integer Shapiro steps has been attributed to a period doubling that can be related to the expected $4\pi$-periodicity of the CPR of a topological Josephson junction~\cite{rokhinsonFractionalAcJosephson2012,wiedenmann4pperiodicJosephsonSupercurrent2016,bocquillonGaplessAndreevBound2017,li4pperiodicAndreevBound2018,dartiailhMissingShapiroSteps2021, elfekyEvolution4pPeriodicSupercurrent2023}.

In this work we present a method for extracting the CPR of a superconducting junction that (i) only involves the junction itself and (ii) allows to extract the full harmonic content of the CPR.
Similar to the Shapiro-step method it relies on ac driving, but, in contrast, it operates in the fast-driving regime and only requires to measure the critical current of the junction.
We test the method in numerical simulations and in an experiment, both showing great promise for this method as an accurate way to measure the CPR of any superconducting junction.
The rest of this paper is structured as follows:
We first present the theory that underpins the basic idea of the method. 
Then we show a proof-of-principle application of the method to numerically generated data, using CPRs with up to four harmonics.
We then present a simplified analytic approach for the case where only two harmonics are significant and demonstrate how it works.
Finally, we apply the method to experimental data measured on a germanium-based Josephson junction and we show how the extracted CPRs agree well with expectations.

\begin{figure}[b!]
    \begin{center}
    \includegraphics[width=0.47\textwidth \vspace{-2.5mm}]{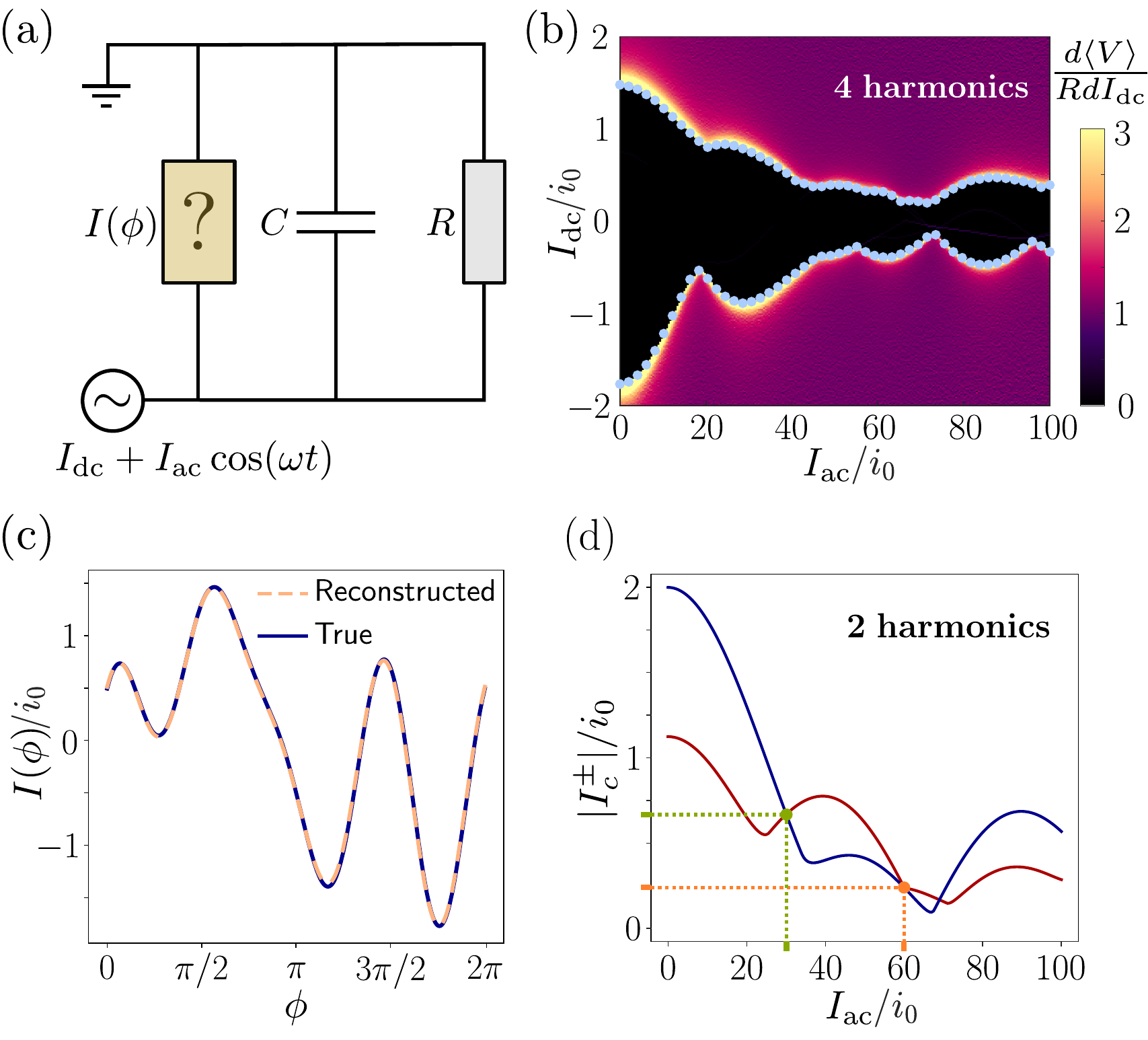}
    \end{center}
    \caption{(a) RCSJ model describing an ac-driven superconducting element with an unknown CPR, characterized by a resistance $R$ and capacitance $C$. (b) Numerical calculation of the average differential resistance $d\langle V \rangle / RdI_\text{dc}$ in the overdamped limit, as a function of $I_\text{ac}$ and $I_\text{dc}$, for a CPR with four harmonics. Blue dots show the boundaries of the zero-voltage region extracted from the numerical data. (c) The reconstructed CPR (dashed orange) after fitting the blue dots to Eqs.~(\ref{boundary_plus},\ref{boundary_minus}) and the true CPR used to generate (b) (dark blue). (d) Illustration of the analytic method to extract a CPR with only two harmonics. The blue and red curves show $I_c^+(I_\text{ac})$ and $-I_c^-(I_\text{ac})$, respectively, and the dashed lines indicate the values of $I_c^\pm$ that need to be read off. The CPRs used in (b) and (d) are not the same and are arbitrarily chosen.}
    \label{fig:IV_boundary}
\end{figure}

\section{Method}

\subsection{Idea}

We model the junction by the electrical circuit sketched in Fig.~\ref{fig:IV_boundary}(a), including the capacitance $C$ and normal-state resistance $R$ of the junction, connecting it to a current source via which both a dc and an ac current can be applied.
We thus describe the dynamics of the junction using the resistively and capacitively shunted junction (RCSJ) model,
\begin{equation} \label{phase_diff_eq}
    \tau_{RC}\ddot \phi + \Dot{\phi}+I(\phi)=I_\text{dc}+I_\text{ac} \cos(\omega t),
\end{equation}
where $\phi$ is the superconducting phase difference over the junction, which is related to the voltage over the junction by the second Josephson relation $V = (\hbar/2e)\dot\phi$, and $\tau_{RC} = RC$ is the $RC$ time of the circuit.
We renormalized all currents by a factor $\hbar/2eR$ such that all terms in Eq.~(\ref{phase_diff_eq}) have dimensions ${\rm s}^{-1}$.
The CPR of the junction $I(\phi)$, which we would like to extract, can be expanded as
\begin{equation}
    I(\phi)=\sum_{m\geq1}^{m_\text{max}}I_m \sin(m\phi+\gamma_m),
\end{equation}
using that it must be $2\pi$-periodic in $\phi$ and assuming that only the lowest $m_\text{max}$ Fourier components are significant.

We analyze the behavior of the junction in the fast-driving regime, which in this case means
\begin{equation}
    \omega \gg \frac{1}{\sqrt 2 \tau_{RC}} \sqrt{\sqrt{1+4(\tau_{RC}^2/\tau_{\rm J}^2)} - 1},
\end{equation}
i.e., the driving period being much shorter than both the $RC$-time of the circuit and $\tau_{\rm J} = \hbar/2eRI_c$, the ``Josephson'' time, i.e., the $LR$ time of the Josephson inductance and the resistance, where $I_c$ is the typical critical current of the junction.
Following the derivation presented in Ref.~\cite{seoanesoutoTuningJosephsonDiode2024}, one can straightforwardly find analytic expressions for the maximally supported supercurrents in the forward ($I_c^+$) and backward ($I_c^-$) directions in this limit,
\begin{align} \label{boundary_plus}
    I_{c}^+(I_\text{ac}) = {} & {} \! \max_{\alpha}\left\{\sum_{m\geq1} \! I_m\sin(m\alpha+ \gamma_m)J_0\left(\! m \frac{I_\text{ac}}{\tilde\omega}\!\right)\!\right\},\!\\
\label{boundary_minus}
    I_{c}^-(I_\text{ac}) = {} & {}\! \min_{\alpha}\left\{\sum_{m\geq1}\! I_m\sin(m\alpha+ \gamma_m)J_0\left( \! m \frac{I_\text{ac}}{\tilde\omega}\! \right)\!\right\},
\end{align}
where $J_0(x)$ is the zeroth Bessel function of the first kind and $\tilde\omega = \omega\sqrt{\tau_{RC}^2\omega^2 + 1}$ is the renormalized driving frequency.
Since the boundaries of the superconducting zero-voltage region in the $( I_\text{ac},I_\text{dc})$-plane thus depend in a non-trivial way on all Fourier coefficients $\{I_m,\gamma_m\}$ of the CPR, we suggest that fitting Eqs.~(\ref{boundary_plus},\ref{boundary_minus}) to an experimentally obtained set of boundaries can provide a way to extract the CPR of the junction.

The method presented only allows determination of the CPR up to a transformation $I(\phi) \to I(\pm\phi+\delta)$, where $\delta$ is an arbitrary phase shift.
Indeed, the boundaries as described by Eqs.~(\ref{boundary_plus},\ref{boundary_minus}) are invariant under a shift $\alpha \to \alpha + \delta$, meaning that we can determine the CPR only up to an arbitrary phase shift $\phi \to \phi + \delta$.
This also implies that it is sufficient to only find $m_\text{max}-1$ of the $\gamma_m$, and we can fix, e.g., $\gamma_1=0$.
Secondly, the boundaries  are also invariant under the change $\alpha \to -\alpha$, which means that we are insensitive to an overall sign of the phase, i.e., we cannot distinguish between $I(\phi)$ and $I(-\phi)$.
Finally, we note that the change $I_m \rightarrow - I_m$ is equivalent to a shift of the phase $\gamma_m \rightarrow \gamma_m-\pi$, which allows us to only consider non-negative $I_m\geq0$ in the fitting procedure and let the sign of $I_m$ be encoded in the phase $\gamma_m$.

\subsection{Proof of principle}

We start by showing a proof of principle of our idea based on numerical simulations.
For simplicity we will consistently work in the overdamped limit from now on, where $\tau_{RC} \ll \tau_{\rm J}$, i.e., we neglect the effect of the capacitance $C$ of the junction and thus set $\tilde \omega \to \omega$.
For cases where the capacitance is large, one can use the exact same approach, substituting $\omega$ by the renormalized frequency $\tilde \omega$.
Equation \eqref{phase_diff_eq} can straightforwardly be solved to obtain the $IV$ characteristics of the circuit as a function of driving frequency $\omega$ and strength $I_\text{ac}$.
In Fig.~\ref{fig:IV_boundary}(b) we show the calculated normalized differential resistance $d\langle V \rangle / RdI_\text{dc}$, where $\langle V \rangle$ is the voltage averaged over many driving periods.
In this particular case, we used $I(\phi)/i_0 = \frac{4}{5}\sin(\phi) + \frac{1}{2}\sin(2\phi-\frac{\pi}{2}) + \frac{7}{10} \sin(3\phi+\frac{\pi}{4}) + \frac{1}{2}\sin(4\phi + \frac{\pi}{2})$ and $\omega = 25 \, i_0 \sim 25 \tau_{\rm J}^{-1}$ to be in the fast-driving regime.
We see a clear superconducting zero-voltage region at small $I_\text{dc}$, sharply transitioning into a region of finite resistance for larger $I_\text{dc}$.

We first extract the boundaries of the zero-voltage region, by thresholding the numerical data on an equally spaced grid of 50 values of $I_\text{ac}$; the blue dots in Fig.~\ref{fig:IV_boundary}(b) show the result.
Next, we fit these 100 points to Eqs.~(\ref{boundary_plus},\ref{boundary_minus}).
To this end, we define the total mean square error of the extracted points compared to the two fitted boundaries as loss function, which then should be minimized.
Since this loss function can feature many local minima for CPRs with more than two harmonics, $m_{\mbox{\footnotesize{max}}} >2$, we choose the gradient-free covariance matrix adaptation evolution strategy (CMA-ES)~\cite{hansenCMAEvolutionStrategy2016} as optimization scheme (implemented via the \texttt{CMA} python package, using its default settings).
For the example data represented by the blue dots in Fig.~\ref{fig:IV_boundary}(b), this yields
\begin{align*}
        I_1 & =0.79\, i_0, & I_2 & =0.49\, i_0, & I_3 & =0.7\, i_0, & I_4 & =0.5\, i_0, \\
        \gamma_1 & = 0, & \gamma_2 & =-0.51\pi, & \gamma_3 & =0.26\pi, & \gamma_4 & =0.5\pi,
\end{align*}
where we fixed $\gamma_1 = 0$.
The resulting ``reconstructed'' CPR is plotted in orange in Fig.~\ref{fig:IV_boundary}(c) and compared to the true CPR underlying the data shown in Fig.~\ref{fig:IV_boundary}(b) (blue curve).
We see that the two CPRs agree very well, indicating that in theory our proposed method is indeed capable of extracting the CPR of a superconducting junction from a single fast-driving experiment.
We find that the method is successful at consistently determining the correct CPR up to at least $m_{\mbox{\footnotesize{max}}}=5$~\footnote{We use three ``restarts'' of the  of the CMA-ES algorithm (i.e., three independent consecutive fitting attempts) for each fit, which leads to convergence to the global minimum around 99 out of 100 times. For larger $m_\text{max}$ or a higher desired success rate, it could be beneficial to increase the number of restarts.}.

\subsection{Analytic method for two harmonics}

Alternatively, if the CPR is dominated by the first two harmonics, then it is possible to connect its Fourier coefficients more directly to the boundaries of the zero-voltage region.
With $z_0$ being the first zero of the Bessel function, $J_0(z_0)=0$, it is straightforward to see from Eqs.~(\ref{boundary_plus},\ref{boundary_minus}) that
\begin{equation} \label{Im_analytical}
    I_1=\frac{I_{c}^\pm(\omega z_0/2)}{|J_0\left(z_0/2\right)|} \quad\text{and}\quad I_2=\frac{I_{c}^\pm(\omega z_0)}{|J_0\left(2 z_0\right)|}.
\end{equation}
Often the $I_\text{ac}$ being driven through the junction cannot be controlled quantitatively (in an experiment).
In that case, the points where $I_\text{ac}=z_0\omega/2, z_0\omega$ can be identified as the first two points where $I^+_c(I_\text{ac})=-I^-_c(I_\text{ac})$~\footnote{This of course assumes that the boundaries are not symmetric around $I_\text{dc}=0$, i.e., $I^+_c(I_\text{ac})=-I^-_c(I_\text{ac})$ for all $I_\text{ac}$, which happens when there is only one dominating harmonic or when $\gamma_2=n\pi$}.
In Fig.~\ref{fig:IV_boundary}(d) we show an example of this simple approach.
The blue and red curves show $I_c^+(I_\text{ac})$ and $-I_c^-(I_\text{ac})$, respectively, using the CPR $I(\phi)/i_0 = \sin(\phi) + \sin(2\phi - \frac{\pi}{2})$, in the same fast-driving limit $\omega = 25\,i_0$.
The green and orange dots indicate the first two crossings of the curves and yield $I_c^+(\omega z_0/2) = 0.67\, i_0$ and $I_c^+(\omega z_0) = 0.24\,i_0$.
Using Eq.~(\ref{Im_analytical}) we find $I_1 = 1.00\, i_0$ and $I_2 = 1.01\,i_0$, both indeed very close to the true values.

The relative phase $\gamma_2$ can then be determined to very good approximation as
\begin{equation}
    \gamma_2 = \pi \left(1\mp \frac{1}{2} \right) + 2\arcsin \left( \pm \frac{8 I_c^\pm(\tilde I_\text{ac})}{7\tilde I} - \frac{9}{7} \right),\label{eq:analgamma2}
\end{equation}
where $\tilde I_\text{ac}$ is the driving strength for which $I_1 J_0(\tilde I_\text{ac}/\omega) = I_2 J_0(2\tilde I_\text{ac}/\omega) \equiv \tilde I$.
Whether one picks the upper or lower sign in Eq.~(\ref{eq:analgamma2}) makes in principle no difference; in practice, the sign for which the argument of the arcsin is closest to zero, i.e., for which $|I_c^\pm(\tilde I_\text{ac})|/\tilde I$ is closest to $\frac{9}{8}$, makes the outcome the least sensitive to errors.
For the example shown in Fig.~\ref{fig:IV_boundary}(d) we find from Eq.~(\ref{eq:analgamma2}) $\gamma_2 = 1.48\,\pi$, again very close to the true value.
See the Appendix.~\ref{appendix:A} for more details and a derivation of Eq.~(\ref{eq:analgamma2}).

\begin{figure}[t!]
    \begin{center}
    \includegraphics[width=0.48\textwidth \vspace{-2.5mm}]{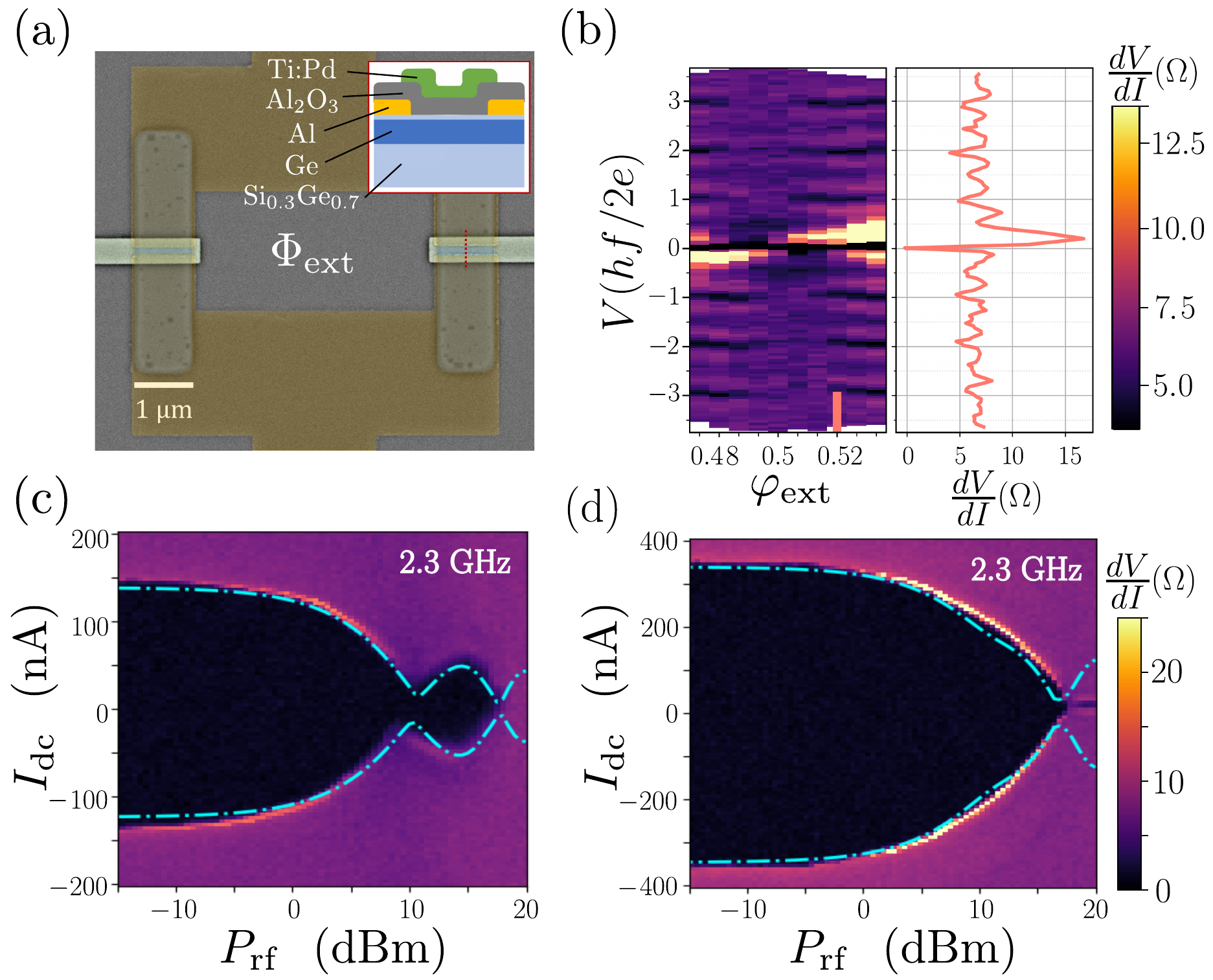}
    \end{center}
    \caption{(a) SEM image of the planar Ge SQUID used for the experiment.
    Inset shows the material stack used to realize the junctions.
    (b) Differential resistance measurement of the balanced SQUID at a constant drive power ($P_{\text{rf}} = 15\,\mathrm{dBm}$) in the slow driving regime ($f=230\,\mathrm{MHz}$) as a function of $V_\text{dc}$ and $\varphi_\text{ext}$, around half flux quantum ($0.5\,\Phi_0$) threading the loop. Dips in resistance corresponding to integer and half-integer Shapiro steps are seen, with the $\frac{1}{2}$ step becoming prominent close to $\varphi_{\rm ext} = 0.5$. A 1D slice taken at $\varphi_{\rm ext} = 0.52$ is shown. (c,d) Differential resistance measured as a function of $I_\text{dc}$ and $P_\text{rf}$: Fast-driving spectroscopy with $f=2.3\;\mathrm{GHz}$ at $\varphi_\text{ext} = 0.5$ and $\varphi_\text{ext} = 0.38$, respectively. The cyan dash-dotted curves show fits to the re-trapping current envelopes according to Eqs.~(\ref{boundary_plus},\ref{boundary_minus}).}
    \label{fig:exp}
\end{figure}

\section{Experimental implementation}

We test the experimental applicability of our method by studying the CPR of a germanium-based SQUID, comprised of two gate-tunable Josephson field effect transistors (JoFETs), see Fig.~\ref{fig:exp}(a).
We emphasize that in this case the SQUID itself is the element we want to determine the CPR of; it is not used as an interferometric tool to determine the CPR of one of the JoFETs.
We use Al top contacts on an ultra-shallow Ge quantum well [inset in Fig.~\ref{fig:exp}(a)], similar to what was done in Ref.~\cite{valentiniParityconservingCooperpairTransport2024}.
This device offers two key advantages: (i) the high-transparency superconducting contacts lead to a sizable second-harmonic contribution to the single-JoFET CPR and (ii) the combined gate- and flux-tunability of the device allows for a high degree of control over the effective CPR of the SQUID~\cite{leblancNonreciprocalCharge4eSupercurrent2024}.
The two JoFETs forming the SQUID can be tuned to a configuration where the first harmonic components of their individual CPRs are equal~\cite{valentiniParityconservingCooperpairTransport2024}.
In this configuration the total CPR of the SQUID can be tuned to be predominantly $\sim \sin(2\phi)$, by threading half a flux quantum through the SQUID loop ($\varphi_\text{ext} = \Phi_\text{ext}/\Phi_0 = 0.5$), causing the balanced first harmonics to interfere destructively.

Around $\varphi_\text{ext} = 0.5$, we first perform a flux-biased measurement in the slow-driving regime ($\omega\tau_{\rm J}<1$).
We operate the device in the overdamped limit by using an additional resistive shunt $R_{s} = 10\,\Omega$ ($\ll R$) and induce an ac current in the device via an antenna close to the chip, see the Appendix~\ref{appendix:D} for more details.
Fig.~\ref{fig:exp}(b) shows the measured differential resistance as a function of $V_\text{dc}$ and $\varphi_\text{ext}$~\footnote{In practice, we did a current-biased measurement and numerically reconstructed $V_\text{dc}(I_\text{dc})=\int_0^{I_\text{dc}} \left( dV/dI \right)_\text{meas} dI$. Plotted as a function of $V_\text{dc}$, Shapiro steps are easy to identify as horizontal resistance dips in the measured differential resistance map.}, for a constant applied power $P_\text{rf} = 15\,\mathrm{dBm}$ at frequency $f=230\,\mathrm{MHz}$ (implying $\omega\tau_{\rm J}\simeq 0.32$~\footnote{$\omega\tau_\text{J}=hf/2eI_cR\simeq 0.32$, for $I_c\simeq 150\,\mathrm{nA}$, $R=10\,\Omega$, and $f = 230\,\mathrm{MHz}$, which is the case for half a flux quantum ($\varphi_{\text{ext}} = 0.5$) threading the loop}).
Integer and half-integer Shapiro steps manifest as dips in the differential resistance at voltages $V_n=\frac{1}{2}nhf/2e$, with $n\in \mathbb{Z}$, indicating a bi-harmonic CPR.
At $\varphi_\text{ext} = 0.5$, a half-integer Shapiro step is visible, while the integer steps become dominant away from this point. 
This is consistent with the picture that the second harmonic of the SQUID CPR dominates at $\varphi_\text{ext} = 0.5$, as explained above, while the CPR has a dominating first harmonic away form this balanced point.

We then turn to fast rf driving with $f=2.3\,\mathrm{GHz}$ ($\omega\tau_\text{J}\simeq 3.2$) to test the method for CPR extraction proposed in this work~\footnote{We also tried even higher frequencies and noticed that for $f \gtrsim 5.5\,$GHz the technique starts to fail. We believe that this could be related to Landau--Zener transitions induced by the driving, see the Appendix~\ref{appendix:D} for more details.}.
In Fig.~\ref{fig:exp}(c,d) we show the measured differential resistance as a function of rf power and dc bias current, at a flux bias of (c) $\varphi_\text{ext} = 0.5$ and (d) $\varphi_\text{ext} = 0.38$.
We extract the re-trapping current boundaries using thresholding, analogous to what was done in Fig.~\ref{fig:IV_boundary}(b).
These values are then fit to Eqs.~(\ref{boundary_plus},\ref{boundary_minus}) using again the CMA-ES algorithm, focusing on two harmonics.
In addition to the $\{I_m,\gamma_m\}$, we fit a scaling parameter $\beta$ that converts $P_\text{rf}$ (in $\mathrm{dBm}$) to $I_\text{ac}$ (in nA) as $I_\text{ac} = \beta\, 10^{P_\text{rf}/20}$.
Since the driving frequency is the same in the two data sets, we assume that the same value of $P_\text{rf}$ corresponds to the same $I_\mathrm{ac}$ in the two sets and thus use $\beta$ as a common fitting parameter for both sets.
This yielded the following reconstructed SQUID CPRs
\begin{align}
\begin{split}
    \text{(c):~~} I(\phi) = {} & {} 0.24\sin(\phi) + 1.15\sin(2\phi-0.93),\\
    \text{(d):~~} I(\phi) = {} & {} 2.78\sin(\phi) + 1.16\sin(2\phi+0.07),
\end{split}
    \label{experimental CPRs}
\end{align}
where all currents are expressed in units of $100\,\mathrm{nA}$, and $\beta = 0.36\,\text{nA}$. We see that at the balanced point (c) the first harmonic is indeed strongly suppressed, whereas it dominates at lower flux (d), consistent with our expectations and with the Shapiro steps observed in Fig.~\ref{fig:exp}(b) (see Appendix~\ref{appendix:C} for a more quantitative interpretation).
The cyan dash-dotted curves in Fig.~\ref{fig:exp}(c,d) show the resulting fast-driving critical currents calculated using Eqs.~(\ref{boundary_plus},\ref{boundary_minus}) using the fit parameters given in (\ref{experimental CPRs}), confirming good agreement of the fit.

\section{Discussion and Conclusion}

We presented a minimally invasive driving-based method for extracting the full harmonic content of a CPR of a superconducting junction.
The method is based on fitting the measured critical currents of the driven junction as a function of driving power to an analytic expression that is valid in the fast-driving regime.
We tested the method both using numerical simulations and in experiment, and we show that it works accurately.
Although the method allows extraction of an arbitrary number of harmonics in theory, it works best up to $\sim 5$ harmonics in practice.
Since the leading contribution to the $n^{\text{th}}$ harmonic is proportional to $T^n$, where $0\leq T \leq 1$ characterizes the transmission of the junction, our method works best for junctions with non-perfect transparency $(T<1)$ \footnote{Taylor expanding the Beenakker formula for ABS energy $E(\varphi)=\Delta\sqrt{1-T\sin^2(\phi/2)}$, for $T=0.7$, typical for SNS junctions, gives $E(\phi)\propto[\cos(\phi) - 0.072\cos(2\phi)+0.011\cos(3\phi)+...\,]$, with $I(\phi)=(2e/h)\partial_\phi E(\phi)$. Here, truncation to two harmonics is fairly accurate. Even for $T=0.95$, the CMA-ES fit can be implemented if errors of $\sim I_6/I_1=4.7\%$ are tolerable.}.

At higher powers, dissipation ($\sim I_\text{ac}^2R$)~\cite{seoanesoutoTuningJosephsonDiode2024} may start having effects on the effective sample temperature.
In practice, this often manifests as smearing of higher-order Shapiro lobes.
For small critical currents, even within the zeroth Shapiro lobe, related effects may smear out the superconducting-to-normal transition, making it challenging to estimate the critical current envelopes.
This can be seen as a source of systematic noise, which leads to an underestimation of the critical current envelopes at higher power [which could be related to the high-power mismatch between the experimental and fitted boundaries in Figs.\ref{fig:exp}(c,d)].
Potential high-power sources of noise can be circumvented by only considering the low-power data in the first Shapiro lobe.
In the Appendix~\ref{appendix:B} we demonstrate this approach, extracting a reasonably accurate reconstruction of the CPR using a truncated set of data points.

Finally, we comment on our use of a shunt resistor $R_s$, which has two purposes:
(i) It brought us to the overdamped limit, where the voltage--current characteristic of the junction is non-hysteretic, with the switching and retrapping currents overlapping.
This greatly simplifies the estimation of the critical current, which for a hysteretic (underdamped) junction would require careful measurement of the switching statistics~\cite{Switching_current_wallraff}.
(ii) Reaching fast driving requires, $2eI_c R < hf$, where the equivalent resistance entering the RCSJ model is $R^{-1} = R_\text{qp}^{-1} + R_s^{-1}$ with $R_\text{qp}$ the quasiparticle resistance contributed by unpaired carriers \cite{Gloos1999}. For high-transparency SNS junctions, $R_\text{qp}\sim R_N-10R_N$, $R_N$ being the junction's normal-state resistance, and $R_\text{qp}$ increases for decreasing interface transparency. For very low transparency SIS junctions $R_\text{qp}$ can be a few orders of magnitudes larger than $R_N$ at $\text{mK}$ temperatures. Although direct measurement of $R_\text{qp}$ is challenging, $R_N$ can be used as a lower bound to estimate frequencies needed to reach fast driving.
For short SNS junctions $2eI_cR_N\sim\Delta$~\cite{beenakkerJosephsonCurrentSuperconducting1991}, meaning that fast driving would imply $hf\gtrsim \Delta$ without a shunt, rendering the RCSJ model inadequate to describe the junction dynamics, as Landau--Zener tunneling~\cite{ac_Josephson_Effect_Averin} and other transitions between the two Andreev bands become relevant for describing the junction response~\cite{Josephson_dynamics_Galaktionov}. 
Thus, with a small shunt $R_s \ll R_N$ we make $R \approx R_s$, and for the critical currents measured for our devices the fast-driving regime $hf \gg I_cR_s$ could then be reached at much lower frequencies of the order of $\sim 1\,\text{GHz}\ll 2\Delta/h$.
\\

\begin{acknowledgments}

The authors thank Afonso Oliveira, Stefano Calcaterra, Daniel Chrastina and Giovanni Isella for providing the Ge/SiGe heterostructure. This work was supported by the Research Council of Norway through its Centers of Excellence funding scheme, Project No.~262633, ``QuSpin'', the FWF Project with DOI:10.55776/PAT7682124, OeAD ``ISTA Austria: ISTernship Summer Programme, Non EU countries'' (Ref. No.: MPC-2025-00945), the Spanish Comunidad de Madrid (CM) ``Talento Program'' (Project No. 2022-T1/IND-24070), and the Spanish Ministry of Science, Innovation, and Universities through Grants No.~CEX2024-001445-S (Severo Ochoa Centres of Excellence program) and No.~PID2022-140552NA-I00, funded by MCIN/AEI/10.13039/501100011033.

\end{acknowledgments}

\section*{Data availability}

The data that support the findings of this article are openly available at \href{https://doi.org/10.5281/zenodo.22646275}{https://doi.org/10.5281/zenodo.22646275}.

\appendix

\section{Analytic method for determining $\gamma_2$ for two harmonics}
\label{appendix:A}

In Eq.~\eqref{Im_analytical} we point out how $I_{1,2}$ can be directly read off from the boundaries of the zero-voltage region, for the case of two harmonics.
Knowing the coefficients $I_1$ and $I_2$, one can then solve the equation $I_1 J_0(\tilde I_\text{ac}/\omega) = I_2 J_0(2\tilde I_\text{ac}/\omega)$ numerically for $\tilde I_\text{ac}$.
At this particular driving strength $\tilde I_\text{ac}$, the critical currents are
\begin{widetext}
\begin{align}
    I_c^+(\tilde I_\text{ac}) = {} & {} \max_\alpha \tilde I\, [ \sin(\alpha) + \sin(2\alpha + \gamma_2) ]
    \approx \frac{9}{8}\tilde I + \frac{7}{8} \tilde I\left| \sin \left( \frac{\gamma_2}{2} - \frac{\pi}{4} \right) \right|,\\
    I_c^-(\tilde I_\text{ac}) = {} & {} \min_\alpha \tilde I \,[ \sin(\alpha) + \sin(2\alpha + \gamma_2) ]
    \approx - \frac{9}{8}\tilde I - \frac{7}{8}\tilde I \left| \sin \left( \frac{\gamma_2}{2} - \frac{3\pi}{4} \right) \right|,
\end{align}
\end{widetext}
where $\tilde I = I_1 J_0(\tilde I_\text{ac}/\omega) = I_2 J_0(2\tilde I_\text{ac}/\omega)$.
The approximate analytic expressions for $I_c^\pm(\tilde I_\text{ac})$ allow to extract
\begin{align}\label{eq:gamma2+}
    \gamma_2 = {} & {} \frac{\pi}{2} \pm 2\arcsin \left( \frac{8 I_c^+(\tilde I_\text{ac})}{7\tilde I} - \frac{9}{7} \right),\\
    \gamma_2 = {} & {} \frac{3\pi}{2} \pm 2\arcsin \left( -\frac{8I_c^-(\tilde I_\text{ac})}{7\tilde I} - \frac{9}{7} \right).\label{eq:gamma2-}
\end{align}
Whether one uses Eq.~(\ref{eq:gamma2+}) or (\ref{eq:gamma2-}) makes in principle no difference; in practice one should pick the expression where the argument of the arcsin is closest to zero, i.e., where $|I_c^\pm(\tilde I_\text{ac})|/\tilde I$ is closest to $\frac{9}{8}$, to make the outcome the least sensitive to errors.
The two distinct solutions in each expression yield identical CPRs up to the transformation $\phi \to \pi - \phi$, as can be easily checked by considering, e.g.,
\begin{align*}
    \begin{split}
        I_1 \sin(\phi)+ I_2 &\sin(2\phi + \tfrac{\pi}{2} + \delta) \\\xrightarrow{\phi \to \pi-\phi}& I_1 \sin(\phi)- I_2 \sin(2\phi - \tfrac{\pi}{2} - \delta) \\=& I_1 \sin(\phi)+ I_2 \sin(2\phi + \tfrac{\pi}{2} - \delta),
    \end{split}
\end{align*}
and similarly around $\gamma_2 = \tfrac{3\pi}{2}$.
Phases $\gamma_2 = \tfrac{\pi}{2} \pm \delta$ and $\gamma_2 = \tfrac{3\pi}{2} \pm \delta$ are thus indistinguishable based on Eqs.~(\ref{boundary_plus},\ref{boundary_minus}), as explained at the end of the Section ``Idea''.

In the example shown in Fig.~\ref{fig:IV_boundary}(d), we had determined $I_1 = 1.00\,i_0$ and $I_2 = 1.01\,i_0$, and we thus need to solve the equation
\begin{equation*}
    1.00\, J_0(\tilde I_\text{ac}/\omega) = 1.01\, J_0(2\tilde I_\text{ac}/\omega),
\end{equation*}
which yields $\tilde I_\text{ac} = 2.91\,i_0$ and thus $\tilde I = 1.00\,J_0(\tilde I_\text{ac}/\omega) = 0.997 i_0$ (using that $\omega = 25\,i_0$).
At this value of $\tilde I_\text{ac}= 2.91\,i_0$, one reads off from the boundaries shown in Fig.~\ref{fig:IV_boundary}(d) $I_c^+(\tilde I_\text{ac})=2.0\,i_0$ and $I_c^-(\tilde I_\text{ac})=-1.1\,i_0$
It is then easy to check that $|I_c^-(\tilde I_\text{ac})|/\tilde I$ is closer to $\frac{9}{8}$ than $|I_c^+(\tilde I_\text{ac})|/\tilde I$, and we thus use Eq.~(\ref{eq:gamma2-}) to extract $\gamma_2$, which yields $\gamma_2 = 1.48\,\pi$ or $\gamma_2 = 1.52\,\pi$ (depending on which sign you pick).
Since $\gamma_2$ should of course be read modulo $2\pi$, this is again very close to the true value $\gamma_2 = -\frac{\pi}{2}$ used in the simulations.

\section{Robustness to noise}
\label{appendix:B}

\subsection{Random fluctuations of the extracted $I_c^\pm$}\label{sec:noise}

\begin{figure}[t!]
    \begin{center}
    \includegraphics[width=0.49\textwidth \vspace{-2.5mm}]{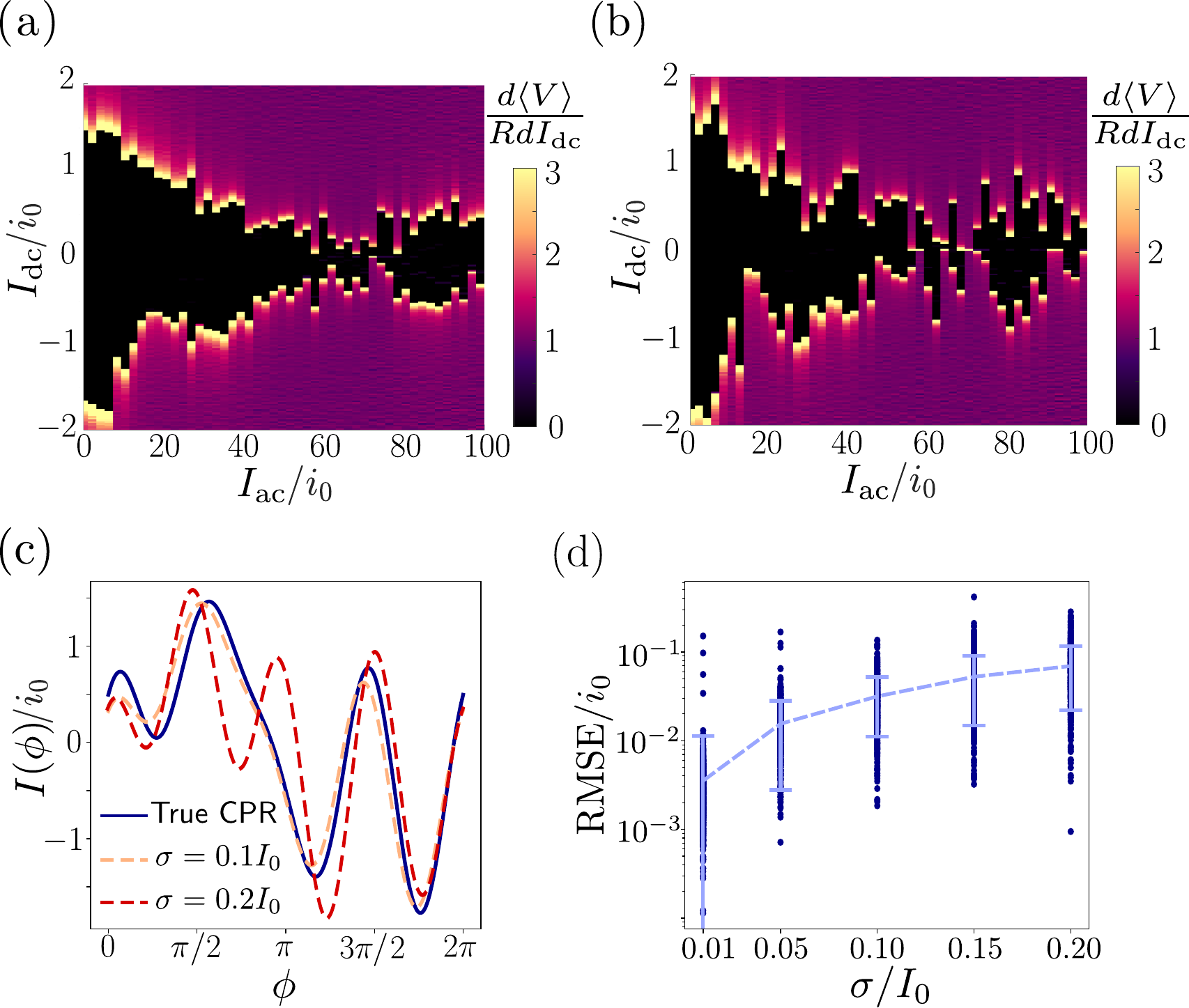}
    \end{center}
    \caption{Simulations of Shapiro spectroscopy in the presence of noise, where we used the same CPR as in Fig.~\ref{fig:IV_boundary}(b). For each $I_\text{ac}$ in the calculated differential resistance in (a,b) a random offset was added to $I_\text{dc}$, drawn from a zero-mean normal distribution with (a) $\sigma = 0.1\,I_0$ and (b) $\sigma = 0.2\,I_0$. (c) the corresponding reconstructed CPRs for $\sigma = 0.1\,I_0$ (dashed orange) and $\sigma = 0.2\,I_0$ (dashed red) compared to the true CPR (solid blue). (d) Error plot as a function of the noise intensity for $3$ restarts with $600$ measurements per noise point.}
    \label{fig:noise1}
\end{figure}

Next, we test the robustness of our method to noise, which is unavoidable in realistic experimental data.
First, we reproduce the numerical data shown in Fig.~\ref{fig:IV_boundary}(b), but now add random offsets $\delta I_\text{dc}$ to all $I_c^+$ and $I_c^-$ separately. This is done for each of the 50 values of $I_\text{ac}$, illustrated by the color plots shown in Fig.~\ref{fig:noise1}(a,b).
We pick the $\delta I_\text{dc}$ from a normal distribution with zero mean and used (a) $\sigma = 0.1\, I_0$ and (b) $\sigma = 0.2\, I_0$, where $I_0=\sqrt{\sum_m I_m^2}$, which equals $\approx 1.28\,i_0$ in this case \footnote{We use $I_0$ instead of $i_0$ for scaling $\sigma$ since we require it to scale with the randomly generated $I_m$ in Fig.~\ref{fig:noise1} for consistency.}.
For both data sets we again extract the (noisy) boundaries by thresholding the data and then fit these values to Eqs.~(\ref{boundary_plus},\ref{boundary_minus}) using the CMA-ES algorithm.
The reconstructed CPRs are shown in Fig.~\ref{fig:noise1}(c) for $\sigma = 0.1\,I_0$ (dashed orange) and $\sigma = 0.2\,I_0$ (dashed red) compared to the true CPR (solid blue). We see that in both cases the reconstructed CPR still resembles the true one, although with $\sigma = 0.2\,I_0$ the deviations start to become significant.

To study the effect of fluctuations in the extracted $I_c^\pm$ more systematically, we repeat the fitting procedure based on numerical data many times using a different CPR with three harmonics in each simulation.
The coefficients $I_m/i_0$ are drawn from a uniform distribution on $[0,1]$ and the $\gamma_m$ from a uniform distribution on $[0,2\pi)$.
The results are summarized in Fig.~\ref{fig:noise1}(d): 
We plot the root mean square error of the three estimated Fourier components $I_m$ for five different noise strengths $\sigma$, with 600 simulations per $\sigma$ (blue dots). 
The average error is highlighted with the dashed blue line and the error bars indicate sample standard deviation of the errors for each $\sigma$.
We focus on the errors in the $I_m$ since the errors in the $\gamma_m$ do not always correlate well with the accuracy of the estimated CPR, as explained in section \ref{section:phase_offsets}.
We see that the mean error stays below $0.1\,i_0$ up to $\sigma=0.2\,I_0$, and for $\sigma=0.1\,I_0$ all 600 simulations yielded results with an accuracy of $\lesssim 0.1\,i_0$.

\subsection{Heating}

\begin{figure}[t]
    \begin{center}
    \includegraphics[width=0.49\textwidth \vspace{-2.5mm}]{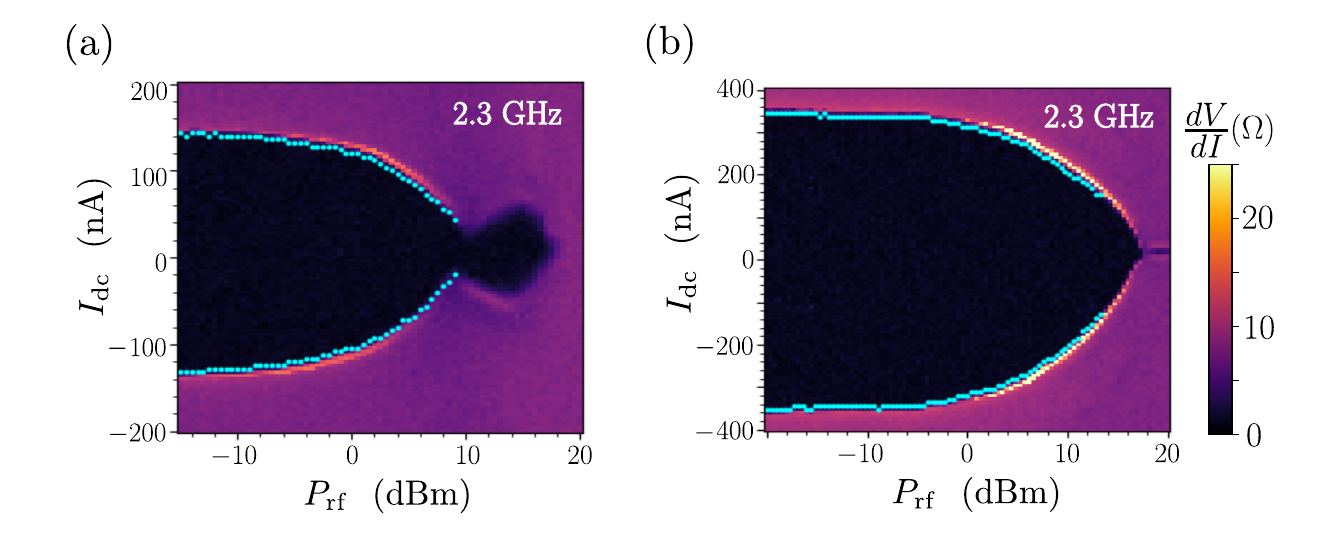}
    \end{center}
    \caption{(a,b) Differential resistance measured as a function of $I_\text{dc}$ and $P_\text{rf}$ using fast driving, $f=2.3\;\mathrm{GHz}$, at (a) $\varphi_\text{ext} = 0.5$ and (b) $\varphi_\text{ext} = 0.38$, corresponding to Figs.~\ref{fig:exp}(c,d), but with data points truncated at higher powers. Each cyan dot corresponds to a data point used for fitting.}
    \label{fig:noise2}
\end{figure}

Another systematic source of noise is presented by dissipation due to the ac driving ($\sim I_\text{ac}^2R$), which may start having effects on the effective sample temperature at higher powers~\cite{seoanesoutoTuningJosephsonDiode2024}.
In practice, this often manifests as smearing of higher-order Shapiro lobes.
For small critical currents, even within the zeroth Shapiro lobe, related effects may smear out the superconducting-to-normal transition, making it challenging to estimate the critical current envelopes.
This leads to an underestimation of the critical current envelopes.
Although this type of noise is taken into account by considering random offsets in $I_\text{dc}$ to some extent, it is clear that measurements at higher powers are more unreliable.
This issue can be circumvented by simply not including higher-power measurements in the fitting procedure.
For accurate CPR extraction one typically does not need several Shapiro lobes in the fitting procedure; somewhere between a half and a whole lobe is often sufficient.
To demonstrate this, we remove some of the highest-power data points from our experimental measurements before performing the fitting procedure, keeping only the data points highlighted in cyan in Fig.~\ref{fig:noise2}.
Using this set of data points, we extract the CPR
\begin{align}
    \varphi_\text{ext} = 0.50: \quad & I(\phi) = 0.16\,\sin(\phi) + 1.24\,\sin(2\phi+1.57),\\
    \varphi_\text{ext} = 0.38: \quad & I(\phi) = 2.89\,\sin(\phi) + 1.05\,\sin(2\phi - 0.06).
\end{align}
We see that the fitted harmonic content is still close to the results presented in \eqref{experimental CPRs}, the only significant deviation being the fitted phase $\gamma_2$ at $\varphi_{\rm ext} = 0.50$, which becomes indeed ill-defined when $I_1 \ll I_2$, see also below.

\subsection{Errors in the phase offsets $\gamma_m$} \label{section:phase_offsets}

When analyzing the robustness of our method to noise in Sec.~\ref{sec:noise} we only considered the error in the Fourier coefficients $I_m$. The reason for doing this is that the CPR becomes insensitive to deviations in a particular $\gamma_i$ when the corresponding coefficient $I_i$ is much smaller than one or more of the other Fourier coefficients (i.e., when $I_i\ll I_0$).
As a result, it is generally difficult to draw conclusions concerning the quality of the fitted CPR from considering the errors in the phases $\gamma_m$.

\begin{figure}[h!]
    \begin{center}
    \includegraphics[width=0.25\textwidth \vspace{-2.5mm}]{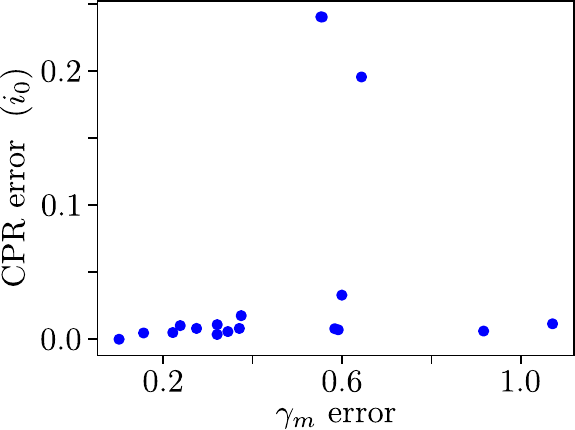}
    \end{center}
    \caption{The error in the CPR as a function of the corresponding $\gamma_m$ error for 18 CPRs with large errors in one or more $\gamma_m$.}
    \label{fig:gamma_m}
\end{figure}

In order to investigate this point more rigorously, we simulated our method again for $300$ randomly generated CPRs with three harmonics, and selected the $18$ fits that contained the $\gamma_m$ with the largest error.
We then compared the corresponding fitted CPRs $\hat I(\phi)$ to the true CPRs $I(\phi)$ by calculating the following error,
\begin{equation}
    \text{error}\left\{ \hat I(\phi),I(\phi)\right\} =\min_{\substack{\delta \in [0,2\pi) \\\sigma=\pm}}\left\{\mbox{mse}\left[\hat I(\phi),I(\sigma\phi+\delta)\right]\right\}.
\end{equation}
The result is shown in Fig.~\ref{fig:gamma_m}, where we see that error in the CPR is relatively small for most of the large-error $\gamma_m$.
We verified that the large-error $\gamma_m$ with a low CPR error indeed have a small corresponding Fourier amplitude $I_m$.
The three points in Fig.~\ref{fig:gamma_m} that show a large CPR error are in fact cases where the error in the $I_m$ is also large, that is, the few simulations (out of the 300) where the CMA-ES algorithm did not converge correctly.

\section{JoFET CPRs from fitting the SQUID interference pattern}
\label{appendix:C}
\begin{figure}[b!]
    \begin{center}
    \includegraphics[width=0.49\textwidth \vspace{-2.5mm}]{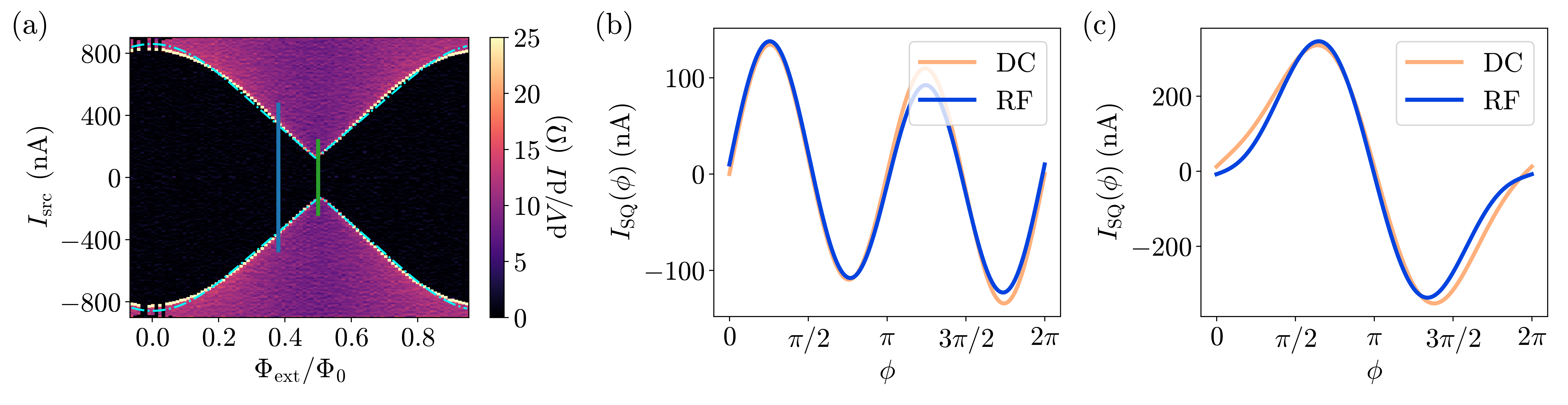}
    \end{center}
    \caption{(a) SQUID interference pattern in the absence of driving: differential resistance measured as a function of flux ($\Phi_{\text{ext}}$) threading the SQUID loop and dc bias current.
    The cyan dash-dotted curves show a fit to the SQUID interference pattern assuming biharmonic CPRs for the two JoFETs, as given in Eq.~(\ref{eq:biham}).
    (b,c) Extracted SQUID CPR from the dc SQUID pattern (orange) and the proposed rf Shapiro spectroscopy (blue), at (b) $\varphi_{\text{ext}} = 0.5$ and (c) $\varphi_{\text{ext}} = 0.38$.}
    \label{fig:SQUID_SM}
\end{figure}

Exploiting the sensitivity of the SQUID to the flux $\varphi_\text{ext}$, we can also try to use the measured SQUID interference pattern in the absence of driving, as shown in Fig.~\ref{fig:SQUID_SM}(a), to extract the CPR of the device.
This provides, in principle, an independent method for obtaining the CPR, the result of which can be compared to the fits presented in \eqref{experimental CPRs}, as a qualitative ``sanity check''.
For simplicity, we assume that the two JoFETs in the SQUID have the CPRs
\begin{align}
    I^{i}(\phi) = I_1^i\sin(\phi) + I_2^i\sin(2\phi), \quad \text{where }i\in\{L,R\}.\label{eq:biham}
\end{align}
The assumption of a predominantly biharmonic CPR is motivated by the slow-drive Shapiro measurement shown in Fig.~\ref{fig:exp}(b) and by prior works on similar devices using germanium SNS junctions~\cite{leblancGateFluxTunable2024,valentiniParityconservingCooperpairTransport2024}. 
In this case, the SQUID current as a function of the external flux and phase bias across the device can be written as
\begin{widetext}
\begin{align}
    I_{SQ}(\phi,\varphi_\text{ext})=I_1^L\sin(\phi)+I_1^R\sin(\phi+\varphi_{\text{ext}}2\pi) + I_2^L\sin(2\phi)+I_2^R\sin(2\phi+\varphi_{\text{ext}}4\pi).
    \label{eq:SQUID_CPR}
\end{align}
The flux-dependent SQUID critical currents can then be obtained as the extrema of $I_{SQ}$ over $\phi\in(-\pi,\pi]$,
\begin{equation}
    I_{SQ}^+ (\varphi_{\text{ext}})={} {} \max_\phi [I_{SQ}(\phi, \varphi_{\text{ext}})]
    \quad\text{and}\quad
    I_{SQ}^- (\varphi_{\text{ext}})={} {} \min_\phi [I_{SQ}(\phi, \varphi_{\text{ext}})].
    \label{eq:SQUID_critical_currents}
\end{equation}
\end{widetext}
These critical currents should describe the boundary of the zero-voltage region in Fig.~\ref{fig:SQUID_SM}(a), which shows the measured differential resistance as a function of external applied flux and applied dc current.
We thus use Eq.~(\ref{eq:SQUID_critical_currents}) to fit the re-trapping critical-current envelopes [cyan dash--dotted curves in Fig.~\ref{fig:SQUID_SM}(a)], which yields
\begin{equation}
    \begin{split}
        &I^L_1 = 421.7\,\text{nA}, \quad I^R_1=404.2\,\text{nA},\\&I^L_2= 63.1\,\text{nA},\quad I^R_2= 58.9\,\text{nA}.
    \end{split}
    \label{eq:SQUID_CPR_amplitudes}
\end{equation}
Using Eq.~(\ref{eq:SQUID_CPR}) we then find for the SQUID CPR at $\varphi_\text{ext}=0.5$ [green marker in Fig.~\ref{fig:SQUID_SM}(a)]  and $\varphi_\text{ext}=0.38$ [blue marker in Fig.~\ref{fig:SQUID_SM}(a)]
\begin{align}
    \varphi_\text{ext} = 0.50: \quad {} & {} I_{SQ}(\phi) = 0.175\,\sin(\phi) + 1.22\,\sin(2\phi),\label{fit1}\\
    \varphi_\text{ext} = 0.38: \quad {} & {} I_{SQ}(\phi) = 3.05\,\sin(\phi) + 0.89\,\sin(2\phi - 0.15),\label{fit2}
\end{align}
again in units of 100$\,$nA.
For visual comparison, we plot the CPRs obtained using the dc SQUID pattern (orange) and the proposed rf fast-driving spectroscopy technique (blue) in Figs.~\ref{fig:SQUID_SM}(b,c), where (b) presents the fits at $\varphi_\text{ext} = 0.5$ and (c) at $\varphi_\text{ext} = 0.38$
(as the global phase of the CPR cannot be extracted using the rf technique, phase shifts $\phi \rightarrow \pm \phi + \delta$ have been applied to maximize the overlap between the orange and blue curves).

Comparing the fits (\ref{fit1},\ref{fit2}) to the fitting results presented in \eqref{experimental CPRs}, we notice good agreement between the two methods: (i) The first harmonic is strongly suppressed at the balanced point $\varphi_\text{ext} = 0.5$, (ii) the weight of the second harmonic is roughly equal at the two flux biases, and (iii) the relative phase offset of the second harmonic ($\gamma_2$) at $\varphi_\text{ext} = 0.38$ is very small (it was 0.07 in the fit in \eqref{experimental CPRs}).
The apparent discrepancy between the phase shifts of the second harmonic at $\varphi_\text{ext}=0.5$ (zero here, $-0.93$ in the fit) is due to the problem of fitting the $\gamma_m$ of small Fourier components discussed in Sec.~\ref{section:phase_offsets}:
In this case, $\gamma_1$, which was set to zero by hand, can in fact not be determined reliably since $I_1$ is much smaller than $I_2$.
The phase $\gamma_2 = -0.93$ reported in \eqref{experimental CPRs} is thus arbitrary and could also be set to zero.
The slightly worse agreement at $\varphi_\text{ext} = 0.38$ [see Fig.~\ref{fig:SQUID_SM}(c)] could be related to a third-harmonic correction which destructively interferes at $\varphi_\text{ext} = 0.5$.
Additionally, the series inductance of the aluminum (of the order of $10\,\text{pH}$) might also affect the fitting to the dc SQUID pattern.

\section{Experimental considerations to include a shunt resistor}
\label{appendix:D}

In our setup we include a $R_s = 10\,\Omega$ shunt resistor placed in parallel to the junction.
We stress, however, that the resistor could be replaced by a commercial HEMT, to have a tunable shunt, which when pinched-off allows one to probe the device in isolation.
The reasons to use a small shunt resistance are twofold:

(i) The additional shunt resistor pushes the circuit effectively into the overdamped limit, where the voltage--current characteristic is non-hysteretic, i.e., the switching and retrapping currents coincide, $I_\text{ret}\simeq I_\text{sw}$.
This greatly simplifies the estimation of the critical current, which, for a hysteretic (underdamped) junction would require careful measurement of the switching statistics~\cite{Switching_current_wallraff}.
In our experiment, we measure retrapping envelopes.

(ii) We need to reach the fast-driving regime where $2eI_c R \ll hf$, with $R^{-1} = R_s^{-1} + R_\text{qp}^{-1}$ being the total resistance entering the RCSJ model, created by the quasiparticle resistance of the device $R_\text{qp}$ \cite{Gloos1999} in parallel with the shunt $R_s$. Typically, $R_\text{qp}$ for SNS junctions is larger than $R_N$, the normal state resistance. For low transparency SIS junctions $R_\text{qp}$ can be as high as few orders of magnitude of $R_N$ at mK temperatures. Thus, the easily measurable $R_N$ value can be used to estimate the lower bound of frequencies needed to reach fast-driving in absence of a shunt ~\cite{SNS_IcRn_Chrestin,furusakiTheoryQuantumConduction1991}. This means that in our case reaching fast driving without a shunt would require at least $hf\sim \Delta$, which is where the simple RCSJ model breaks down, as Landau--Zener tunneling~\cite{ac_Josephson_Effect_Averin} and other transitions between the two Andreev bands become relevant for describing the junction response~\cite{Josephson_dynamics_Galaktionov}.
Additionally, at very high frequencies the $RC$-properties of the setup might become frequency-dependent.
With a small shunt $R_s \ll R_N$ one has $R\approx R_s$, and for the critical currents measured for our devices, the fast-driving regime $hf \gtrsim I_cR_s$ can then be reached at much lower microwave frequencies of the order of $\sim 1\,\text{GHz}\ll 2\Delta/h$ \footnote{A typical induced gap in shallow Ge quantum wells is of the order of $\Delta \simeq 150\,\mathrm{\mu eV}$ \cite{valentiniParityconservingCooperpairTransport2024}, such that $2\Delta\equiv 72~\mathrm{GHz}$}, expected to be within the RCSJ regime.


\begin{thebibliography}{107}%
\makeatletter
\providecommand \@ifxundefined [1]{%
 \@ifx{#1\undefined}
}%
\providecommand \@ifnum [1]{%
 \ifnum #1\expandafter \@firstoftwo
 \else \expandafter \@secondoftwo
 \fi
}%
\providecommand \@ifx [1]{%
 \ifx #1\expandafter \@firstoftwo
 \else \expandafter \@secondoftwo
 \fi
}%
\providecommand \natexlab [1]{#1}%
\providecommand \enquote  [1]{``#1''}%
\providecommand \bibnamefont  [1]{#1}%
\providecommand \bibfnamefont [1]{#1}%
\providecommand \citenamefont [1]{#1}%
\providecommand \href@noop [0]{\@secondoftwo}%
\providecommand \href [0]{\begingroup \@sanitize@url \@href}%
\providecommand \@href[1]{\@@startlink{#1}\@@href}%
\providecommand \@@href[1]{\endgroup#1\@@endlink}%
\providecommand \@sanitize@url [0]{\catcode `\\12\catcode `\$12\catcode `\&12\catcode `\#12\catcode `\^12\catcode `\_12\catcode `\%12\relax}%
\providecommand \@@startlink[1]{}%
\providecommand \@@endlink[0]{}%
\providecommand \url  [0]{\begingroup\@sanitize@url \@url }%
\providecommand \@url [1]{\endgroup\@href {#1}{\urlprefix }}%
\providecommand \urlprefix  [0]{URL }%
\providecommand \Eprint [0]{\href }%
\providecommand \doibase [0]{https://doi.org/}%
\providecommand \selectlanguage [0]{\@gobble}%
\providecommand \bibinfo  [0]{\@secondoftwo}%
\providecommand \bibfield  [0]{\@secondoftwo}%
\providecommand \translation [1]{[#1]}%
\providecommand \BibitemOpen [0]{}%
\providecommand \bibitemStop [0]{}%
\providecommand \bibitemNoStop [0]{.\EOS\space}%
\providecommand \EOS [0]{\spacefactor3000\relax}%
\providecommand \BibitemShut  [1]{\csname bibitem#1\endcsname}%
\let\auto@bib@innerbib\@empty
\bibitem [{\citenamefont {Clarke}\ and\ \citenamefont {Wilhelm}(2008)}]{clarkeSuperconductingQuantumBits2008}%
  \BibitemOpen
  \bibfield  {author} {\bibinfo {author} {\bibfnamefont {J.}~\bibnamefont {Clarke}}\ and\ \bibinfo {author} {\bibfnamefont {F.~K.}\ \bibnamefont {Wilhelm}},\ }\bibfield  {title} {\bibinfo {title} {Superconducting quantum bits},\ }\href {https://doi.org/10.1038/nature07128} {\bibfield  {journal} {\bibinfo  {journal} {Nature}\ }\textbf {\bibinfo {volume} {453}},\ \bibinfo {pages} {1031} (\bibinfo {year} {2008})}\BibitemShut {NoStop}%
\bibitem [{\citenamefont {Wendin}(2017)}]{wendinQuantumInformationProcessing2017}%
  \BibitemOpen
  \bibfield  {author} {\bibinfo {author} {\bibfnamefont {G.}~\bibnamefont {Wendin}},\ }\bibfield  {title} {\bibinfo {title} {Quantum information processing with superconducting circuits: A review},\ }\href {https://doi.org/10.1088/1361-6633/aa7e1a} {\bibfield  {journal} {\bibinfo  {journal} {Rep. Prog. Phys.}\ }\textbf {\bibinfo {volume} {80}},\ \bibinfo {pages} {106001} (\bibinfo {year} {2017})}\BibitemShut {NoStop}%
\bibitem [{\citenamefont {Krantz}\ \emph {et~al.}(2019)\citenamefont {Krantz}, \citenamefont {Kjaergaard}, \citenamefont {Yan}, \citenamefont {Orlando}, \citenamefont {Gustavsson},\ and\ \citenamefont {Oliver}}]{krantzQuantumEngineerGuide2019}%
  \BibitemOpen
  \bibfield  {author} {\bibinfo {author} {\bibfnamefont {P.}~\bibnamefont {Krantz}}, \bibinfo {author} {\bibfnamefont {M.}~\bibnamefont {Kjaergaard}}, \bibinfo {author} {\bibfnamefont {F.}~\bibnamefont {Yan}}, \bibinfo {author} {\bibfnamefont {T.~P.}\ \bibnamefont {Orlando}}, \bibinfo {author} {\bibfnamefont {S.}~\bibnamefont {Gustavsson}},\ and\ \bibinfo {author} {\bibfnamefont {W.~D.}\ \bibnamefont {Oliver}},\ }\bibfield  {title} {\bibinfo {title} {A quantum engineer's guide to superconducting qubits},\ }\href {https://doi.org/10.1063/1.5089550} {\bibfield  {journal} {\bibinfo  {journal} {Appl. Phys. Rev.}\ }\textbf {\bibinfo {volume} {6}},\ \bibinfo {pages} {021318} (\bibinfo {year} {2019})}\BibitemShut {NoStop}%
\bibitem [{\citenamefont {Blais}\ \emph {et~al.}(2021)\citenamefont {Blais}, \citenamefont {Grimsmo}, \citenamefont {Girvin},\ and\ \citenamefont {Wallraff}}]{blaisCircuitQuantumElectrodynamics2020}%
  \BibitemOpen
  \bibfield  {author} {\bibinfo {author} {\bibfnamefont {A.}~\bibnamefont {Blais}}, \bibinfo {author} {\bibfnamefont {A.~L.}\ \bibnamefont {Grimsmo}}, \bibinfo {author} {\bibfnamefont {S.~M.}\ \bibnamefont {Girvin}},\ and\ \bibinfo {author} {\bibfnamefont {A.}~\bibnamefont {Wallraff}},\ }\bibfield  {title} {\bibinfo {title} {Circuit quantum electrodynamics},\ }\href {https://doi.org/10.1103/RevModPhys.93.025005} {\bibfield  {journal} {\bibinfo  {journal} {Rev. Mod. Phys.}\ }\textbf {\bibinfo {volume} {93}},\ \bibinfo {pages} {025005} (\bibinfo {year} {2021})}\BibitemShut {NoStop}%
\bibitem [{\citenamefont {Gramm}\ \emph {et~al.}(1976)\citenamefont {Gramm}, \citenamefont {Lundgren},\ and\ \citenamefont {Beckman}}]{kgrammSQUIDMagnetometerMangetization1976}%
  \BibitemOpen
  \bibfield  {author} {\bibinfo {author} {\bibfnamefont {K.}~\bibnamefont {Gramm}}, \bibinfo {author} {\bibfnamefont {L.}~\bibnamefont {Lundgren}},\ and\ \bibinfo {author} {\bibfnamefont {O.}~\bibnamefont {Beckman}},\ }\bibfield  {title} {\bibinfo {title} {{SQUID} magnetometer for magnetization measurements},\ }\href {https://doi.org/10.1088/0031-8949/13/2/004} {\bibfield  {journal} {\bibinfo  {journal} {Phys. Scr.}\ }\textbf {\bibinfo {volume} {13}},\ \bibinfo {pages} {93} (\bibinfo {year} {1976})}\BibitemShut {NoStop}%
\bibitem [{\citenamefont {Fagaly}(2006)}]{fagalySuperconductingQuantumInterference2006}%
  \BibitemOpen
  \bibfield  {author} {\bibinfo {author} {\bibfnamefont {R.~L.}\ \bibnamefont {Fagaly}},\ }\bibfield  {title} {\bibinfo {title} {Superconducting quantum interference device instruments and applications},\ }\href {https://doi.org/10.1063/1.2354545} {\bibfield  {journal} {\bibinfo  {journal} {Rev. Sci. Instrum.}\ }\textbf {\bibinfo {volume} {77}},\ \bibinfo {pages} {101101} (\bibinfo {year} {2006})}\BibitemShut {NoStop}%
\bibitem [{\citenamefont {Granata}\ and\ \citenamefont {Vettoliere}(2016)}]{granataNanoSuperconductingQuantum2016}%
  \BibitemOpen
  \bibfield  {author} {\bibinfo {author} {\bibfnamefont {C.}~\bibnamefont {Granata}}\ and\ \bibinfo {author} {\bibfnamefont {A.}~\bibnamefont {Vettoliere}},\ }\bibfield  {title} {\bibinfo {title} {Nano superconducting quantum interference device: A powerful tool for nanoscale investigations},\ }\href {https://doi.org/10.1016/j.physrep.2015.12.001} {\bibfield  {journal} {\bibinfo  {journal} {Phys. Rep.}\ }\textbf {\bibinfo {volume} {614}},\ \bibinfo {pages} {1} (\bibinfo {year} {2016})}\BibitemShut {NoStop}%
\bibitem [{\citenamefont {Degen}\ \emph {et~al.}(2017)\citenamefont {Degen}, \citenamefont {Reinhard},\ and\ \citenamefont {Cappellaro}}]{degenQuantumSensing2017}%
  \BibitemOpen
  \bibfield  {author} {\bibinfo {author} {\bibfnamefont {C.~L.}\ \bibnamefont {Degen}}, \bibinfo {author} {\bibfnamefont {F.}~\bibnamefont {Reinhard}},\ and\ \bibinfo {author} {\bibfnamefont {P.}~\bibnamefont {Cappellaro}},\ }\bibfield  {title} {\bibinfo {title} {Quantum sensing},\ }\href {https://doi.org/10.1103/RevModPhys.89.035002} {\bibfield  {journal} {\bibinfo  {journal} {Rev. Mod. Phys.}\ }\textbf {\bibinfo {volume} {89}},\ \bibinfo {pages} {035002} (\bibinfo {year} {2017})}\BibitemShut {NoStop}%
\bibitem [{\citenamefont {Benz}\ and\ \citenamefont {Hamilton}(2004)}]{benzApplicationJosephsonEffect2004}%
  \BibitemOpen
  \bibfield  {author} {\bibinfo {author} {\bibfnamefont {S.~P.}\ \bibnamefont {Benz}}\ and\ \bibinfo {author} {\bibfnamefont {C.~A.}\ \bibnamefont {Hamilton}},\ }\bibfield  {title} {\bibinfo {title} {Application of the {Josephson} effect to voltage metrology},\ }\href {https://doi.org/10.1109/JPROC.2004.833671} {\bibfield  {journal} {\bibinfo  {journal} {Proceedings of the IEEE}\ }\textbf {\bibinfo {volume} {92}},\ \bibinfo {pages} {1617} (\bibinfo {year} {2004})}\BibitemShut {NoStop}%
\bibitem [{\citenamefont {Scherer}\ and\ \citenamefont {Camarota}(2012)}]{schererQuantumMetrologyTriangle2012}%
  \BibitemOpen
  \bibfield  {author} {\bibinfo {author} {\bibfnamefont {H.}~\bibnamefont {Scherer}}\ and\ \bibinfo {author} {\bibfnamefont {B.}~\bibnamefont {Camarota}},\ }\bibfield  {title} {\bibinfo {title} {Quantum metrology triangle experiments: A status review},\ }\href {https://doi.org/10.1088/0957-0233/23/12/124010} {\bibfield  {journal} {\bibinfo  {journal} {Meas. Sci. Technol.}\ }\textbf {\bibinfo {volume} {23}},\ \bibinfo {pages} {124010} (\bibinfo {year} {2012})}\BibitemShut {NoStop}%
\bibitem [{\citenamefont {Tafuri}(2019)}]{tafuriFundamentalsFrontiersJosephson2019}%
  \BibitemOpen
  \bibinfo {editor} {\bibfnamefont {F.}~\bibnamefont {Tafuri}},\ ed.,\ \href {https://doi.org/10.1007/978-3-030-20726-7} {\emph {\bibinfo {title} {Fundamentals and {{Frontiers}} of the {{Josephson Effect}}}}},\ \bibinfo {series} {Springer {{Series}} in {{Materials Science}}}, Vol.\ \bibinfo {volume} {286}\ (\bibinfo  {publisher} {Springer International Publishing},\ \bibinfo {year} {2019})\BibitemShut {NoStop}%
\bibitem [{\citenamefont {{Castellanos-Beltran}}\ and\ \citenamefont {Lehnert}(2007)}]{castellanos-beltranWidelyTunableParametric2007}%
  \BibitemOpen
  \bibfield  {author} {\bibinfo {author} {\bibfnamefont {M.~A.}\ \bibnamefont {{Castellanos-Beltran}}}\ and\ \bibinfo {author} {\bibfnamefont {K.~W.}\ \bibnamefont {Lehnert}},\ }\bibfield  {title} {\bibinfo {title} {Widely tunable parametric amplifier based on a superconducting quantum interference device array resonator},\ }\href {https://doi.org/10.1063/1.2773988} {\bibfield  {journal} {\bibinfo  {journal} {Appl. Phys. Lett.}\ }\textbf {\bibinfo {volume} {91}},\ \bibinfo {pages} {83509} (\bibinfo {year} {2007})}\BibitemShut {NoStop}%
\bibitem [{\citenamefont {Braginski}(2019)}]{Braginski_JSNM2019}%
  \BibitemOpen
  \bibfield  {author} {\bibinfo {author} {\bibfnamefont {A.~I.}\ \bibnamefont {Braginski}},\ }\bibfield  {title} {\bibinfo {title} {Superconductor electronics: Status and outlook},\ }\href {https://doi.org/10.1007/s10948-018-4884-4} {\bibfield  {journal} {\bibinfo  {journal} {J. Supercond. Nov. Magn.}\ }\textbf {\bibinfo {volume} {32}},\ \bibinfo {pages} {23} (\bibinfo {year} {2019})}\BibitemShut {NoStop}%
\bibitem [{\citenamefont {Josephson}(1962)}]{josephsonPossibleNewEffects1962}%
  \BibitemOpen
  \bibfield  {author} {\bibinfo {author} {\bibfnamefont {B.~D.}\ \bibnamefont {Josephson}},\ }\bibfield  {title} {\bibinfo {title} {Possible new effects in superconductive tunnelling},\ }\href {https://doi.org/10.1016/0031-9163(62)91369-0} {\bibfield  {journal} {\bibinfo  {journal} {Phys. Lett.}\ }\textbf {\bibinfo {volume} {1}},\ \bibinfo {pages} {251} (\bibinfo {year} {1962})}\BibitemShut {NoStop}%
\bibitem [{\citenamefont {Galaktionov}\ and\ \citenamefont {Zaikin}(2023)}]{Josephson_dynamics_Galaktionov}%
  \BibitemOpen
  \bibfield  {author} {\bibinfo {author} {\bibfnamefont {A.~V.}\ \bibnamefont {Galaktionov}}\ and\ \bibinfo {author} {\bibfnamefont {A.~D.}\ \bibnamefont {Zaikin}},\ }\bibfield  {title} {\bibinfo {title} {Josephson dynamics at high transmissions: Perturbation theory},\ }\href {https://doi.org/10.1103/PhysRevB.107.214507} {\bibfield  {journal} {\bibinfo  {journal} {Phys. Rev. B}\ }\textbf {\bibinfo {volume} {107}},\ \bibinfo {pages} {214507} (\bibinfo {year} {2023})}\BibitemShut {NoStop}%
\bibitem [{\citenamefont {Willsch}\ \emph {et~al.}(2024)\citenamefont {Willsch}, \citenamefont {Rieger}, \citenamefont {Winkel}, \citenamefont {Willsch}, \citenamefont {Dickel}, \citenamefont {Krause}, \citenamefont {Ando}, \citenamefont {Lescanne}, \citenamefont {Leghtas}, \citenamefont {Bronn}, \citenamefont {Deb}, \citenamefont {Lanes}, \citenamefont {Minev}, \citenamefont {Dennig}, \citenamefont {Geisert}, \citenamefont {G{\"u}nzler}, \citenamefont {Ihssen}, \citenamefont {Paluch}, \citenamefont {Reisinger}, \citenamefont {Hanna}, \citenamefont {Bae}, \citenamefont {Sch{\"u}ffelgen}, \citenamefont {Gr{\"u}tzmacher}, \citenamefont {{Buimaga-Iarinca}}, \citenamefont {Morari}, \citenamefont {Wernsdorfer}, \citenamefont {DiVincenzo}, \citenamefont {Michielsen}, \citenamefont {Catelani},\ and\ \citenamefont {Pop}}]{willschObservationJosephsonHarmonics2024}%
  \BibitemOpen
  \bibfield  {author} {\bibinfo {author} {\bibfnamefont {D.}~\bibnamefont {Willsch}}, \bibinfo {author} {\bibfnamefont {D.}~\bibnamefont {Rieger}}, \bibinfo {author} {\bibfnamefont {P.}~\bibnamefont {Winkel}}, \bibinfo {author} {\bibfnamefont {M.}~\bibnamefont {Willsch}}, \bibinfo {author} {\bibfnamefont {C.}~\bibnamefont {Dickel}}, \bibinfo {author} {\bibfnamefont {J.}~\bibnamefont {Krause}}, \bibinfo {author} {\bibfnamefont {Y.}~\bibnamefont {Ando}}, \bibinfo {author} {\bibfnamefont {R.}~\bibnamefont {Lescanne}}, \bibinfo {author} {\bibfnamefont {Z.}~\bibnamefont {Leghtas}}, \bibinfo {author} {\bibfnamefont {N.~T.}\ \bibnamefont {Bronn}}, \bibinfo {author} {\bibfnamefont {P.}~\bibnamefont {Deb}}, \bibinfo {author} {\bibfnamefont {O.}~\bibnamefont {Lanes}}, \bibinfo {author} {\bibfnamefont {Z.~K.}\ \bibnamefont {Minev}}, \bibinfo {author} {\bibfnamefont {B.}~\bibnamefont {Dennig}}, \bibinfo {author} {\bibfnamefont {S.}~\bibnamefont {Geisert}}, \bibinfo {author} {\bibfnamefont {S.}~\bibnamefont
  {G{\"u}nzler}}, \bibinfo {author} {\bibfnamefont {S.}~\bibnamefont {Ihssen}}, \bibinfo {author} {\bibfnamefont {P.}~\bibnamefont {Paluch}}, \bibinfo {author} {\bibfnamefont {T.}~\bibnamefont {Reisinger}}, \bibinfo {author} {\bibfnamefont {R.}~\bibnamefont {Hanna}}, \bibinfo {author} {\bibfnamefont {J.~H.}\ \bibnamefont {Bae}}, \bibinfo {author} {\bibfnamefont {P.}~\bibnamefont {Sch{\"u}ffelgen}}, \bibinfo {author} {\bibfnamefont {D.}~\bibnamefont {Gr{\"u}tzmacher}}, \bibinfo {author} {\bibfnamefont {L.}~\bibnamefont {{Buimaga-Iarinca}}}, \bibinfo {author} {\bibfnamefont {C.}~\bibnamefont {Morari}}, \bibinfo {author} {\bibfnamefont {W.}~\bibnamefont {Wernsdorfer}}, \bibinfo {author} {\bibfnamefont {D.~P.}\ \bibnamefont {DiVincenzo}}, \bibinfo {author} {\bibfnamefont {K.}~\bibnamefont {Michielsen}}, \bibinfo {author} {\bibfnamefont {G.}~\bibnamefont {Catelani}},\ and\ \bibinfo {author} {\bibfnamefont {I.~M.}\ \bibnamefont {Pop}},\ }\bibfield  {title} {\bibinfo {title} {Observation of {Josephson} harmonics in
  tunnel junctions},\ }\href {https://doi.org/10.1038/s41567-024-02400-8} {\bibfield  {journal} {\bibinfo  {journal} {Nat. Phys.}\ }\textbf {\bibinfo {volume} {20}},\ \bibinfo {pages} {815} (\bibinfo {year} {2024})}\BibitemShut {NoStop}%
\bibitem [{\citenamefont {Baselmans}\ \emph {et~al.}(1999)\citenamefont {Baselmans}, \citenamefont {Morpurgo}, \citenamefont {{van Wees}},\ and\ \citenamefont {Klapwijk}}]{baselmansReversingDirectionSupercurrent1999}%
  \BibitemOpen
  \bibfield  {author} {\bibinfo {author} {\bibfnamefont {J.~J.~A.}\ \bibnamefont {Baselmans}}, \bibinfo {author} {\bibfnamefont {A.~F.}\ \bibnamefont {Morpurgo}}, \bibinfo {author} {\bibfnamefont {B.~J.}\ \bibnamefont {{van Wees}}},\ and\ \bibinfo {author} {\bibfnamefont {T.~M.}\ \bibnamefont {Klapwijk}},\ }\bibfield  {title} {\bibinfo {title} {Reversing the direction of the supercurrent in a controllable {Josephson} junction},\ }\href {https://doi.org/10.1038/16204} {\bibfield  {journal} {\bibinfo  {journal} {Nature}\ }\textbf {\bibinfo {volume} {397}},\ \bibinfo {pages} {43} (\bibinfo {year} {1999})}\BibitemShut {NoStop}%
\bibitem [{\citenamefont {Golubov}\ \emph {et~al.}(2004)\citenamefont {Golubov}, \citenamefont {Kupriyanov},\ and\ \citenamefont {Il'ichev}}]{golubovCurrentphaseRelationJosephson2004}%
  \BibitemOpen
  \bibfield  {author} {\bibinfo {author} {\bibfnamefont {A.~A.}\ \bibnamefont {Golubov}}, \bibinfo {author} {\bibfnamefont {M.~{\relax Yu}.}\ \bibnamefont {Kupriyanov}},\ and\ \bibinfo {author} {\bibfnamefont {E.}~\bibnamefont {Il'ichev}},\ }\bibfield  {title} {\bibinfo {title} {The current-phase relation in {{Josephson}} junctions},\ }\href {https://doi.org/10.1103/RevModPhys.76.411} {\bibfield  {journal} {\bibinfo  {journal} {Rev. Mod. Phys.}\ }\textbf {\bibinfo {volume} {76}},\ \bibinfo {pages} {411} (\bibinfo {year} {2004})}\BibitemShut {NoStop}%
\bibitem [{\citenamefont {Buzdin}(2005)}]{buzdinProximityEffectsSuperconductorferromagnet2005}%
  \BibitemOpen
  \bibfield  {author} {\bibinfo {author} {\bibfnamefont {A.~I.}\ \bibnamefont {Buzdin}},\ }\bibfield  {title} {\bibinfo {title} {Proximity effects in superconductor-ferromagnet heterostructures},\ }\href {https://doi.org/10.1103/revmodphys.77.935} {\bibfield  {journal} {\bibinfo  {journal} {Rev. Mod. Phys.}\ }\textbf {\bibinfo {volume} {77}},\ \bibinfo {pages} {935} (\bibinfo {year} {2005})}\BibitemShut {NoStop}%
\bibitem [{\citenamefont {Bergeret}\ \emph {et~al.}(2005)\citenamefont {Bergeret}, \citenamefont {Volkov},\ and\ \citenamefont {Efetov}}]{bergeretOddTripletSuperconductivity2005}%
  \BibitemOpen
  \bibfield  {author} {\bibinfo {author} {\bibfnamefont {F.~S.}\ \bibnamefont {Bergeret}}, \bibinfo {author} {\bibfnamefont {A.~F.}\ \bibnamefont {Volkov}},\ and\ \bibinfo {author} {\bibfnamefont {K.~B.}\ \bibnamefont {Efetov}},\ }\bibfield  {title} {\bibinfo {title} {Odd triplet superconductivity and related phenomena in superconductor-ferromagnet structures},\ }\href {https://doi.org/10.1103/RevModPhys.77.1321} {\bibfield  {journal} {\bibinfo  {journal} {Rev. Mod. Phys.}\ }\textbf {\bibinfo {volume} {77}},\ \bibinfo {pages} {1321} (\bibinfo {year} {2005})}\BibitemShut {NoStop}%
\bibitem [{\citenamefont {Linder}\ and\ \citenamefont {Robinson}(2015)}]{linderSuperconductingSpintronics2015}%
  \BibitemOpen
  \bibfield  {author} {\bibinfo {author} {\bibfnamefont {J.}~\bibnamefont {Linder}}\ and\ \bibinfo {author} {\bibfnamefont {J.~W.~A.}\ \bibnamefont {Robinson}},\ }\bibfield  {title} {\bibinfo {title} {Superconducting spintronics},\ }\href {https://doi.org/10.1038/nphys3242} {\bibfield  {journal} {\bibinfo  {journal} {Nat. Phys.}\ }\textbf {\bibinfo {volume} {11}},\ \bibinfo {pages} {307} (\bibinfo {year} {2015})}\BibitemShut {NoStop}%
\bibitem [{\citenamefont {Ryazanov}\ \emph {et~al.}(2001)\citenamefont {Ryazanov}, \citenamefont {Oboznov}, \citenamefont {Rusanov}, \citenamefont {Veretennikov}, \citenamefont {Golubov},\ and\ \citenamefont {Aarts}}]{ryazanovCouplingTwoSuperconductors2001}%
  \BibitemOpen
  \bibfield  {author} {\bibinfo {author} {\bibfnamefont {V.~V.}\ \bibnamefont {Ryazanov}}, \bibinfo {author} {\bibfnamefont {V.~A.}\ \bibnamefont {Oboznov}}, \bibinfo {author} {\bibfnamefont {A.~{\relax Yu}.}\ \bibnamefont {Rusanov}}, \bibinfo {author} {\bibfnamefont {A.~V.}\ \bibnamefont {Veretennikov}}, \bibinfo {author} {\bibfnamefont {A.~A.}\ \bibnamefont {Golubov}},\ and\ \bibinfo {author} {\bibfnamefont {J.}~\bibnamefont {Aarts}},\ }\bibfield  {title} {\bibinfo {title} {Coupling of two superconductors through a ferromagnet: {{Evidence}} for a {$\pi$} junction},\ }\href {https://doi.org/10.1103/physrevlett.86.2427} {\bibfield  {journal} {\bibinfo  {journal} {Phys. Rev. Lett.}\ }\textbf {\bibinfo {volume} {86}},\ \bibinfo {pages} {2427} (\bibinfo {year} {2001})}\BibitemShut {NoStop}%
\bibitem [{\citenamefont {Sellier}\ \emph {et~al.}(2004)\citenamefont {Sellier}, \citenamefont {Baraduc}, \citenamefont {Lefloch},\ and\ \citenamefont {Calemczuk}}]{sellierHalfIntegerShapiroSteps2004}%
  \BibitemOpen
  \bibfield  {author} {\bibinfo {author} {\bibfnamefont {H.}~\bibnamefont {Sellier}}, \bibinfo {author} {\bibfnamefont {C.}~\bibnamefont {Baraduc}}, \bibinfo {author} {\bibfnamefont {F.}~\bibnamefont {Lefloch}},\ and\ \bibinfo {author} {\bibfnamefont {R.}~\bibnamefont {Calemczuk}},\ }\bibfield  {title} {\bibinfo {title} {Half-integer {Shapiro} steps at the 0--{$\pi$} crossover of a ferromagnetic {Josephson} junction},\ }\href {https://doi.org/10.1103/PhysRevLett.92.257005} {\bibfield  {journal} {\bibinfo  {journal} {Phys. Rev. Lett.}\ }\textbf {\bibinfo {volume} {92}},\ \bibinfo {pages} {257005} (\bibinfo {year} {2004})}\BibitemShut {NoStop}%
\bibitem [{\citenamefont {Frolov}\ \emph {et~al.}(2006)\citenamefont {Frolov}, \citenamefont {Van~Harlingen}, \citenamefont {Bolginov}, \citenamefont {Oboznov},\ and\ \citenamefont {Ryazanov}}]{frolovJosephsonInterferometryShapiro2006}%
  \BibitemOpen
  \bibfield  {author} {\bibinfo {author} {\bibfnamefont {S.~M.}\ \bibnamefont {Frolov}}, \bibinfo {author} {\bibfnamefont {D.~J.}\ \bibnamefont {Van~Harlingen}}, \bibinfo {author} {\bibfnamefont {V.~V.}\ \bibnamefont {Bolginov}}, \bibinfo {author} {\bibfnamefont {V.~A.}\ \bibnamefont {Oboznov}},\ and\ \bibinfo {author} {\bibfnamefont {V.~V.}\ \bibnamefont {Ryazanov}},\ }\bibfield  {title} {\bibinfo {title} {Josephson interferometry and {{Shapiro}} step measurements of superconductor-ferromagnet-superconductor 0--{$\pi$} junctions},\ }\href {https://doi.org/10.1103/PhysRevB.74.020503} {\bibfield  {journal} {\bibinfo  {journal} {Phys. Rev. B}\ }\textbf {\bibinfo {volume} {74}},\ \bibinfo {pages} {020503} (\bibinfo {year} {2006})}\BibitemShut {NoStop}%
\bibitem [{\citenamefont {Stoutimore}\ \emph {et~al.}(2018)\citenamefont {Stoutimore}, \citenamefont {Rossolenko}, \citenamefont {Bolginov}, \citenamefont {Oboznov}, \citenamefont {Rusanov}, \citenamefont {Baranov}, \citenamefont {Pugach}, \citenamefont {Frolov}, \citenamefont {Ryazanov},\ and\ \citenamefont {Van~Harlingen}}]{stoutimoreSecondHarmonicCurrentPhaseRelation2018}%
  \BibitemOpen
  \bibfield  {author} {\bibinfo {author} {\bibfnamefont {M.~J.~A.}\ \bibnamefont {Stoutimore}}, \bibinfo {author} {\bibfnamefont {A.~N.}\ \bibnamefont {Rossolenko}}, \bibinfo {author} {\bibfnamefont {V.~V.}\ \bibnamefont {Bolginov}}, \bibinfo {author} {\bibfnamefont {V.~A.}\ \bibnamefont {Oboznov}}, \bibinfo {author} {\bibfnamefont {A.~Y.}\ \bibnamefont {Rusanov}}, \bibinfo {author} {\bibfnamefont {D.~S.}\ \bibnamefont {Baranov}}, \bibinfo {author} {\bibfnamefont {N.}~\bibnamefont {Pugach}}, \bibinfo {author} {\bibfnamefont {S.~M.}\ \bibnamefont {Frolov}}, \bibinfo {author} {\bibfnamefont {V.~V.}\ \bibnamefont {Ryazanov}},\ and\ \bibinfo {author} {\bibfnamefont {D.~J.}\ \bibnamefont {Van~Harlingen}},\ }\bibfield  {title} {\bibinfo {title} {Second-harmonic current-phase relation in {Josephson} junctions with ferromagnetic barriers},\ }\href {https://doi.org/10.1103/PhysRevLett.121.177702} {\bibfield  {journal} {\bibinfo  {journal} {Phys. Rev. Lett.}\ }\textbf {\bibinfo {volume} {121}},\ \bibinfo {pages}
  {177702} (\bibinfo {year} {2018})}\BibitemShut {NoStop}%
\bibitem [{\citenamefont {Lee}\ and\ \citenamefont {Lee}(2018)}]{leeProximityCouplingSuperconductorgraphene2018}%
  \BibitemOpen
  \bibfield  {author} {\bibinfo {author} {\bibfnamefont {G.-H.}\ \bibnamefont {Lee}}\ and\ \bibinfo {author} {\bibfnamefont {H.-J.}\ \bibnamefont {Lee}},\ }\bibfield  {title} {\bibinfo {title} {Proximity coupling in superconductor-graphene heterostructures},\ }\href {https://doi.org/10.1088/1361-6633/aaafe1} {\bibfield  {journal} {\bibinfo  {journal} {Rep. Prog. Phys.}\ }\textbf {\bibinfo {volume} {81}},\ \bibinfo {pages} {56502} (\bibinfo {year} {2018})}\BibitemShut {NoStop}%
\bibitem [{\citenamefont {Heersche}\ \emph {et~al.}(2007)\citenamefont {Heersche}, \citenamefont {{Jarillo-Herrero}}, \citenamefont {Oostinga}, \citenamefont {Vandersypen},\ and\ \citenamefont {Morpurgo}}]{heerscheBipolarSupercurrentGraphene2007}%
  \BibitemOpen
  \bibfield  {author} {\bibinfo {author} {\bibfnamefont {H.~B.}\ \bibnamefont {Heersche}}, \bibinfo {author} {\bibfnamefont {P.}~\bibnamefont {{Jarillo-Herrero}}}, \bibinfo {author} {\bibfnamefont {J.~B.}\ \bibnamefont {Oostinga}}, \bibinfo {author} {\bibfnamefont {L.~M.~K.}\ \bibnamefont {Vandersypen}},\ and\ \bibinfo {author} {\bibfnamefont {A.~F.}\ \bibnamefont {Morpurgo}},\ }\bibfield  {title} {\bibinfo {title} {Bipolar supercurrent in graphene},\ }\href {https://doi.org/10.1038/nature05555} {\bibfield  {journal} {\bibinfo  {journal} {Nature}\ }\textbf {\bibinfo {volume} {446}},\ \bibinfo {pages} {56} (\bibinfo {year} {2007})}\BibitemShut {NoStop}%
\bibitem [{\citenamefont {English}\ \emph {et~al.}(2016)\citenamefont {English}, \citenamefont {Hamilton}, \citenamefont {Chialvo}, \citenamefont {Moraru}, \citenamefont {Mason},\ and\ \citenamefont {Van~Harlingen}}]{englishObservationNonsinusoidalCurrentphase2016}%
  \BibitemOpen
  \bibfield  {author} {\bibinfo {author} {\bibfnamefont {C.~D.}\ \bibnamefont {English}}, \bibinfo {author} {\bibfnamefont {D.~R.}\ \bibnamefont {Hamilton}}, \bibinfo {author} {\bibfnamefont {C.}~\bibnamefont {Chialvo}}, \bibinfo {author} {\bibfnamefont {I.~C.}\ \bibnamefont {Moraru}}, \bibinfo {author} {\bibfnamefont {N.}~\bibnamefont {Mason}},\ and\ \bibinfo {author} {\bibfnamefont {D.~J.}\ \bibnamefont {Van~Harlingen}},\ }\bibfield  {title} {\bibinfo {title} {Observation of nonsinusoidal current-phase relation in graphene {{Josephson}} junctions},\ }\href {https://doi.org/10.1103/PhysRevB.94.115435} {\bibfield  {journal} {\bibinfo  {journal} {Phys. Rev. B}\ }\textbf {\bibinfo {volume} {94}},\ \bibinfo {pages} {115435} (\bibinfo {year} {2016})}\BibitemShut {NoStop}%
\bibitem [{\citenamefont {Nanda}\ \emph {et~al.}(2017)\citenamefont {Nanda}, \citenamefont {{Aguilera-Servin}}, \citenamefont {Rakyta}, \citenamefont {Korm{\'a}nyos}, \citenamefont {Kleiner}, \citenamefont {Koelle}, \citenamefont {Watanabe}, \citenamefont {Taniguchi}, \citenamefont {Vandersypen},\ and\ \citenamefont {Goswami}}]{nandaCurrentPhaseRelationBallistic2017}%
  \BibitemOpen
  \bibfield  {author} {\bibinfo {author} {\bibfnamefont {G.}~\bibnamefont {Nanda}}, \bibinfo {author} {\bibfnamefont {J.~L.}\ \bibnamefont {{Aguilera-Servin}}}, \bibinfo {author} {\bibfnamefont {P.}~\bibnamefont {Rakyta}}, \bibinfo {author} {\bibfnamefont {A.}~\bibnamefont {Korm{\'a}nyos}}, \bibinfo {author} {\bibfnamefont {R.}~\bibnamefont {Kleiner}}, \bibinfo {author} {\bibfnamefont {D.}~\bibnamefont {Koelle}}, \bibinfo {author} {\bibfnamefont {K.}~\bibnamefont {Watanabe}}, \bibinfo {author} {\bibfnamefont {T.}~\bibnamefont {Taniguchi}}, \bibinfo {author} {\bibfnamefont {L.~M.~K.}\ \bibnamefont {Vandersypen}},\ and\ \bibinfo {author} {\bibfnamefont {S.}~\bibnamefont {Goswami}},\ }\bibfield  {title} {\bibinfo {title} {Current-phase relation of ballistic graphene {Josephson} junctions},\ }\href {https://doi.org/10.1021/acs.nanolett.7b00097} {\bibfield  {journal} {\bibinfo  {journal} {Nano Lett.}\ }\textbf {\bibinfo {volume} {17}},\ \bibinfo {pages} {3396} (\bibinfo {year} {2017})}\BibitemShut {NoStop}%
\bibitem [{\citenamefont {Manjarr{\'e}s}\ \emph {et~al.}(2020)\citenamefont {Manjarr{\'e}s}, \citenamefont {G{\'o}mez~P{\'a}ez},\ and\ \citenamefont {Herrera}}]{manjarresSkewnessCriticalCurrent2020}%
  \BibitemOpen
  \bibfield  {author} {\bibinfo {author} {\bibfnamefont {D.~A.}\ \bibnamefont {Manjarr{\'e}s}}, \bibinfo {author} {\bibfnamefont {S.}~\bibnamefont {G{\'o}mez~P{\'a}ez}},\ and\ \bibinfo {author} {\bibfnamefont {W.~J.}\ \bibnamefont {Herrera}},\ }\bibfield  {title} {\bibinfo {title} {Skewness and critical current behavior in a graphene {{Josephson}} junction},\ }\href {https://doi.org/10.1103/PhysRevB.101.064503} {\bibfield  {journal} {\bibinfo  {journal} {Phys. Rev. B}\ }\textbf {\bibinfo {volume} {101}},\ \bibinfo {pages} {064503} (\bibinfo {year} {2020})}\BibitemShut {NoStop}%
\bibitem [{\citenamefont {Fu}\ and\ \citenamefont {Kane}(2009)}]{fuJosephsonCurrentNoise2009}%
  \BibitemOpen
  \bibfield  {author} {\bibinfo {author} {\bibfnamefont {L.}~\bibnamefont {Fu}}\ and\ \bibinfo {author} {\bibfnamefont {C.~L.}\ \bibnamefont {Kane}},\ }\bibfield  {title} {\bibinfo {title} {Josephson current and noise at a superconductor/quantum-spin-{{Hall-insulator}}/superconductor junction},\ }\href {https://doi.org/10.1103/physrevb.79.161408} {\bibfield  {journal} {\bibinfo  {journal} {Phys. Rev. B}\ }\textbf {\bibinfo {volume} {79}},\ \bibinfo {pages} {161408} (\bibinfo {year} {2009})}\BibitemShut {NoStop}%
\bibitem [{\citenamefont {Kayyalha}\ \emph {et~al.}(2020)\citenamefont {Kayyalha}, \citenamefont {Kazakov}, \citenamefont {Miotkowski}, \citenamefont {Khlebnikov}, \citenamefont {Rokhinson},\ and\ \citenamefont {Chen}}]{kayyalhaHighlySkewedCurrent2020}%
  \BibitemOpen
  \bibfield  {author} {\bibinfo {author} {\bibfnamefont {M.}~\bibnamefont {Kayyalha}}, \bibinfo {author} {\bibfnamefont {A.}~\bibnamefont {Kazakov}}, \bibinfo {author} {\bibfnamefont {I.}~\bibnamefont {Miotkowski}}, \bibinfo {author} {\bibfnamefont {S.}~\bibnamefont {Khlebnikov}}, \bibinfo {author} {\bibfnamefont {L.~P.}\ \bibnamefont {Rokhinson}},\ and\ \bibinfo {author} {\bibfnamefont {Y.~P.}\ \bibnamefont {Chen}},\ }\bibfield  {title} {\bibinfo {title} {Highly skewed current-phase relation in superconductor-topological insulator-superconductor {{Josephson}} junctions},\ }\href {https://doi.org/10.1038/s41535-020-0209-5} {\bibfield  {journal} {\bibinfo  {journal} {npj Quant. Mater.}\ }\textbf {\bibinfo {volume} {5}},\ \bibinfo {pages} {7} (\bibinfo {year} {2020})}\BibitemShut {NoStop}%
\bibitem [{\citenamefont {{van Dam}}\ \emph {et~al.}(2006)\citenamefont {{van Dam}}, \citenamefont {Nazarov}, \citenamefont {Bakkers}, \citenamefont {De~Franceschi},\ and\ \citenamefont {Kouwenhoven}}]{vandamSupercurrentReversalQuantum2006}%
  \BibitemOpen
  \bibfield  {author} {\bibinfo {author} {\bibfnamefont {J.~A.}\ \bibnamefont {{van Dam}}}, \bibinfo {author} {\bibfnamefont {Y.~V.}\ \bibnamefont {Nazarov}}, \bibinfo {author} {\bibfnamefont {E.~P. A.~M.}\ \bibnamefont {Bakkers}}, \bibinfo {author} {\bibfnamefont {S.}~\bibnamefont {De~Franceschi}},\ and\ \bibinfo {author} {\bibfnamefont {L.~P.}\ \bibnamefont {Kouwenhoven}},\ }\bibfield  {title} {\bibinfo {title} {Supercurrent reversal in quantum dots},\ }\href {https://doi.org/10.1038/nature05018} {\bibfield  {journal} {\bibinfo  {journal} {Nature}\ }\textbf {\bibinfo {volume} {442}},\ \bibinfo {pages} {667} (\bibinfo {year} {2006})}\BibitemShut {NoStop}%
\bibitem [{\citenamefont {Spanton}\ \emph {et~al.}(2017)\citenamefont {Spanton}, \citenamefont {Deng}, \citenamefont {Vaitiek{\.e}nas}, \citenamefont {Krogstrup}, \citenamefont {Nyg{\aa}rd}, \citenamefont {Marcus},\ and\ \citenamefont {Moler}}]{spantonCurrentPhaseRelations2017}%
  \BibitemOpen
  \bibfield  {author} {\bibinfo {author} {\bibfnamefont {E.~M.}\ \bibnamefont {Spanton}}, \bibinfo {author} {\bibfnamefont {M.}~\bibnamefont {Deng}}, \bibinfo {author} {\bibfnamefont {S.}~\bibnamefont {Vaitiek{\.e}nas}}, \bibinfo {author} {\bibfnamefont {P.}~\bibnamefont {Krogstrup}}, \bibinfo {author} {\bibfnamefont {J.}~\bibnamefont {Nyg{\aa}rd}}, \bibinfo {author} {\bibfnamefont {C.~M.}\ \bibnamefont {Marcus}},\ and\ \bibinfo {author} {\bibfnamefont {K.~A.}\ \bibnamefont {Moler}},\ }\bibfield  {title} {\bibinfo {title} {Current-phase relations of few-mode {{InAs}} nanowire {Josephson} junctions},\ }\href {https://doi.org/10.1038/nphys4224} {\bibfield  {journal} {\bibinfo  {journal} {Nat. Phys.}\ }\textbf {\bibinfo {volume} {13}},\ \bibinfo {pages} {1177} (\bibinfo {year} {2017})}\BibitemShut {NoStop}%
\bibitem [{\citenamefont {Hart}\ \emph {et~al.}(2019)\citenamefont {Hart}, \citenamefont {Cui}, \citenamefont {M{\'e}nard}, \citenamefont {Deng}, \citenamefont {Antipov}, \citenamefont {Lutchyn}, \citenamefont {Krogstrup}, \citenamefont {Marcus},\ and\ \citenamefont {Moler}}]{hartCurrentphaseRelationsInAs2019}%
  \BibitemOpen
  \bibfield  {author} {\bibinfo {author} {\bibfnamefont {S.}~\bibnamefont {Hart}}, \bibinfo {author} {\bibfnamefont {Z.}~\bibnamefont {Cui}}, \bibinfo {author} {\bibfnamefont {G.}~\bibnamefont {M{\'e}nard}}, \bibinfo {author} {\bibfnamefont {M.}~\bibnamefont {Deng}}, \bibinfo {author} {\bibfnamefont {A.~E.}\ \bibnamefont {Antipov}}, \bibinfo {author} {\bibfnamefont {R.~M.}\ \bibnamefont {Lutchyn}}, \bibinfo {author} {\bibfnamefont {P.}~\bibnamefont {Krogstrup}}, \bibinfo {author} {\bibfnamefont {C.~M.}\ \bibnamefont {Marcus}},\ and\ \bibinfo {author} {\bibfnamefont {K.~A.}\ \bibnamefont {Moler}},\ }\bibfield  {title} {\bibinfo {title} {Current-phase relations of {{InAs}} nanowire {{Josephson}} junctions: {{From}} interacting to multimode regimes},\ }\href {https://doi.org/10.1103/PhysRevB.100.064523} {\bibfield  {journal} {\bibinfo  {journal} {Phys. Rev. B}\ }\textbf {\bibinfo {volume} {100}},\ \bibinfo {pages} {064523} (\bibinfo {year} {2019})}\BibitemShut {NoStop}%
\bibitem [{\citenamefont {Nichele}\ \emph {et~al.}(2020)\citenamefont {Nichele}, \citenamefont {Portol{\'e}s}, \citenamefont {Fornieri}, \citenamefont {Whiticar}, \citenamefont {Drachmann}, \citenamefont {Gronin}, \citenamefont {Wang}, \citenamefont {Gardner}, \citenamefont {Thomas}, \citenamefont {Hatke}, \citenamefont {Manfra},\ and\ \citenamefont {Marcus}}]{nicheleRelatingAndreevBound2020}%
  \BibitemOpen
  \bibfield  {author} {\bibinfo {author} {\bibfnamefont {F.}~\bibnamefont {Nichele}}, \bibinfo {author} {\bibfnamefont {E.}~\bibnamefont {Portol{\'e}s}}, \bibinfo {author} {\bibfnamefont {A.}~\bibnamefont {Fornieri}}, \bibinfo {author} {\bibfnamefont {A.~M.}\ \bibnamefont {Whiticar}}, \bibinfo {author} {\bibfnamefont {A.~C.~C.}\ \bibnamefont {Drachmann}}, \bibinfo {author} {\bibfnamefont {S.}~\bibnamefont {Gronin}}, \bibinfo {author} {\bibfnamefont {T.}~\bibnamefont {Wang}}, \bibinfo {author} {\bibfnamefont {G.~C.}\ \bibnamefont {Gardner}}, \bibinfo {author} {\bibfnamefont {C.}~\bibnamefont {Thomas}}, \bibinfo {author} {\bibfnamefont {A.~T.}\ \bibnamefont {Hatke}}, \bibinfo {author} {\bibfnamefont {M.~J.}\ \bibnamefont {Manfra}},\ and\ \bibinfo {author} {\bibfnamefont {C.~M.}\ \bibnamefont {Marcus}},\ }\bibfield  {title} {\bibinfo {title} {Relating {Andreev} bound states and supercurrents in hybrid {Josephson} junctions},\ }\href {https://doi.org/10.1103/PhysRevLett.124.226801} {\bibfield  {journal} {\bibinfo
   {journal} {Phys. Rev. Lett.}\ }\textbf {\bibinfo {volume} {124}},\ \bibinfo {pages} {226801} (\bibinfo {year} {2020})}\BibitemShut {NoStop}%
\bibitem [{\citenamefont {Ueda}\ \emph {et~al.}(2020)\citenamefont {Ueda}, \citenamefont {Matsuo}, \citenamefont {Kamata}, \citenamefont {Sato}, \citenamefont {Takeshige}, \citenamefont {Li}, \citenamefont {Samuelson}, \citenamefont {Xu},\ and\ \citenamefont {Tarucha}}]{uedaEvidenceHalfintegerShapiro2020}%
  \BibitemOpen
  \bibfield  {author} {\bibinfo {author} {\bibfnamefont {K.}~\bibnamefont {Ueda}}, \bibinfo {author} {\bibfnamefont {S.}~\bibnamefont {Matsuo}}, \bibinfo {author} {\bibfnamefont {H.}~\bibnamefont {Kamata}}, \bibinfo {author} {\bibfnamefont {Y.}~\bibnamefont {Sato}}, \bibinfo {author} {\bibfnamefont {Y.}~\bibnamefont {Takeshige}}, \bibinfo {author} {\bibfnamefont {K.}~\bibnamefont {Li}}, \bibinfo {author} {\bibfnamefont {L.}~\bibnamefont {Samuelson}}, \bibinfo {author} {\bibfnamefont {H.}~\bibnamefont {Xu}},\ and\ \bibinfo {author} {\bibfnamefont {S.}~\bibnamefont {Tarucha}},\ }\bibfield  {title} {\bibinfo {title} {Evidence of half-integer {{Shapiro}} steps originated from nonsinusoidal current phase relation in a short ballistic {{InAs}} nanowire {{Josephson}} junction},\ }\href {https://doi.org/10.1103/PhysRevResearch.2.033435} {\bibfield  {journal} {\bibinfo  {journal} {Phys. Rev. Res.}\ }\textbf {\bibinfo {volume} {2}},\ \bibinfo {pages} {033435} (\bibinfo {year} {2020})}\BibitemShut {NoStop}%
\bibitem [{\citenamefont {Aggarwal}\ \emph {et~al.}(2021)\citenamefont {Aggarwal}, \citenamefont {Hofmann}, \citenamefont {Jirovec}, \citenamefont {Prieto}, \citenamefont {Sammak}, \citenamefont {Botifoll}, \citenamefont {{Mart{\'i}-S{\'a}nchez}}, \citenamefont {Veldhorst}, \citenamefont {Arbiol}, \citenamefont {Scappucci}, \citenamefont {Danon},\ and\ \citenamefont {Katsaros}}]{aggarwalEnhancementProximityinducedSuperconductivity2021}%
  \BibitemOpen
  \bibfield  {author} {\bibinfo {author} {\bibfnamefont {K.}~\bibnamefont {Aggarwal}}, \bibinfo {author} {\bibfnamefont {A.}~\bibnamefont {Hofmann}}, \bibinfo {author} {\bibfnamefont {D.}~\bibnamefont {Jirovec}}, \bibinfo {author} {\bibfnamefont {I.}~\bibnamefont {Prieto}}, \bibinfo {author} {\bibfnamefont {A.}~\bibnamefont {Sammak}}, \bibinfo {author} {\bibfnamefont {M.}~\bibnamefont {Botifoll}}, \bibinfo {author} {\bibfnamefont {S.}~\bibnamefont {{Mart{\'i}-S{\'a}nchez}}}, \bibinfo {author} {\bibfnamefont {M.}~\bibnamefont {Veldhorst}}, \bibinfo {author} {\bibfnamefont {J.}~\bibnamefont {Arbiol}}, \bibinfo {author} {\bibfnamefont {G.}~\bibnamefont {Scappucci}}, \bibinfo {author} {\bibfnamefont {J.}~\bibnamefont {Danon}},\ and\ \bibinfo {author} {\bibfnamefont {G.}~\bibnamefont {Katsaros}},\ }\bibfield  {title} {\bibinfo {title} {Enhancement of proximity-induced superconductivity in a planar {{Ge}} hole gas},\ }\href {https://doi.org/10.1103/physrevresearch.3.l022005} {\bibfield  {journal} {\bibinfo
  {journal} {Phys. Rev. Res.}\ }\textbf {\bibinfo {volume} {3}},\ \bibinfo {pages} {L022005} (\bibinfo {year} {2021})}\BibitemShut {NoStop}%
\bibitem [{\citenamefont {Valentini}\ \emph {et~al.}(2024)\citenamefont {Valentini}, \citenamefont {Sagi}, \citenamefont {Baghumyan}, \citenamefont {De~Gijsel}, \citenamefont {Jung}, \citenamefont {Calcaterra}, \citenamefont {Ballabio}, \citenamefont {Aguilera~Servin}, \citenamefont {Aggarwal}, \citenamefont {Janik}, \citenamefont {Adletzberger}, \citenamefont {Seoane~Souto}, \citenamefont {Leijnse}, \citenamefont {Danon}, \citenamefont {Schrade}, \citenamefont {Bakkers}, \citenamefont {Chrastina}, \citenamefont {Isella},\ and\ \citenamefont {Katsaros}}]{valentiniParityconservingCooperpairTransport2024}%
  \BibitemOpen
  \bibfield  {author} {\bibinfo {author} {\bibfnamefont {M.}~\bibnamefont {Valentini}}, \bibinfo {author} {\bibfnamefont {O.}~\bibnamefont {Sagi}}, \bibinfo {author} {\bibfnamefont {L.}~\bibnamefont {Baghumyan}}, \bibinfo {author} {\bibfnamefont {T.}~\bibnamefont {De~Gijsel}}, \bibinfo {author} {\bibfnamefont {J.}~\bibnamefont {Jung}}, \bibinfo {author} {\bibfnamefont {S.}~\bibnamefont {Calcaterra}}, \bibinfo {author} {\bibfnamefont {A.}~\bibnamefont {Ballabio}}, \bibinfo {author} {\bibfnamefont {J.}~\bibnamefont {Aguilera~Servin}}, \bibinfo {author} {\bibfnamefont {K.}~\bibnamefont {Aggarwal}}, \bibinfo {author} {\bibfnamefont {M.}~\bibnamefont {Janik}}, \bibinfo {author} {\bibfnamefont {T.}~\bibnamefont {Adletzberger}}, \bibinfo {author} {\bibfnamefont {R.}~\bibnamefont {Seoane~Souto}}, \bibinfo {author} {\bibfnamefont {M.}~\bibnamefont {Leijnse}}, \bibinfo {author} {\bibfnamefont {J.}~\bibnamefont {Danon}}, \bibinfo {author} {\bibfnamefont {C.}~\bibnamefont {Schrade}}, \bibinfo {author} {\bibfnamefont
  {E.}~\bibnamefont {Bakkers}}, \bibinfo {author} {\bibfnamefont {D.}~\bibnamefont {Chrastina}}, \bibinfo {author} {\bibfnamefont {G.}~\bibnamefont {Isella}},\ and\ \bibinfo {author} {\bibfnamefont {G.}~\bibnamefont {Katsaros}},\ }\bibfield  {title} {\bibinfo {title} {Parity-conserving {{Cooper-pair}} transport and ideal superconducting diode in planar germanium},\ }\href {https://doi.org/10.1038/s41467-023-44114-0} {\bibfield  {journal} {\bibinfo  {journal} {Nat. Commun.}\ }\textbf {\bibinfo {volume} {15}},\ \bibinfo {pages} {169} (\bibinfo {year} {2024})}\BibitemShut {NoStop}%
\bibitem [{\citenamefont {Leblanc}\ \emph {et~al.}(2024{\natexlab{a}})\citenamefont {Leblanc}, \citenamefont {Tangchingchai}, \citenamefont {Momtaz}, \citenamefont {Kiyooka}, \citenamefont {Hartmann}, \citenamefont {Gustavo}, \citenamefont {Thomassin}, \citenamefont {Brun}, \citenamefont {Schmitt}, \citenamefont {Zihlmann}, \citenamefont {Maurand}, \citenamefont {Dumur}, \citenamefont {De~Franceschi},\ and\ \citenamefont {Lefloch}}]{leblancGateFluxTunable2024}%
  \BibitemOpen
  \bibfield  {author} {\bibinfo {author} {\bibfnamefont {A.}~\bibnamefont {Leblanc}}, \bibinfo {author} {\bibfnamefont {C.}~\bibnamefont {Tangchingchai}}, \bibinfo {author} {\bibfnamefont {Z.~S.}\ \bibnamefont {Momtaz}}, \bibinfo {author} {\bibfnamefont {E.}~\bibnamefont {Kiyooka}}, \bibinfo {author} {\bibfnamefont {J.-M.}\ \bibnamefont {Hartmann}}, \bibinfo {author} {\bibfnamefont {F.}~\bibnamefont {Gustavo}}, \bibinfo {author} {\bibfnamefont {J.-L.}\ \bibnamefont {Thomassin}}, \bibinfo {author} {\bibfnamefont {B.}~\bibnamefont {Brun}}, \bibinfo {author} {\bibfnamefont {V.}~\bibnamefont {Schmitt}}, \bibinfo {author} {\bibfnamefont {S.}~\bibnamefont {Zihlmann}}, \bibinfo {author} {\bibfnamefont {R.}~\bibnamefont {Maurand}}, \bibinfo {author} {\bibfnamefont {E.}~\bibnamefont {Dumur}}, \bibinfo {author} {\bibfnamefont {S.}~\bibnamefont {De~Franceschi}},\ and\ \bibinfo {author} {\bibfnamefont {F.}~\bibnamefont {Lefloch}},\ }\bibfield  {title} {\bibinfo {title} {Gate and flux tunable $\sin(2\varphi)$ {Josephson}
  element in proximitized junctions},\ }\href {https://arxiv.org/abs/2405.14695} {\bibfield  {journal} {\bibinfo  {journal} {arXiv:2405.14695}\ } (\bibinfo {year} {2024}{\natexlab{a}})}\BibitemShut {NoStop}%
\bibitem [{\citenamefont {Scher{\"u}bl}\ \emph {et~al.}(2025)\citenamefont {Scher{\"u}bl}, \citenamefont {S{\"u}t{\H o}}, \citenamefont {K{\'o}ti}, \citenamefont {T{\'o}v{\'a}ri}, \citenamefont {Horv{\'a}th}, \citenamefont {Kalm{\'a}r}, \citenamefont {Vasas}, \citenamefont {Berke}, \citenamefont {Kirti}, \citenamefont {Biasiol}, \citenamefont {Csonka}, \citenamefont {Makk},\ and\ \citenamefont {F{\"u}l{\"o}p}}]{scherublDeterminationCurrentphaseRelation2025}%
  \BibitemOpen
  \bibfield  {author} {\bibinfo {author} {\bibfnamefont {Z.}~\bibnamefont {Scher{\"u}bl}}, \bibinfo {author} {\bibfnamefont {M.}~\bibnamefont {S{\"u}t{\H o}}}, \bibinfo {author} {\bibfnamefont {D.}~\bibnamefont {K{\'o}ti}}, \bibinfo {author} {\bibfnamefont {E.}~\bibnamefont {T{\'o}v{\'a}ri}}, \bibinfo {author} {\bibfnamefont {C.}~\bibnamefont {Horv{\'a}th}}, \bibinfo {author} {\bibfnamefont {T.}~\bibnamefont {Kalm{\'a}r}}, \bibinfo {author} {\bibfnamefont {B.}~\bibnamefont {Vasas}}, \bibinfo {author} {\bibfnamefont {M.}~\bibnamefont {Berke}}, \bibinfo {author} {\bibfnamefont {M.}~\bibnamefont {Kirti}}, \bibinfo {author} {\bibfnamefont {G.}~\bibnamefont {Biasiol}}, \bibinfo {author} {\bibfnamefont {S.}~\bibnamefont {Csonka}}, \bibinfo {author} {\bibfnamefont {P.}~\bibnamefont {Makk}},\ and\ \bibinfo {author} {\bibfnamefont {G.}~\bibnamefont {F{\"u}l{\"o}p}},\ }\bibfield  {title} {\bibinfo {title} {Determination of the current-phase relation of an {{InAs 2DEG Josephson}} junction with a microwave resonator},\
  }\href {https://doi.org/10.1103/PhysRevResearch.7.023173} {\bibfield  {journal} {\bibinfo  {journal} {Phys. Rev. Res.}\ }\textbf {\bibinfo {volume} {7}},\ \bibinfo {pages} {023173} (\bibinfo {year} {2025})}\BibitemShut {NoStop}%
\bibitem [{\citenamefont {Banszerus}\ \emph {et~al.}(2025)\citenamefont {Banszerus}, \citenamefont {Andersson}, \citenamefont {Marshall}, \citenamefont {Lindemann}, \citenamefont {Manfra}, \citenamefont {Marcus},\ and\ \citenamefont {Vaitiek{\.e}nas}}]{banszerusHybridJosephsonRhombus2025}%
  \BibitemOpen
  \bibfield  {author} {\bibinfo {author} {\bibfnamefont {L.}~\bibnamefont {Banszerus}}, \bibinfo {author} {\bibfnamefont {C.~W.}\ \bibnamefont {Andersson}}, \bibinfo {author} {\bibfnamefont {W.}~\bibnamefont {Marshall}}, \bibinfo {author} {\bibfnamefont {T.}~\bibnamefont {Lindemann}}, \bibinfo {author} {\bibfnamefont {M.~J.}\ \bibnamefont {Manfra}}, \bibinfo {author} {\bibfnamefont {C.~M.}\ \bibnamefont {Marcus}},\ and\ \bibinfo {author} {\bibfnamefont {S.}~\bibnamefont {Vaitiek{\.e}nas}},\ }\bibfield  {title} {\bibinfo {title} {Hybrid {Josephson} rhombus: A superconducting element with tailored current-phase relation},\ }\href {https://doi.org/10.1103/PhysRevX.15.011021} {\bibfield  {journal} {\bibinfo  {journal} {Phys. Rev. X}\ }\textbf {\bibinfo {volume} {15}},\ \bibinfo {pages} {011021} (\bibinfo {year} {2025})}\BibitemShut {NoStop}%
\bibitem [{\citenamefont {Ando}\ \emph {et~al.}(2020)\citenamefont {Ando}, \citenamefont {Miyasaka}, \citenamefont {Li}, \citenamefont {Ishizuka}, \citenamefont {Arakawa}, \citenamefont {Shiota}, \citenamefont {Moriyama}, \citenamefont {Yanase},\ and\ \citenamefont {Ono}}]{andoObservationSuperconductingDiode2020}%
  \BibitemOpen
  \bibfield  {author} {\bibinfo {author} {\bibfnamefont {F.}~\bibnamefont {Ando}}, \bibinfo {author} {\bibfnamefont {Y.}~\bibnamefont {Miyasaka}}, \bibinfo {author} {\bibfnamefont {T.}~\bibnamefont {Li}}, \bibinfo {author} {\bibfnamefont {J.}~\bibnamefont {Ishizuka}}, \bibinfo {author} {\bibfnamefont {T.}~\bibnamefont {Arakawa}}, \bibinfo {author} {\bibfnamefont {Y.}~\bibnamefont {Shiota}}, \bibinfo {author} {\bibfnamefont {T.}~\bibnamefont {Moriyama}}, \bibinfo {author} {\bibfnamefont {Y.}~\bibnamefont {Yanase}},\ and\ \bibinfo {author} {\bibfnamefont {T.}~\bibnamefont {Ono}},\ }\bibfield  {title} {\bibinfo {title} {Observation of superconducting diode effect},\ }\href {https://doi.org/10.1038/s41586-020-2590-4} {\bibfield  {journal} {\bibinfo  {journal} {Nature}\ }\textbf {\bibinfo {volume} {584}},\ \bibinfo {pages} {373} (\bibinfo {year} {2020})}\BibitemShut {NoStop}%
\bibitem [{\citenamefont {Bauriedl}\ \emph {et~al.}(2022)\citenamefont {Bauriedl}, \citenamefont {B{\"a}uml}, \citenamefont {Fuchs}, \citenamefont {Baumgartner}, \citenamefont {Paulik}, \citenamefont {Bauer}, \citenamefont {Lin}, \citenamefont {Lupton}, \citenamefont {Taniguchi}, \citenamefont {Watanabe}, \citenamefont {Strunk},\ and\ \citenamefont {Paradiso}}]{bauriedlSupercurrentDiodeEffect2022}%
  \BibitemOpen
  \bibfield  {author} {\bibinfo {author} {\bibfnamefont {L.}~\bibnamefont {Bauriedl}}, \bibinfo {author} {\bibfnamefont {C.}~\bibnamefont {B{\"a}uml}}, \bibinfo {author} {\bibfnamefont {L.}~\bibnamefont {Fuchs}}, \bibinfo {author} {\bibfnamefont {C.}~\bibnamefont {Baumgartner}}, \bibinfo {author} {\bibfnamefont {N.}~\bibnamefont {Paulik}}, \bibinfo {author} {\bibfnamefont {J.~M.}\ \bibnamefont {Bauer}}, \bibinfo {author} {\bibfnamefont {K.-Q.}\ \bibnamefont {Lin}}, \bibinfo {author} {\bibfnamefont {J.~M.}\ \bibnamefont {Lupton}}, \bibinfo {author} {\bibfnamefont {T.}~\bibnamefont {Taniguchi}}, \bibinfo {author} {\bibfnamefont {K.}~\bibnamefont {Watanabe}}, \bibinfo {author} {\bibfnamefont {C.}~\bibnamefont {Strunk}},\ and\ \bibinfo {author} {\bibfnamefont {N.}~\bibnamefont {Paradiso}},\ }\bibfield  {title} {\bibinfo {title} {Supercurrent diode effect and magnetochiral anisotropy in few-layer {{NbSe2}}},\ }\href {https://doi.org/10.1038/s41467-022-31954-5} {\bibfield  {journal} {\bibinfo  {journal} {Nat.
  Commun.}\ }\textbf {\bibinfo {volume} {13}},\ \bibinfo {pages} {4266} (\bibinfo {year} {2022})}\BibitemShut {NoStop}%
\bibitem [{\citenamefont {Baumgartner}\ \emph {et~al.}(2022)\citenamefont {Baumgartner}, \citenamefont {Fuchs}, \citenamefont {Costa}, \citenamefont {Reinhardt}, \citenamefont {Gronin}, \citenamefont {Gardner}, \citenamefont {Lindemann}, \citenamefont {Manfra}, \citenamefont {Faria~Junior}, \citenamefont {Kochan}, \citenamefont {Fabian}, \citenamefont {Paradiso},\ and\ \citenamefont {Strunk}}]{baumgartnerSupercurrentRectificationMagnetochiral2022a}%
  \BibitemOpen
  \bibfield  {author} {\bibinfo {author} {\bibfnamefont {C.}~\bibnamefont {Baumgartner}}, \bibinfo {author} {\bibfnamefont {L.}~\bibnamefont {Fuchs}}, \bibinfo {author} {\bibfnamefont {A.}~\bibnamefont {Costa}}, \bibinfo {author} {\bibfnamefont {S.}~\bibnamefont {Reinhardt}}, \bibinfo {author} {\bibfnamefont {S.}~\bibnamefont {Gronin}}, \bibinfo {author} {\bibfnamefont {G.~C.}\ \bibnamefont {Gardner}}, \bibinfo {author} {\bibfnamefont {T.}~\bibnamefont {Lindemann}}, \bibinfo {author} {\bibfnamefont {M.~J.}\ \bibnamefont {Manfra}}, \bibinfo {author} {\bibfnamefont {P.~E.}\ \bibnamefont {Faria~Junior}}, \bibinfo {author} {\bibfnamefont {D.}~\bibnamefont {Kochan}}, \bibinfo {author} {\bibfnamefont {J.}~\bibnamefont {Fabian}}, \bibinfo {author} {\bibfnamefont {N.}~\bibnamefont {Paradiso}},\ and\ \bibinfo {author} {\bibfnamefont {C.}~\bibnamefont {Strunk}},\ }\bibfield  {title} {\bibinfo {title} {Supercurrent rectification and magnetochiral effects in symmetric {{Josephson}} junctions},\ }\href
  {https://doi.org/10.1038/s41565-021-01009-9} {\bibfield  {journal} {\bibinfo  {journal} {Nat. Nanotech.}\ }\textbf {\bibinfo {volume} {17}},\ \bibinfo {pages} {39} (\bibinfo {year} {2022})}\BibitemShut {NoStop}%
\bibitem [{\citenamefont {Nadeem}\ \emph {et~al.}(2023)\citenamefont {Nadeem}, \citenamefont {Fuhrer},\ and\ \citenamefont {Wang}}]{nadeemSuperconductingDiodeEffect2023a}%
  \BibitemOpen
  \bibfield  {author} {\bibinfo {author} {\bibfnamefont {M.}~\bibnamefont {Nadeem}}, \bibinfo {author} {\bibfnamefont {M.~S.}\ \bibnamefont {Fuhrer}},\ and\ \bibinfo {author} {\bibfnamefont {X.}~\bibnamefont {Wang}},\ }\bibfield  {title} {\bibinfo {title} {The superconducting diode effect},\ }\href {https://doi.org/10.1038/s42254-023-00632-w} {\bibfield  {journal} {\bibinfo  {journal} {Nat. Rev. Phys.}\ }\textbf {\bibinfo {volume} {5}},\ \bibinfo {pages} {558} (\bibinfo {year} {2023})}\BibitemShut {NoStop}%
\bibitem [{\citenamefont {Fracassi}\ \emph {et~al.}(2024)\citenamefont {Fracassi}, \citenamefont {Traverso}, \citenamefont {Traverso~Ziani}, \citenamefont {Carrega}, \citenamefont {Heun},\ and\ \citenamefont {Sassetti}}]{fracassiAnomalousSupercurrentDiode2024b}%
  \BibitemOpen
  \bibfield  {author} {\bibinfo {author} {\bibfnamefont {S.}~\bibnamefont {Fracassi}}, \bibinfo {author} {\bibfnamefont {S.}~\bibnamefont {Traverso}}, \bibinfo {author} {\bibfnamefont {N.}~\bibnamefont {Traverso~Ziani}}, \bibinfo {author} {\bibfnamefont {M.}~\bibnamefont {Carrega}}, \bibinfo {author} {\bibfnamefont {S.}~\bibnamefont {Heun}},\ and\ \bibinfo {author} {\bibfnamefont {M.}~\bibnamefont {Sassetti}},\ }\bibfield  {title} {\bibinfo {title} {Anomalous supercurrent and diode effect in locally perturbed topological {{Josephson}} junctions},\ }\href {https://doi.org/10.1063/5.0210660} {\bibfield  {journal} {\bibinfo  {journal} {Applied Physics Letters}\ }\textbf {\bibinfo {volume} {124}},\ \bibinfo {pages} {242601} (\bibinfo {year} {2024})}\BibitemShut {NoStop}%
\bibitem [{\citenamefont {Larsen}\ \emph {et~al.}(2015)\citenamefont {Larsen}, \citenamefont {Petersson}, \citenamefont {Kuemmeth}, \citenamefont {Jespersen}, \citenamefont {Krogstrup}, \citenamefont {Nyg{\aa}rd},\ and\ \citenamefont {Marcus}}]{larsenSemiconductorNanowireBasedSuperconductingQubit2015}%
  \BibitemOpen
  \bibfield  {author} {\bibinfo {author} {\bibfnamefont {T.~W.}\ \bibnamefont {Larsen}}, \bibinfo {author} {\bibfnamefont {K.~D.}\ \bibnamefont {Petersson}}, \bibinfo {author} {\bibfnamefont {F.}~\bibnamefont {Kuemmeth}}, \bibinfo {author} {\bibfnamefont {T.~S.}\ \bibnamefont {Jespersen}}, \bibinfo {author} {\bibfnamefont {P.}~\bibnamefont {Krogstrup}}, \bibinfo {author} {\bibfnamefont {J.}~\bibnamefont {Nyg{\aa}rd}},\ and\ \bibinfo {author} {\bibfnamefont {C.~M.}\ \bibnamefont {Marcus}},\ }\bibfield  {title} {\bibinfo {title} {Semiconductor-nanowire-based superconducting qubit},\ }\href {https://doi.org/10.1103/PhysRevLett.115.127001} {\bibfield  {journal} {\bibinfo  {journal} {Phys. Rev. Lett.}\ }\textbf {\bibinfo {volume} {115}},\ \bibinfo {pages} {127001} (\bibinfo {year} {2015})}\BibitemShut {NoStop}%
\bibitem [{\citenamefont {{de Lange}}\ \emph {et~al.}(2015)\citenamefont {{de Lange}}, \citenamefont {{van Heck}}, \citenamefont {Bruno}, \citenamefont {{van Woerkom}}, \citenamefont {Geresdi}, \citenamefont {Plissard}, \citenamefont {Bakkers}, \citenamefont {Akhmerov},\ and\ \citenamefont {DiCarlo}}]{delangeRealizationMicrowaveQuantum2015}%
  \BibitemOpen
  \bibfield  {author} {\bibinfo {author} {\bibfnamefont {G.}~\bibnamefont {{de Lange}}}, \bibinfo {author} {\bibfnamefont {B.}~\bibnamefont {{van Heck}}}, \bibinfo {author} {\bibfnamefont {A.}~\bibnamefont {Bruno}}, \bibinfo {author} {\bibfnamefont {D.~J.}\ \bibnamefont {{van Woerkom}}}, \bibinfo {author} {\bibfnamefont {A.}~\bibnamefont {Geresdi}}, \bibinfo {author} {\bibfnamefont {S.~R.}\ \bibnamefont {Plissard}}, \bibinfo {author} {\bibfnamefont {E.~P. A.~M.}\ \bibnamefont {Bakkers}}, \bibinfo {author} {\bibfnamefont {A.~R.}\ \bibnamefont {Akhmerov}},\ and\ \bibinfo {author} {\bibfnamefont {L.}~\bibnamefont {DiCarlo}},\ }\bibfield  {title} {\bibinfo {title} {Realization of microwave quantum circuits using hybrid superconducting-semiconducting nanowire {Josephson} elements},\ }\href {https://doi.org/10.1103/PhysRevLett.115.127002} {\bibfield  {journal} {\bibinfo  {journal} {Phys. Rev. Lett.}\ }\textbf {\bibinfo {volume} {115}},\ \bibinfo {pages} {127002} (\bibinfo {year} {2015})}\BibitemShut {NoStop}%
\bibitem [{\citenamefont {Casparis}\ \emph {et~al.}(2018)\citenamefont {Casparis}, \citenamefont {Connolly}, \citenamefont {Kjaergaard}, \citenamefont {Pearson}, \citenamefont {Kringh{\o}j}, \citenamefont {Larsen}, \citenamefont {Kuemmeth}, \citenamefont {Wang}, \citenamefont {Thomas}, \citenamefont {Gronin}, \citenamefont {Gardner}, \citenamefont {Manfra}, \citenamefont {Marcus},\ and\ \citenamefont {Petersson}}]{casparisSuperconductingGatemonQubit2018}%
  \BibitemOpen
  \bibfield  {author} {\bibinfo {author} {\bibfnamefont {L.}~\bibnamefont {Casparis}}, \bibinfo {author} {\bibfnamefont {M.~R.}\ \bibnamefont {Connolly}}, \bibinfo {author} {\bibfnamefont {M.}~\bibnamefont {Kjaergaard}}, \bibinfo {author} {\bibfnamefont {N.~J.}\ \bibnamefont {Pearson}}, \bibinfo {author} {\bibfnamefont {A.}~\bibnamefont {Kringh{\o}j}}, \bibinfo {author} {\bibfnamefont {T.~W.}\ \bibnamefont {Larsen}}, \bibinfo {author} {\bibfnamefont {F.}~\bibnamefont {Kuemmeth}}, \bibinfo {author} {\bibfnamefont {T.}~\bibnamefont {Wang}}, \bibinfo {author} {\bibfnamefont {C.}~\bibnamefont {Thomas}}, \bibinfo {author} {\bibfnamefont {S.}~\bibnamefont {Gronin}}, \bibinfo {author} {\bibfnamefont {G.~C.}\ \bibnamefont {Gardner}}, \bibinfo {author} {\bibfnamefont {M.~J.}\ \bibnamefont {Manfra}}, \bibinfo {author} {\bibfnamefont {C.~M.}\ \bibnamefont {Marcus}},\ and\ \bibinfo {author} {\bibfnamefont {K.~D.}\ \bibnamefont {Petersson}},\ }\bibfield  {title} {\bibinfo {title} {Superconducting gatemon qubit based on a
  proximitized two-dimensional electron gas},\ }\href {https://doi.org/10.1038/s41565-018-0207-y} {\bibfield  {journal} {\bibinfo  {journal} {Nat. Nanotech.}\ }\textbf {\bibinfo {volume} {13}},\ \bibinfo {pages} {915} (\bibinfo {year} {2018})}\BibitemShut {NoStop}%
\bibitem [{\citenamefont {Hertel}\ \emph {et~al.}(2022)\citenamefont {Hertel}, \citenamefont {Eichinger}, \citenamefont {Andersen}, \citenamefont {Van~Zanten}, \citenamefont {Kallatt}, \citenamefont {Scarlino}, \citenamefont {Kringh{\o}j}, \citenamefont {{Chavez-Garcia}}, \citenamefont {Gardner}, \citenamefont {Gronin}, \citenamefont {Manfra}, \citenamefont {Gyenis}, \citenamefont {Kjaergaard}, \citenamefont {Marcus},\ and\ \citenamefont {Petersson}}]{hertelGatetunableTransmonUsing2022}%
  \BibitemOpen
  \bibfield  {author} {\bibinfo {author} {\bibfnamefont {A.}~\bibnamefont {Hertel}}, \bibinfo {author} {\bibfnamefont {M.}~\bibnamefont {Eichinger}}, \bibinfo {author} {\bibfnamefont {L.~O.}\ \bibnamefont {Andersen}}, \bibinfo {author} {\bibfnamefont {D.~M.}\ \bibnamefont {Van~Zanten}}, \bibinfo {author} {\bibfnamefont {S.}~\bibnamefont {Kallatt}}, \bibinfo {author} {\bibfnamefont {P.}~\bibnamefont {Scarlino}}, \bibinfo {author} {\bibfnamefont {A.}~\bibnamefont {Kringh{\o}j}}, \bibinfo {author} {\bibfnamefont {J.~M.}\ \bibnamefont {{Chavez-Garcia}}}, \bibinfo {author} {\bibfnamefont {G.~C.}\ \bibnamefont {Gardner}}, \bibinfo {author} {\bibfnamefont {S.}~\bibnamefont {Gronin}}, \bibinfo {author} {\bibfnamefont {M.~J.}\ \bibnamefont {Manfra}}, \bibinfo {author} {\bibfnamefont {A.}~\bibnamefont {Gyenis}}, \bibinfo {author} {\bibfnamefont {M.}~\bibnamefont {Kjaergaard}}, \bibinfo {author} {\bibfnamefont {C.~M.}\ \bibnamefont {Marcus}},\ and\ \bibinfo {author} {\bibfnamefont {K.~D.}\ \bibnamefont {Petersson}},\
  }\bibfield  {title} {\bibinfo {title} {Gate-tunable transmon using selective-area-grown superconductor-semiconductor hybrid structures on silicon},\ }\href {https://doi.org/10.1103/PhysRevApplied.18.034042} {\bibfield  {journal} {\bibinfo  {journal} {Phys. Rev. Appl.}\ }\textbf {\bibinfo {volume} {18}},\ \bibinfo {pages} {034042} (\bibinfo {year} {2022})}\BibitemShut {NoStop}%
\bibitem [{\citenamefont {Sagi}\ \emph {et~al.}(2024)\citenamefont {Sagi}, \citenamefont {Crippa}, \citenamefont {Valentini}, \citenamefont {Janik}, \citenamefont {Baghumyan}, \citenamefont {Fabris}, \citenamefont {Kapoor}, \citenamefont {Hassani}, \citenamefont {Fink}, \citenamefont {Calcaterra}, \citenamefont {Chrastina}, \citenamefont {Isella},\ and\ \citenamefont {Katsaros}}]{sagiGateTunableTransmon2024a}%
  \BibitemOpen
  \bibfield  {author} {\bibinfo {author} {\bibfnamefont {O.}~\bibnamefont {Sagi}}, \bibinfo {author} {\bibfnamefont {A.}~\bibnamefont {Crippa}}, \bibinfo {author} {\bibfnamefont {M.}~\bibnamefont {Valentini}}, \bibinfo {author} {\bibfnamefont {M.}~\bibnamefont {Janik}}, \bibinfo {author} {\bibfnamefont {L.}~\bibnamefont {Baghumyan}}, \bibinfo {author} {\bibfnamefont {G.}~\bibnamefont {Fabris}}, \bibinfo {author} {\bibfnamefont {L.}~\bibnamefont {Kapoor}}, \bibinfo {author} {\bibfnamefont {F.}~\bibnamefont {Hassani}}, \bibinfo {author} {\bibfnamefont {J.}~\bibnamefont {Fink}}, \bibinfo {author} {\bibfnamefont {S.}~\bibnamefont {Calcaterra}}, \bibinfo {author} {\bibfnamefont {D.}~\bibnamefont {Chrastina}}, \bibinfo {author} {\bibfnamefont {G.}~\bibnamefont {Isella}},\ and\ \bibinfo {author} {\bibfnamefont {G.}~\bibnamefont {Katsaros}},\ }\bibfield  {title} {\bibinfo {title} {A gate tunable transmon qubit in planar {{Ge}}},\ }\href {https://doi.org/10.1038/s41467-024-50763-6} {\bibfield  {journal} {\bibinfo
  {journal} {Nat. Commun.}\ }\textbf {\bibinfo {volume} {15}},\ \bibinfo {pages} {6400} (\bibinfo {year} {2024})}\BibitemShut {NoStop}%
\bibitem [{\citenamefont {Kiyooka}\ \emph {et~al.}(2025)\citenamefont {Kiyooka}, \citenamefont {Tangchingchai}, \citenamefont {Noirot}, \citenamefont {Leblanc}, \citenamefont {Brun}, \citenamefont {Zihlmann}, \citenamefont {Maurand}, \citenamefont {Schmitt}, \citenamefont {Dumur}, \citenamefont {Hartmann}, \citenamefont {Lefloch},\ and\ \citenamefont {De~Franceschi}}]{kiyookaGatemonQubitGermanium2025}%
  \BibitemOpen
  \bibfield  {author} {\bibinfo {author} {\bibfnamefont {E.}~\bibnamefont {Kiyooka}}, \bibinfo {author} {\bibfnamefont {C.}~\bibnamefont {Tangchingchai}}, \bibinfo {author} {\bibfnamefont {L.}~\bibnamefont {Noirot}}, \bibinfo {author} {\bibfnamefont {A.}~\bibnamefont {Leblanc}}, \bibinfo {author} {\bibfnamefont {B.}~\bibnamefont {Brun}}, \bibinfo {author} {\bibfnamefont {S.}~\bibnamefont {Zihlmann}}, \bibinfo {author} {\bibfnamefont {R.}~\bibnamefont {Maurand}}, \bibinfo {author} {\bibfnamefont {V.}~\bibnamefont {Schmitt}}, \bibinfo {author} {\bibfnamefont {{\'E}.}~\bibnamefont {Dumur}}, \bibinfo {author} {\bibfnamefont {J.-M.}\ \bibnamefont {Hartmann}}, \bibinfo {author} {\bibfnamefont {F.}~\bibnamefont {Lefloch}},\ and\ \bibinfo {author} {\bibfnamefont {S.}~\bibnamefont {De~Franceschi}},\ }\bibfield  {title} {\bibinfo {title} {Gatemon qubit on a germanium quantum-well heterostructure},\ }\href {https://doi.org/10.1021/acs.nanolett.4c05539} {\bibfield  {journal} {\bibinfo  {journal} {Nano Lett.}\ }\textbf
  {\bibinfo {volume} {25}},\ \bibinfo {pages} {562} (\bibinfo {year} {2025})}\BibitemShut {NoStop}%
\bibitem [{\citenamefont {Bell}\ \emph {et~al.}(2014)\citenamefont {Bell}, \citenamefont {Paramanandam}, \citenamefont {Ioffe},\ and\ \citenamefont {Gershenson}}]{bellProtectedJosephsonRhombus2014}%
  \BibitemOpen
  \bibfield  {author} {\bibinfo {author} {\bibfnamefont {M.~T.}\ \bibnamefont {Bell}}, \bibinfo {author} {\bibfnamefont {J.}~\bibnamefont {Paramanandam}}, \bibinfo {author} {\bibfnamefont {L.~B.}\ \bibnamefont {Ioffe}},\ and\ \bibinfo {author} {\bibfnamefont {M.~E.}\ \bibnamefont {Gershenson}},\ }\bibfield  {title} {\bibinfo {title} {Protected {Josephson} rhombus chains},\ }\href {https://doi.org/10.1103/PhysRevLett.112.167001} {\bibfield  {journal} {\bibinfo  {journal} {Phys. Rev. Lett.}\ }\textbf {\bibinfo {volume} {112}},\ \bibinfo {pages} {167001} (\bibinfo {year} {2014})}\BibitemShut {NoStop}%
\bibitem [{\citenamefont {Larsen}\ \emph {et~al.}(2020)\citenamefont {Larsen}, \citenamefont {Gershenson}, \citenamefont {Casparis}, \citenamefont {Kringh{\o}j}, \citenamefont {Pearson}, \citenamefont {McNeil}, \citenamefont {Kuemmeth}, \citenamefont {Krogstrup}, \citenamefont {Petersson},\ and\ \citenamefont {Marcus}}]{larsenParityprotectedSuperconductorsemiconductorQubit2020}%
  \BibitemOpen
  \bibfield  {author} {\bibinfo {author} {\bibfnamefont {T.~W.}\ \bibnamefont {Larsen}}, \bibinfo {author} {\bibfnamefont {M.~E.}\ \bibnamefont {Gershenson}}, \bibinfo {author} {\bibfnamefont {L.}~\bibnamefont {Casparis}}, \bibinfo {author} {\bibfnamefont {A.}~\bibnamefont {Kringh{\o}j}}, \bibinfo {author} {\bibfnamefont {N.~J.}\ \bibnamefont {Pearson}}, \bibinfo {author} {\bibfnamefont {R.~P.~G.}\ \bibnamefont {McNeil}}, \bibinfo {author} {\bibfnamefont {F.}~\bibnamefont {Kuemmeth}}, \bibinfo {author} {\bibfnamefont {P.}~\bibnamefont {Krogstrup}}, \bibinfo {author} {\bibfnamefont {K.~D.}\ \bibnamefont {Petersson}},\ and\ \bibinfo {author} {\bibfnamefont {C.~M.}\ \bibnamefont {Marcus}},\ }\bibfield  {title} {\bibinfo {title} {Parity-protected superconductor-semiconductor qubit},\ }\href {https://doi.org/10.1103/physrevlett.125.056801} {\bibfield  {journal} {\bibinfo  {journal} {Phys. Rev. Lett.}\ }\textbf {\bibinfo {volume} {125}},\ \bibinfo {pages} {056801} (\bibinfo {year} {2020})}\BibitemShut {NoStop}%
\bibitem [{\citenamefont {Smith}\ \emph {et~al.}(2020)\citenamefont {Smith}, \citenamefont {Kou}, \citenamefont {Xiao}, \citenamefont {Vool},\ and\ \citenamefont {Devoret}}]{smithSuperconductingCircuitProtected2020}%
  \BibitemOpen
  \bibfield  {author} {\bibinfo {author} {\bibfnamefont {W.~C.}\ \bibnamefont {Smith}}, \bibinfo {author} {\bibfnamefont {A.}~\bibnamefont {Kou}}, \bibinfo {author} {\bibfnamefont {X.}~\bibnamefont {Xiao}}, \bibinfo {author} {\bibfnamefont {U.}~\bibnamefont {Vool}},\ and\ \bibinfo {author} {\bibfnamefont {M.~H.}\ \bibnamefont {Devoret}},\ }\bibfield  {title} {\bibinfo {title} {Superconducting circuit protected by two-{{Cooper-pair}} tunneling},\ }\href {https://doi.org/10.1038/s41534-019-0231-2} {\bibfield  {journal} {\bibinfo  {journal} {npj Quant. Inf.}\ }\textbf {\bibinfo {volume} {6}},\ \bibinfo {pages} {1} (\bibinfo {year} {2020})}\BibitemShut {NoStop}%
\bibitem [{\citenamefont {Schrade}\ \emph {et~al.}(2022)\citenamefont {Schrade}, \citenamefont {Marcus},\ and\ \citenamefont {Gyenis}}]{schradeProtectedHybridSuperconducting2021}%
  \BibitemOpen
  \bibfield  {author} {\bibinfo {author} {\bibfnamefont {C.}~\bibnamefont {Schrade}}, \bibinfo {author} {\bibfnamefont {C.~M.}\ \bibnamefont {Marcus}},\ and\ \bibinfo {author} {\bibfnamefont {A.}~\bibnamefont {Gyenis}},\ }\bibfield  {title} {\bibinfo {title} {Protected hybrid superconducting qubit in an array of gate-tunable {Josephson} interferometers},\ }\href {https://doi.org/10.1103/PRXQuantum.3.030303} {\bibfield  {journal} {\bibinfo  {journal} {PRX Quantum}\ }\textbf {\bibinfo {volume} {3}},\ \bibinfo {pages} {030303} (\bibinfo {year} {2022})}\BibitemShut {NoStop}%
\bibitem [{\citenamefont {Danon}\ \emph {et~al.}(2021)\citenamefont {Danon}, \citenamefont {Chatterjee}, \citenamefont {Gyenis},\ and\ \citenamefont {Kuemmeth}}]{danonProtectedSolidstateQubits2021}%
  \BibitemOpen
  \bibfield  {author} {\bibinfo {author} {\bibfnamefont {J.}~\bibnamefont {Danon}}, \bibinfo {author} {\bibfnamefont {A.}~\bibnamefont {Chatterjee}}, \bibinfo {author} {\bibfnamefont {A.}~\bibnamefont {Gyenis}},\ and\ \bibinfo {author} {\bibfnamefont {F.}~\bibnamefont {Kuemmeth}},\ }\bibfield  {title} {\bibinfo {title} {Protected solid-state qubits},\ }\href {https://doi.org/10.1063/5.0073945} {\bibfield  {journal} {\bibinfo  {journal} {Appl. Phys. Lett.}\ }\textbf {\bibinfo {volume} {119}},\ \bibinfo {pages} {260502} (\bibinfo {year} {2021})}\BibitemShut {NoStop}%
\bibitem [{\citenamefont {Chtchelkatchev}\ and\ \citenamefont {Nazarov}(2003)}]{chtchelkatchevAndreevQuantumDots2003a}%
  \BibitemOpen
  \bibfield  {author} {\bibinfo {author} {\bibfnamefont {N.~M.}\ \bibnamefont {Chtchelkatchev}}\ and\ \bibinfo {author} {\bibfnamefont {{\relax Yu}.~V.}\ \bibnamefont {Nazarov}},\ }\bibfield  {title} {\bibinfo {title} {Andreev quantum dots for spin manipulation},\ }\href {https://doi.org/10.1103/PhysRevLett.90.226806} {\bibfield  {journal} {\bibinfo  {journal} {Phys. Rev. Lett.}\ }\textbf {\bibinfo {volume} {90}},\ \bibinfo {pages} {226806} (\bibinfo {year} {2003})}\BibitemShut {NoStop}%
\bibitem [{\citenamefont {Padurariu}\ and\ \citenamefont {Nazarov}(2010)}]{padurariuTheoreticalProposalSuperconducting2010}%
  \BibitemOpen
  \bibfield  {author} {\bibinfo {author} {\bibfnamefont {C.}~\bibnamefont {Padurariu}}\ and\ \bibinfo {author} {\bibfnamefont {{\relax Yu}.~V.}\ \bibnamefont {Nazarov}},\ }\bibfield  {title} {\bibinfo {title} {Theoretical proposal for superconducting spin qubits},\ }\href {https://doi.org/10.1103/PhysRevB.81.144519} {\bibfield  {journal} {\bibinfo  {journal} {Phys. Rev. B}\ }\textbf {\bibinfo {volume} {81}},\ \bibinfo {pages} {144519} (\bibinfo {year} {2010})}\BibitemShut {NoStop}%
\bibitem [{\citenamefont {Hays}\ \emph {et~al.}(2021)\citenamefont {Hays}, \citenamefont {Fatemi}, \citenamefont {Bouman}, \citenamefont {Cerrillo}, \citenamefont {Diamond}, \citenamefont {Serniak}, \citenamefont {Connolly}, \citenamefont {Krogstrup}, \citenamefont {Nyg{\aa}rd}, \citenamefont {Yeyati}, \citenamefont {Geresdi},\ and\ \citenamefont {Devoret}}]{haysCoherentManipulationAndreev2021}%
  \BibitemOpen
  \bibfield  {author} {\bibinfo {author} {\bibfnamefont {M.}~\bibnamefont {Hays}}, \bibinfo {author} {\bibfnamefont {V.}~\bibnamefont {Fatemi}}, \bibinfo {author} {\bibfnamefont {D.}~\bibnamefont {Bouman}}, \bibinfo {author} {\bibfnamefont {J.}~\bibnamefont {Cerrillo}}, \bibinfo {author} {\bibfnamefont {S.}~\bibnamefont {Diamond}}, \bibinfo {author} {\bibfnamefont {K.}~\bibnamefont {Serniak}}, \bibinfo {author} {\bibfnamefont {T.}~\bibnamefont {Connolly}}, \bibinfo {author} {\bibfnamefont {P.}~\bibnamefont {Krogstrup}}, \bibinfo {author} {\bibfnamefont {J.}~\bibnamefont {Nyg{\aa}rd}}, \bibinfo {author} {\bibfnamefont {A.~L.}\ \bibnamefont {Yeyati}}, \bibinfo {author} {\bibfnamefont {A.}~\bibnamefont {Geresdi}},\ and\ \bibinfo {author} {\bibfnamefont {M.~H.}\ \bibnamefont {Devoret}},\ }\bibfield  {title} {\bibinfo {title} {Coherent manipulation of an {{Andreev}} spin qubit},\ }\href {https://doi.org/10.1126/science.abf0345} {\bibfield  {journal} {\bibinfo  {journal} {Science}\ }\textbf {\bibinfo {volume}
  {373}},\ \bibinfo {pages} {430} (\bibinfo {year} {2021})}\BibitemShut {NoStop}%
\bibitem [{\citenamefont {{Pita-Vidal}}\ \emph {et~al.}(2023)\citenamefont {{Pita-Vidal}}, \citenamefont {Bargerbos}, \citenamefont {{\v Z}itko}, \citenamefont {Splitthoff}, \citenamefont {Gr{\"u}nhaupt}, \citenamefont {Wesdorp}, \citenamefont {Liu}, \citenamefont {Kouwenhoven}, \citenamefont {Aguado}, \citenamefont {van Heck}, \citenamefont {Kou},\ and\ \citenamefont {Andersen}}]{pita-vidalDirectManipulationSuperconducting2023}%
  \BibitemOpen
  \bibfield  {author} {\bibinfo {author} {\bibfnamefont {M.}~\bibnamefont {{Pita-Vidal}}}, \bibinfo {author} {\bibfnamefont {A.}~\bibnamefont {Bargerbos}}, \bibinfo {author} {\bibfnamefont {R.}~\bibnamefont {{\v Z}itko}}, \bibinfo {author} {\bibfnamefont {L.~J.}\ \bibnamefont {Splitthoff}}, \bibinfo {author} {\bibfnamefont {L.}~\bibnamefont {Gr{\"u}nhaupt}}, \bibinfo {author} {\bibfnamefont {J.~J.}\ \bibnamefont {Wesdorp}}, \bibinfo {author} {\bibfnamefont {Y.}~\bibnamefont {Liu}}, \bibinfo {author} {\bibfnamefont {L.~P.}\ \bibnamefont {Kouwenhoven}}, \bibinfo {author} {\bibfnamefont {R.}~\bibnamefont {Aguado}}, \bibinfo {author} {\bibfnamefont {B.}~\bibnamefont {van Heck}}, \bibinfo {author} {\bibfnamefont {A.}~\bibnamefont {Kou}},\ and\ \bibinfo {author} {\bibfnamefont {C.~K.}\ \bibnamefont {Andersen}},\ }\bibfield  {title} {\bibinfo {title} {Direct manipulation of a superconducting spin qubit strongly coupled to a transmon qubit},\ }\href {https://doi.org/10.1038/s41567-023-02071-x} {\bibfield  {journal}
  {\bibinfo  {journal} {Nat. Phys.}\ }\textbf {\bibinfo {volume} {19}},\ \bibinfo {pages} {1110} (\bibinfo {year} {2023})}\BibitemShut {NoStop}%
\bibitem [{\citenamefont {Della~Rocca}\ \emph {et~al.}(2007)\citenamefont {Della~Rocca}, \citenamefont {Chauvin}, \citenamefont {Huard}, \citenamefont {Pothier}, \citenamefont {Esteve},\ and\ \citenamefont {Urbina}}]{dellaroccaMeasurementCurrentPhaseRelation2007}%
  \BibitemOpen
  \bibfield  {author} {\bibinfo {author} {\bibfnamefont {M.~L.}\ \bibnamefont {Della~Rocca}}, \bibinfo {author} {\bibfnamefont {M.}~\bibnamefont {Chauvin}}, \bibinfo {author} {\bibfnamefont {B.}~\bibnamefont {Huard}}, \bibinfo {author} {\bibfnamefont {H.}~\bibnamefont {Pothier}}, \bibinfo {author} {\bibfnamefont {D.}~\bibnamefont {Esteve}},\ and\ \bibinfo {author} {\bibfnamefont {C.}~\bibnamefont {Urbina}},\ }\bibfield  {title} {\bibinfo {title} {Measurement of the current-phase relation of superconducting atomic contacts},\ }\href {https://doi.org/10.1103/PhysRevLett.99.127005} {\bibfield  {journal} {\bibinfo  {journal} {Phys. Rev. Lett.}\ }\textbf {\bibinfo {volume} {99}},\ \bibinfo {pages} {127005} (\bibinfo {year} {2007})}\BibitemShut {NoStop}%
\bibitem [{\citenamefont {Lee}\ \emph {et~al.}(2015)\citenamefont {Lee}, \citenamefont {Kim}, \citenamefont {Jhi},\ and\ \citenamefont {Lee}}]{leeUltimatelyShortBallistic2015}%
  \BibitemOpen
  \bibfield  {author} {\bibinfo {author} {\bibfnamefont {G.-H.}\ \bibnamefont {Lee}}, \bibinfo {author} {\bibfnamefont {S.}~\bibnamefont {Kim}}, \bibinfo {author} {\bibfnamefont {S.-H.}\ \bibnamefont {Jhi}},\ and\ \bibinfo {author} {\bibfnamefont {H.-J.}\ \bibnamefont {Lee}},\ }\bibfield  {title} {\bibinfo {title} {Ultimately short ballistic vertical graphene {{Josephson}} junctions},\ }\href {https://doi.org/10.1038/ncomms7181} {\bibfield  {journal} {\bibinfo  {journal} {Nat. Commun.}\ }\textbf {\bibinfo {volume} {6}},\ \bibinfo {pages} {6181} (\bibinfo {year} {2015})}\BibitemShut {NoStop}%
\bibitem [{\citenamefont {Murani}\ \emph {et~al.}(2017)\citenamefont {Murani}, \citenamefont {Kasumov}, \citenamefont {Sengupta}, \citenamefont {Kasumov}, \citenamefont {Volkov}, \citenamefont {Khodos}, \citenamefont {Brisset}, \citenamefont {Delagrange}, \citenamefont {Chepelianskii}, \citenamefont {Deblock}, \citenamefont {Bouchiat},\ and\ \citenamefont {Gu{\'e}ron}}]{muraniBallisticEdgeStates2017}%
  \BibitemOpen
  \bibfield  {author} {\bibinfo {author} {\bibfnamefont {A.}~\bibnamefont {Murani}}, \bibinfo {author} {\bibfnamefont {A.}~\bibnamefont {Kasumov}}, \bibinfo {author} {\bibfnamefont {S.}~\bibnamefont {Sengupta}}, \bibinfo {author} {\bibfnamefont {Y.~A.}\ \bibnamefont {Kasumov}}, \bibinfo {author} {\bibfnamefont {V.~T.}\ \bibnamefont {Volkov}}, \bibinfo {author} {\bibfnamefont {I.~I.}\ \bibnamefont {Khodos}}, \bibinfo {author} {\bibfnamefont {F.}~\bibnamefont {Brisset}}, \bibinfo {author} {\bibfnamefont {R.}~\bibnamefont {Delagrange}}, \bibinfo {author} {\bibfnamefont {A.}~\bibnamefont {Chepelianskii}}, \bibinfo {author} {\bibfnamefont {R.}~\bibnamefont {Deblock}}, \bibinfo {author} {\bibfnamefont {H.}~\bibnamefont {Bouchiat}},\ and\ \bibinfo {author} {\bibfnamefont {S.}~\bibnamefont {Gu{\'e}ron}},\ }\bibfield  {title} {\bibinfo {title} {Ballistic edge states in bismuth nanowires revealed by {{SQUID}} interferometry},\ }\href {https://doi.org/10.1038/ncomms15941} {\bibfield  {journal} {\bibinfo  {journal} {Nat.
  Commun.}\ }\textbf {\bibinfo {volume} {8}},\ \bibinfo {pages} {15941} (\bibinfo {year} {2017})}\BibitemShut {NoStop}%
\bibitem [{\citenamefont {Ginzburg}\ \emph {et~al.}(2018)\citenamefont {Ginzburg}, \citenamefont {Batov}, \citenamefont {Bol'ginov}, \citenamefont {Egorov}, \citenamefont {Chichkov}, \citenamefont {Shchegolev}, \citenamefont {Klenov}, \citenamefont {Soloviev}, \citenamefont {Bakurskiy},\ and\ \citenamefont {Kupriyanov}}]{ginzburgDeterminationCurrentPhase2018}%
  \BibitemOpen
  \bibfield  {author} {\bibinfo {author} {\bibfnamefont {L.~V.}\ \bibnamefont {Ginzburg}}, \bibinfo {author} {\bibfnamefont {I.~E.}\ \bibnamefont {Batov}}, \bibinfo {author} {\bibfnamefont {V.~V.}\ \bibnamefont {Bol'ginov}}, \bibinfo {author} {\bibfnamefont {S.~V.}\ \bibnamefont {Egorov}}, \bibinfo {author} {\bibfnamefont {V.~I.}\ \bibnamefont {Chichkov}}, \bibinfo {author} {\bibfnamefont {A.~E.}\ \bibnamefont {Shchegolev}}, \bibinfo {author} {\bibfnamefont {N.~V.}\ \bibnamefont {Klenov}}, \bibinfo {author} {\bibfnamefont {I.~I.}\ \bibnamefont {Soloviev}}, \bibinfo {author} {\bibfnamefont {S.~V.}\ \bibnamefont {Bakurskiy}},\ and\ \bibinfo {author} {\bibfnamefont {M.~{\relax Yu}.}\ \bibnamefont {Kupriyanov}},\ }\bibfield  {title} {\bibinfo {title} {Determination of the current-phase relation in {Josephson} junctions by means of an asymmetric two-junction {SQUID}},\ }\href {https://doi.org/10.1134/S0021364018010058} {\bibfield  {journal} {\bibinfo  {journal} {JETP Lett.}\ }\textbf {\bibinfo {volume} {107}},\
  \bibinfo {pages} {48} (\bibinfo {year} {2018})}\BibitemShut {NoStop}%
\bibitem [{\citenamefont {Li}\ \emph {et~al.}(2019)\citenamefont {Li}, \citenamefont {{de Ronde}}, \citenamefont {{de Boer}}, \citenamefont {Ridderbos}, \citenamefont {Zwanenburg}, \citenamefont {Huang}, \citenamefont {Golubov},\ and\ \citenamefont {Brinkman}}]{liZeemanEffectInduced$0textensuremathensuremathpi$Transitions2019}%
  \BibitemOpen
  \bibfield  {author} {\bibinfo {author} {\bibfnamefont {C.}~\bibnamefont {Li}}, \bibinfo {author} {\bibfnamefont {B.}~\bibnamefont {{de Ronde}}}, \bibinfo {author} {\bibfnamefont {J.}~\bibnamefont {{de Boer}}}, \bibinfo {author} {\bibfnamefont {J.}~\bibnamefont {Ridderbos}}, \bibinfo {author} {\bibfnamefont {F.}~\bibnamefont {Zwanenburg}}, \bibinfo {author} {\bibfnamefont {Y.}~\bibnamefont {Huang}}, \bibinfo {author} {\bibfnamefont {A.}~\bibnamefont {Golubov}},\ and\ \bibinfo {author} {\bibfnamefont {A.}~\bibnamefont {Brinkman}},\ }\bibfield  {title} {\bibinfo {title} {Zeeman-effect-induced 0--$\pi$ transitions in ballistic {Dirac} semimetal {Josephson} junctions},\ }\href {https://doi.org/10.1103/PhysRevLett.123.026802} {\bibfield  {journal} {\bibinfo  {journal} {Phys. Rev. Lett.}\ }\textbf {\bibinfo {volume} {123}},\ \bibinfo {pages} {026802} (\bibinfo {year} {2019})}\BibitemShut {NoStop}%
\bibitem [{\citenamefont {Kayyalha}\ \emph {et~al.}(2019)\citenamefont {Kayyalha}, \citenamefont {Kargarian}, \citenamefont {Kazakov}, \citenamefont {Miotkowski}, \citenamefont {Galitski}, \citenamefont {Yakovenko}, \citenamefont {Rokhinson},\ and\ \citenamefont {Chen}}]{kayyalhaAnomalousLowTemperatureEnhancement2019}%
  \BibitemOpen
  \bibfield  {author} {\bibinfo {author} {\bibfnamefont {M.}~\bibnamefont {Kayyalha}}, \bibinfo {author} {\bibfnamefont {M.}~\bibnamefont {Kargarian}}, \bibinfo {author} {\bibfnamefont {A.}~\bibnamefont {Kazakov}}, \bibinfo {author} {\bibfnamefont {I.}~\bibnamefont {Miotkowski}}, \bibinfo {author} {\bibfnamefont {V.~M.}\ \bibnamefont {Galitski}}, \bibinfo {author} {\bibfnamefont {V.~M.}\ \bibnamefont {Yakovenko}}, \bibinfo {author} {\bibfnamefont {L.~P.}\ \bibnamefont {Rokhinson}},\ and\ \bibinfo {author} {\bibfnamefont {Y.~P.}\ \bibnamefont {Chen}},\ }\bibfield  {title} {\bibinfo {title} {Anomalous low-temperature enhancement of supercurrent in topological-insulator nanoribbon {Josephson} junctions: Evidence for low-energy {Andreev} bound states},\ }\href {https://doi.org/10.1103/PhysRevLett.122.047003} {\bibfield  {journal} {\bibinfo  {journal} {Phys. Rev. Lett.}\ }\textbf {\bibinfo {volume} {122}},\ \bibinfo {pages} {047003} (\bibinfo {year} {2019})}\BibitemShut {NoStop}%
\bibitem [{\citenamefont {Endres}\ \emph {et~al.}(2023)\citenamefont {Endres}, \citenamefont {Kononov}, \citenamefont {Arachchige}, \citenamefont {Yan}, \citenamefont {Mandrus}, \citenamefont {Watanabe}, \citenamefont {Taniguchi},\ and\ \citenamefont {Sch{\"o}nenberger}}]{endresCurrentPhaseRelation2023}%
  \BibitemOpen
  \bibfield  {author} {\bibinfo {author} {\bibfnamefont {M.}~\bibnamefont {Endres}}, \bibinfo {author} {\bibfnamefont {A.}~\bibnamefont {Kononov}}, \bibinfo {author} {\bibfnamefont {H.~S.}\ \bibnamefont {Arachchige}}, \bibinfo {author} {\bibfnamefont {J.}~\bibnamefont {Yan}}, \bibinfo {author} {\bibfnamefont {D.}~\bibnamefont {Mandrus}}, \bibinfo {author} {\bibfnamefont {K.}~\bibnamefont {Watanabe}}, \bibinfo {author} {\bibfnamefont {T.}~\bibnamefont {Taniguchi}},\ and\ \bibinfo {author} {\bibfnamefont {C.}~\bibnamefont {Sch{\"o}nenberger}},\ }\bibfield  {title} {\bibinfo {title} {Current-phase relation of a {WTe2} {Josephson} junction},\ }\href {https://doi.org/10.1021/acs.nanolett.3c01416} {\bibfield  {journal} {\bibinfo  {journal} {Nano Lett.}\ }\textbf {\bibinfo {volume} {23}},\ \bibinfo {pages} {4654} (\bibinfo {year} {2023})}\BibitemShut {NoStop}%
\bibitem [{\citenamefont {Babich}\ \emph {et~al.}(2023)\citenamefont {Babich}, \citenamefont {Kudriashov}, \citenamefont {Baranov},\ and\ \citenamefont {Stolyarov}}]{babichLimitationsCurrentPhase2023}%
  \BibitemOpen
  \bibfield  {author} {\bibinfo {author} {\bibfnamefont {I.}~\bibnamefont {Babich}}, \bibinfo {author} {\bibfnamefont {A.}~\bibnamefont {Kudriashov}}, \bibinfo {author} {\bibfnamefont {D.}~\bibnamefont {Baranov}},\ and\ \bibinfo {author} {\bibfnamefont {V.~S.}\ \bibnamefont {Stolyarov}},\ }\bibfield  {title} {\bibinfo {title} {Limitations of the current-phase relation measurements by an asymmetric dc-{SQUID}},\ }\href {https://doi.org/10.1021/acs.nanolett.3c01970} {\bibfield  {journal} {\bibinfo  {journal} {Nano Lett.}\ }\textbf {\bibinfo {volume} {23}},\ \bibinfo {pages} {6713} (\bibinfo {year} {2023})}\BibitemShut {NoStop}%
\bibitem [{\citenamefont {Messelot}\ \emph {et~al.}(2024)\citenamefont {Messelot}, \citenamefont {Aparicio}, \citenamefont {De~Seze}, \citenamefont {Eyraud}, \citenamefont {Coraux}, \citenamefont {Watanabe}, \citenamefont {Taniguchi},\ and\ \citenamefont {Renard}}]{messelotDirectMeasurementSin2024}%
  \BibitemOpen
  \bibfield  {author} {\bibinfo {author} {\bibfnamefont {S.}~\bibnamefont {Messelot}}, \bibinfo {author} {\bibfnamefont {N.}~\bibnamefont {Aparicio}}, \bibinfo {author} {\bibfnamefont {E.}~\bibnamefont {De~Seze}}, \bibinfo {author} {\bibfnamefont {E.}~\bibnamefont {Eyraud}}, \bibinfo {author} {\bibfnamefont {J.}~\bibnamefont {Coraux}}, \bibinfo {author} {\bibfnamefont {K.}~\bibnamefont {Watanabe}}, \bibinfo {author} {\bibfnamefont {T.}~\bibnamefont {Taniguchi}},\ and\ \bibinfo {author} {\bibfnamefont {J.}~\bibnamefont {Renard}},\ }\bibfield  {title} {\bibinfo {title} {Direct measurement of a $\sin(2\varphi)$ current phase relation in a graphene superconducting quantum interference device},\ }\href {https://doi.org/10.1103/PhysRevLett.133.106001} {\bibfield  {journal} {\bibinfo  {journal} {Phys. Rev. Lett.}\ }\textbf {\bibinfo {volume} {133}},\ \bibinfo {pages} {106001} (\bibinfo {year} {2024})}\BibitemShut {NoStop}%
\bibitem [{\citenamefont {Chieppa}\ \emph {et~al.}(2025)\citenamefont {Chieppa}, \citenamefont {Shukla}, \citenamefont {Traverso}, \citenamefont {Bucci}, \citenamefont {Zannier}, \citenamefont {Fracassi}, \citenamefont {Traverso~Ziani}, \citenamefont {Sassetti}, \citenamefont {Carrega}, \citenamefont {Beltram}, \citenamefont {Giazotto}, \citenamefont {Sorba},\ and\ \citenamefont {Heun}}]{chieppaUnveilingCurrentPhaseRelationship2025a}%
  \BibitemOpen
  \bibfield  {author} {\bibinfo {author} {\bibfnamefont {A.}~\bibnamefont {Chieppa}}, \bibinfo {author} {\bibfnamefont {G.}~\bibnamefont {Shukla}}, \bibinfo {author} {\bibfnamefont {S.}~\bibnamefont {Traverso}}, \bibinfo {author} {\bibfnamefont {G.}~\bibnamefont {Bucci}}, \bibinfo {author} {\bibfnamefont {V.}~\bibnamefont {Zannier}}, \bibinfo {author} {\bibfnamefont {S.}~\bibnamefont {Fracassi}}, \bibinfo {author} {\bibfnamefont {N.}~\bibnamefont {Traverso~Ziani}}, \bibinfo {author} {\bibfnamefont {M.}~\bibnamefont {Sassetti}}, \bibinfo {author} {\bibfnamefont {M.}~\bibnamefont {Carrega}}, \bibinfo {author} {\bibfnamefont {F.}~\bibnamefont {Beltram}}, \bibinfo {author} {\bibfnamefont {F.}~\bibnamefont {Giazotto}}, \bibinfo {author} {\bibfnamefont {L.}~\bibnamefont {Sorba}},\ and\ \bibinfo {author} {\bibfnamefont {S.}~\bibnamefont {Heun}},\ }\bibfield  {title} {\bibinfo {title} {Unveiling the {{Current-Phase Relationship}} of {{InSb Nanoflag Josephson Junctions Using}} a {{NanoSQUID Magnetometer}}},\ }\href
  {https://doi.org/10.1021/acs.nanolett.5c03765} {\bibfield  {journal} {\bibinfo  {journal} {Nano Letters}\ }\textbf {\bibinfo {volume} {25}},\ \bibinfo {pages} {14412} (\bibinfo {year} {2025})}\BibitemShut {NoStop}%
\bibitem [{\citenamefont {Strambini}\ \emph {et~al.}(2020)\citenamefont {Strambini}, \citenamefont {Iorio}, \citenamefont {Durante}, \citenamefont {Citro}, \citenamefont {{Sanz-Fern{\'a}ndez}}, \citenamefont {Guarcello}, \citenamefont {Tokatly}, \citenamefont {Braggio}, \citenamefont {Rocci}, \citenamefont {Ligato}, \citenamefont {Zannier}, \citenamefont {Sorba}, \citenamefont {Bergeret},\ and\ \citenamefont {Giazotto}}]{strambiniJosephsonPhaseBattery2020}%
  \BibitemOpen
  \bibfield  {author} {\bibinfo {author} {\bibfnamefont {E.}~\bibnamefont {Strambini}}, \bibinfo {author} {\bibfnamefont {A.}~\bibnamefont {Iorio}}, \bibinfo {author} {\bibfnamefont {O.}~\bibnamefont {Durante}}, \bibinfo {author} {\bibfnamefont {R.}~\bibnamefont {Citro}}, \bibinfo {author} {\bibfnamefont {C.}~\bibnamefont {{Sanz-Fern{\'a}ndez}}}, \bibinfo {author} {\bibfnamefont {C.}~\bibnamefont {Guarcello}}, \bibinfo {author} {\bibfnamefont {I.~V.}\ \bibnamefont {Tokatly}}, \bibinfo {author} {\bibfnamefont {A.}~\bibnamefont {Braggio}}, \bibinfo {author} {\bibfnamefont {M.}~\bibnamefont {Rocci}}, \bibinfo {author} {\bibfnamefont {N.}~\bibnamefont {Ligato}}, \bibinfo {author} {\bibfnamefont {V.}~\bibnamefont {Zannier}}, \bibinfo {author} {\bibfnamefont {L.}~\bibnamefont {Sorba}}, \bibinfo {author} {\bibfnamefont {F.~S.}\ \bibnamefont {Bergeret}},\ and\ \bibinfo {author} {\bibfnamefont {F.}~\bibnamefont {Giazotto}},\ }\bibfield  {title} {\bibinfo {title} {A {{Josephson}} phase battery},\ }\href
  {https://doi.org/10.1038/s41565-020-0712-7} {\bibfield  {journal} {\bibinfo  {journal} {Nature Nanotechnology}\ }\textbf {\bibinfo {volume} {15}},\ \bibinfo {pages} {656} (\bibinfo {year} {2020})}\BibitemShut {NoStop}%
\bibitem [{\citenamefont {Silver}\ and\ \citenamefont {Zimmerman}(1967)}]{silverQuantumStatesTransitions1967}%
  \BibitemOpen
  \bibfield  {author} {\bibinfo {author} {\bibfnamefont {A.~H.}\ \bibnamefont {Silver}}\ and\ \bibinfo {author} {\bibfnamefont {J.~E.}\ \bibnamefont {Zimmerman}},\ }\bibfield  {title} {\bibinfo {title} {Quantum states and transitions in weakly connected superconducting rings},\ }\href {https://doi.org/10.1103/PhysRev.157.317} {\bibfield  {journal} {\bibinfo  {journal} {Phys. Rev.}\ }\textbf {\bibinfo {volume} {157}},\ \bibinfo {pages} {317} (\bibinfo {year} {1967})}\BibitemShut {NoStop}%
\bibitem [{\citenamefont {Rifkin}\ and\ \citenamefont {Deaver}(1976)}]{rifkinCurrentphaseRelationPhasedependent1976}%
  \BibitemOpen
  \bibfield  {author} {\bibinfo {author} {\bibfnamefont {R.}~\bibnamefont {Rifkin}}\ and\ \bibinfo {author} {\bibfnamefont {B.~S.}\ \bibnamefont {Deaver}},\ }\bibfield  {title} {\bibinfo {title} {Current-phase relation and phase-dependent conductance of superconducting point contacts from rf impedance measurements},\ }\href {https://doi.org/10.1103/PhysRevB.13.3894} {\bibfield  {journal} {\bibinfo  {journal} {Phys. Rev. B}\ }\textbf {\bibinfo {volume} {13}},\ \bibinfo {pages} {3894} (\bibinfo {year} {1976})}\BibitemShut {NoStop}%
\bibitem [{\citenamefont {Frolov}\ \emph {et~al.}(2004)\citenamefont {Frolov}, \citenamefont {Van~Harlingen}, \citenamefont {Oboznov}, \citenamefont {Bolginov},\ and\ \citenamefont {Ryazanov}}]{frolovMeasurementCurrentphaseRelation2004}%
  \BibitemOpen
  \bibfield  {author} {\bibinfo {author} {\bibfnamefont {S.~M.}\ \bibnamefont {Frolov}}, \bibinfo {author} {\bibfnamefont {D.~J.}\ \bibnamefont {Van~Harlingen}}, \bibinfo {author} {\bibfnamefont {V.~A.}\ \bibnamefont {Oboznov}}, \bibinfo {author} {\bibfnamefont {V.~V.}\ \bibnamefont {Bolginov}},\ and\ \bibinfo {author} {\bibfnamefont {V.~V.}\ \bibnamefont {Ryazanov}},\ }\bibfield  {title} {\bibinfo {title} {Measurement of the current-phase relation of superconductor/ferromagnet/superconductor $\pi$ {{Josephson}} junctions},\ }\href {https://doi.org/10.1103/PhysRevB.70.144505} {\bibfield  {journal} {\bibinfo  {journal} {Phys. Rev. B}\ }\textbf {\bibinfo {volume} {70}},\ \bibinfo {pages} {144505} (\bibinfo {year} {2004})}\BibitemShut {NoStop}%
\bibitem [{\citenamefont {Troeman}\ \emph {et~al.}(2008)\citenamefont {Troeman}, \citenamefont {Van Der~Ploeg}, \citenamefont {Il'Ichev}, \citenamefont {Meyer}, \citenamefont {Golubov}, \citenamefont {Kupriyanov},\ and\ \citenamefont {Hilgenkamp}}]{troemanTemperatureDependenceMeasurements2008a}%
  \BibitemOpen
  \bibfield  {author} {\bibinfo {author} {\bibfnamefont {A.~G.~P.}\ \bibnamefont {Troeman}}, \bibinfo {author} {\bibfnamefont {S.~H.~W.}\ \bibnamefont {Van Der~Ploeg}}, \bibinfo {author} {\bibfnamefont {E.}~\bibnamefont {Il'Ichev}}, \bibinfo {author} {\bibfnamefont {H.-G.}\ \bibnamefont {Meyer}}, \bibinfo {author} {\bibfnamefont {A.~A.}\ \bibnamefont {Golubov}}, \bibinfo {author} {\bibfnamefont {M.~{\relax Yu}.}\ \bibnamefont {Kupriyanov}},\ and\ \bibinfo {author} {\bibfnamefont {H.}~\bibnamefont {Hilgenkamp}},\ }\bibfield  {title} {\bibinfo {title} {Temperature dependence measurements of the supercurrent-phase relationship in niobium nanobridges},\ }\href {https://doi.org/10.1103/PhysRevB.77.024509} {\bibfield  {journal} {\bibinfo  {journal} {Phys. Rev. B}\ }\textbf {\bibinfo {volume} {77}},\ \bibinfo {pages} {024509} (\bibinfo {year} {2008})}\BibitemShut {NoStop}%
\bibitem [{\citenamefont {Haller}\ \emph {et~al.}(2022)\citenamefont {Haller}, \citenamefont {F\"ul\"op}, \citenamefont {Indolese}, \citenamefont {Ridderbos}, \citenamefont {Kraft}, \citenamefont {Cheung}, \citenamefont {Ungerer}, \citenamefont {Watanabe}, \citenamefont {Taniguchi}, \citenamefont {Beckmann}, \citenamefont {Danneau}, \citenamefont {Virtanen},\ and\ \citenamefont {Sch\"onenberger}}]{hallerPhasedependentMicrowaveResponse2022}%
  \BibitemOpen
  \bibfield  {author} {\bibinfo {author} {\bibfnamefont {R.}~\bibnamefont {Haller}}, \bibinfo {author} {\bibfnamefont {G.}~\bibnamefont {F\"ul\"op}}, \bibinfo {author} {\bibfnamefont {D.}~\bibnamefont {Indolese}}, \bibinfo {author} {\bibfnamefont {J.}~\bibnamefont {Ridderbos}}, \bibinfo {author} {\bibfnamefont {R.}~\bibnamefont {Kraft}}, \bibinfo {author} {\bibfnamefont {L.~Y.}\ \bibnamefont {Cheung}}, \bibinfo {author} {\bibfnamefont {J.~H.}\ \bibnamefont {Ungerer}}, \bibinfo {author} {\bibfnamefont {K.}~\bibnamefont {Watanabe}}, \bibinfo {author} {\bibfnamefont {T.}~\bibnamefont {Taniguchi}}, \bibinfo {author} {\bibfnamefont {D.}~\bibnamefont {Beckmann}}, \bibinfo {author} {\bibfnamefont {R.}~\bibnamefont {Danneau}}, \bibinfo {author} {\bibfnamefont {P.}~\bibnamefont {Virtanen}},\ and\ \bibinfo {author} {\bibfnamefont {C.}~\bibnamefont {Sch\"onenberger}},\ }\bibfield  {title} {\bibinfo {title} {Phase-dependent microwave response of a graphene {Josephson} junction},\ }\href
  {https://doi.org/10.1103/PhysRevResearch.4.013198} {\bibfield  {journal} {\bibinfo  {journal} {Phys. Rev. Res.}\ }\textbf {\bibinfo {volume} {4}},\ \bibinfo {pages} {013198} (\bibinfo {year} {2022})}\BibitemShut {NoStop}%
\bibitem [{\citenamefont {Uhl}\ \emph {et~al.}(2024)\citenamefont {Uhl}, \citenamefont {Hackenbeck}, \citenamefont {Koelle}, \citenamefont {Kleiner},\ and\ \citenamefont {Bothner}}]{uhl2024extracting}%
  \BibitemOpen
  \bibfield  {author} {\bibinfo {author} {\bibfnamefont {K.}~\bibnamefont {Uhl}}, \bibinfo {author} {\bibfnamefont {D.}~\bibnamefont {Hackenbeck}}, \bibinfo {author} {\bibfnamefont {D.}~\bibnamefont {Koelle}}, \bibinfo {author} {\bibfnamefont {R.}~\bibnamefont {Kleiner}},\ and\ \bibinfo {author} {\bibfnamefont {D.}~\bibnamefont {Bothner}},\ }\bibfield  {title} {\bibinfo {title} {Extracting the current-phase relation of a monolithic three-dimensional nanoconstriction using a dc-current-tunable superconducting microwave cavity},\ }\href {https://doi.org/10.1103/PhysRevApplied.22.064052} {\bibfield  {journal} {\bibinfo  {journal} {Phys. Rev. Appl.}\ }\textbf {\bibinfo {volume} {22}},\ \bibinfo {pages} {064052} (\bibinfo {year} {2024})}\BibitemShut {NoStop}%
\bibitem [{\citenamefont {Sochnikov}\ \emph {et~al.}(2013)\citenamefont {Sochnikov}, \citenamefont {Bestwick}, \citenamefont {Williams}, \citenamefont {Lippman}, \citenamefont {Fisher}, \citenamefont {{Goldhaber-Gordon}}, \citenamefont {Kirtley},\ and\ \citenamefont {Moler}}]{sochnikovDirectMeasurementCurrentPhase2013}%
  \BibitemOpen
  \bibfield  {author} {\bibinfo {author} {\bibfnamefont {I.}~\bibnamefont {Sochnikov}}, \bibinfo {author} {\bibfnamefont {A.~J.}\ \bibnamefont {Bestwick}}, \bibinfo {author} {\bibfnamefont {J.~R.}\ \bibnamefont {Williams}}, \bibinfo {author} {\bibfnamefont {T.~M.}\ \bibnamefont {Lippman}}, \bibinfo {author} {\bibfnamefont {I.~R.}\ \bibnamefont {Fisher}}, \bibinfo {author} {\bibfnamefont {D.}~\bibnamefont {{Goldhaber-Gordon}}}, \bibinfo {author} {\bibfnamefont {J.~R.}\ \bibnamefont {Kirtley}},\ and\ \bibinfo {author} {\bibfnamefont {K.~A.}\ \bibnamefont {Moler}},\ }\bibfield  {title} {\bibinfo {title} {Direct measurement of current-phase relations in superconductor/topological insulator/superconductor junctions},\ }\href {https://doi.org/10.1021/nl400997k} {\bibfield  {journal} {\bibinfo  {journal} {Nano Lett.}\ }\textbf {\bibinfo {volume} {13}},\ \bibinfo {pages} {3086} (\bibinfo {year} {2013})}\BibitemShut {NoStop}%
\bibitem [{\citenamefont {Sochnikov}\ \emph {et~al.}(2015)\citenamefont {Sochnikov}, \citenamefont {Maier}, \citenamefont {Watson}, \citenamefont {Kirtley}, \citenamefont {Gould}, \citenamefont {Tkachov}, \citenamefont {Hankiewicz}, \citenamefont {Br{\"u}ne}, \citenamefont {Buhmann}, \citenamefont {Molenkamp},\ and\ \citenamefont {Moler}}]{sochnikovNonsinusoidalCurrentPhaseRelationship2015}%
  \BibitemOpen
  \bibfield  {author} {\bibinfo {author} {\bibfnamefont {I.}~\bibnamefont {Sochnikov}}, \bibinfo {author} {\bibfnamefont {L.}~\bibnamefont {Maier}}, \bibinfo {author} {\bibfnamefont {C.~A.}\ \bibnamefont {Watson}}, \bibinfo {author} {\bibfnamefont {J.~R.}\ \bibnamefont {Kirtley}}, \bibinfo {author} {\bibfnamefont {C.}~\bibnamefont {Gould}}, \bibinfo {author} {\bibfnamefont {G.}~\bibnamefont {Tkachov}}, \bibinfo {author} {\bibfnamefont {E.~M.}\ \bibnamefont {Hankiewicz}}, \bibinfo {author} {\bibfnamefont {C.}~\bibnamefont {Br{\"u}ne}}, \bibinfo {author} {\bibfnamefont {H.}~\bibnamefont {Buhmann}}, \bibinfo {author} {\bibfnamefont {L.~W.}\ \bibnamefont {Molenkamp}},\ and\ \bibinfo {author} {\bibfnamefont {K.~A.}\ \bibnamefont {Moler}},\ }\bibfield  {title} {\bibinfo {title} {Nonsinusoidal current-phase relationship in {Josephson} junctions from the {3D} topological insulator {HgTe}},\ }\href {https://doi.org/10.1103/PhysRevLett.114.066801} {\bibfield  {journal} {\bibinfo  {journal} {Phys. Rev. Lett.}\ }\textbf
  {\bibinfo {volume} {114}},\ \bibinfo {pages} {66801} (\bibinfo {year} {2015})}\BibitemShut {NoStop}%
\bibitem [{\citenamefont {Leblanc}\ \emph {et~al.}(2024{\natexlab{b}})\citenamefont {Leblanc}, \citenamefont {Tangchingchai}, \citenamefont {Momtaz}, \citenamefont {Kiyooka}, \citenamefont {Hartmann}, \citenamefont {{Fernandez-Bada}}, \citenamefont {Scher{\"u}bl}, \citenamefont {Brun}, \citenamefont {Schmitt}, \citenamefont {Zihlmann}, \citenamefont {Maurand}, \citenamefont {Dumur}, \citenamefont {De~Franceschi},\ and\ \citenamefont {Lefloch}}]{leblancNonreciprocalCharge4eSupercurrent2024}%
  \BibitemOpen
  \bibfield  {author} {\bibinfo {author} {\bibfnamefont {A.}~\bibnamefont {Leblanc}}, \bibinfo {author} {\bibfnamefont {C.}~\bibnamefont {Tangchingchai}}, \bibinfo {author} {\bibfnamefont {Z.~S.}\ \bibnamefont {Momtaz}}, \bibinfo {author} {\bibfnamefont {E.}~\bibnamefont {Kiyooka}}, \bibinfo {author} {\bibfnamefont {J.-M.}\ \bibnamefont {Hartmann}}, \bibinfo {author} {\bibfnamefont {G.~T.}\ \bibnamefont {{Fernandez-Bada}}}, \bibinfo {author} {\bibfnamefont {Z.}~\bibnamefont {Scher{\"u}bl}}, \bibinfo {author} {\bibfnamefont {B.}~\bibnamefont {Brun}}, \bibinfo {author} {\bibfnamefont {V.}~\bibnamefont {Schmitt}}, \bibinfo {author} {\bibfnamefont {S.}~\bibnamefont {Zihlmann}}, \bibinfo {author} {\bibfnamefont {R.}~\bibnamefont {Maurand}}, \bibinfo {author} {\bibfnamefont {{\'E}.}~\bibnamefont {Dumur}}, \bibinfo {author} {\bibfnamefont {S.}~\bibnamefont {De~Franceschi}},\ and\ \bibinfo {author} {\bibfnamefont {F.}~\bibnamefont {Lefloch}},\ }\bibfield  {title} {\bibinfo {title} {From nonreciprocal to charge-4e
  supercurrent in {{Ge-based Josephson}} devices with tunable harmonic content},\ }\href {https://doi.org/10.1103/PhysRevResearch.6.033281} {\bibfield  {journal} {\bibinfo  {journal} {Phys. Rev. Res.}\ }\textbf {\bibinfo {volume} {6}},\ \bibinfo {pages} {033281} (\bibinfo {year} {2024}{\natexlab{b}})}\BibitemShut {NoStop}%
\bibitem [{\citenamefont {Zhao}\ \emph {et~al.}(2023)\citenamefont {Zhao}, \citenamefont {Cui}, \citenamefont {Volkov}, \citenamefont {Yoo}, \citenamefont {Lee}, \citenamefont {Gardener}, \citenamefont {Akey}, \citenamefont {Engelke}, \citenamefont {Ronen}, \citenamefont {Zhong}, \citenamefont {Gu}, \citenamefont {Plugge}, \citenamefont {Tummuru}, \citenamefont {Kim}, \citenamefont {Franz}, \citenamefont {Pixley}, \citenamefont {Poccia},\ and\ \citenamefont {Kim}}]{zhaoTimereversalSymmetryBreaking2023}%
  \BibitemOpen
  \bibfield  {author} {\bibinfo {author} {\bibfnamefont {S.~Y.~F.}\ \bibnamefont {Zhao}}, \bibinfo {author} {\bibfnamefont {X.}~\bibnamefont {Cui}}, \bibinfo {author} {\bibfnamefont {P.~A.}\ \bibnamefont {Volkov}}, \bibinfo {author} {\bibfnamefont {H.}~\bibnamefont {Yoo}}, \bibinfo {author} {\bibfnamefont {S.}~\bibnamefont {Lee}}, \bibinfo {author} {\bibfnamefont {J.~A.}\ \bibnamefont {Gardener}}, \bibinfo {author} {\bibfnamefont {A.~J.}\ \bibnamefont {Akey}}, \bibinfo {author} {\bibfnamefont {R.}~\bibnamefont {Engelke}}, \bibinfo {author} {\bibfnamefont {Y.}~\bibnamefont {Ronen}}, \bibinfo {author} {\bibfnamefont {R.}~\bibnamefont {Zhong}}, \bibinfo {author} {\bibfnamefont {G.}~\bibnamefont {Gu}}, \bibinfo {author} {\bibfnamefont {S.}~\bibnamefont {Plugge}}, \bibinfo {author} {\bibfnamefont {T.}~\bibnamefont {Tummuru}}, \bibinfo {author} {\bibfnamefont {M.}~\bibnamefont {Kim}}, \bibinfo {author} {\bibfnamefont {M.}~\bibnamefont {Franz}}, \bibinfo {author} {\bibfnamefont {J.~H.}\ \bibnamefont {Pixley}},
  \bibinfo {author} {\bibfnamefont {N.}~\bibnamefont {Poccia}},\ and\ \bibinfo {author} {\bibfnamefont {P.}~\bibnamefont {Kim}},\ }\bibfield  {title} {\bibinfo {title} {Time-reversal symmetry breaking superconductivity between twisted cuprate superconductors},\ }\href {https://doi.org/10.1126/science.abl8371} {\bibfield  {journal} {\bibinfo  {journal} {Science}\ }\textbf {\bibinfo {volume} {382}},\ \bibinfo {pages} {1422} (\bibinfo {year} {2023})}\BibitemShut {NoStop}%
\bibitem [{\citenamefont {Tsarev}\ and\ \citenamefont {Fominov}(2025)}]{tsarevAllFractionalShapiro2025}%
  \BibitemOpen
  \bibfield  {author} {\bibinfo {author} {\bibfnamefont {P.~N.}\ \bibnamefont {Tsarev}}\ and\ \bibinfo {author} {\bibfnamefont {Y.~V.}\ \bibnamefont {Fominov}},\ }\bibfield  {title} {\bibinfo {title} {All fractional {{Shapiro}} steps in the {{RSJ}} model with two {{Josephson}} harmonics},\ }\href {https://arxiv.org/abs/2505.20502} {\bibfield  {journal} {\bibinfo  {journal} {arXiv:2505.20502}\ } (\bibinfo {year} {2025})}\BibitemShut {NoStop}%
\bibitem [{\citenamefont {Iorio}\ \emph {et~al.}(2023)\citenamefont {Iorio}, \citenamefont {Crippa}, \citenamefont {Turini}, \citenamefont {Salimian}, \citenamefont {Carrega}, \citenamefont {Chirolli}, \citenamefont {Zannier}, \citenamefont {Sorba}, \citenamefont {Strambini}, \citenamefont {Giazotto},\ and\ \citenamefont {Heun}}]{iorioHalfintegerShapiroSteps2023a}%
  \BibitemOpen
  \bibfield  {author} {\bibinfo {author} {\bibfnamefont {A.}~\bibnamefont {Iorio}}, \bibinfo {author} {\bibfnamefont {A.}~\bibnamefont {Crippa}}, \bibinfo {author} {\bibfnamefont {B.}~\bibnamefont {Turini}}, \bibinfo {author} {\bibfnamefont {S.}~\bibnamefont {Salimian}}, \bibinfo {author} {\bibfnamefont {M.}~\bibnamefont {Carrega}}, \bibinfo {author} {\bibfnamefont {L.}~\bibnamefont {Chirolli}}, \bibinfo {author} {\bibfnamefont {V.}~\bibnamefont {Zannier}}, \bibinfo {author} {\bibfnamefont {L.}~\bibnamefont {Sorba}}, \bibinfo {author} {\bibfnamefont {E.}~\bibnamefont {Strambini}}, \bibinfo {author} {\bibfnamefont {F.}~\bibnamefont {Giazotto}},\ and\ \bibinfo {author} {\bibfnamefont {S.}~\bibnamefont {Heun}},\ }\bibfield  {title} {\bibinfo {title} {Half-integer {{Shapiro}} steps in highly transmissive {{InSb}} nanoflag {{Josephson}} junctions},\ }\href {https://doi.org/10.1103/PhysRevResearch.5.033015} {\bibfield  {journal} {\bibinfo  {journal} {Physical Review Research}\ }\textbf {\bibinfo {volume} {5}},\
  \bibinfo {pages} {033015} (\bibinfo {year} {2023})}\BibitemShut {NoStop}%
\bibitem [{\citenamefont {Rokhinson}\ \emph {et~al.}(2012)\citenamefont {Rokhinson}, \citenamefont {Liu},\ and\ \citenamefont {Furdyna}}]{rokhinsonFractionalAcJosephson2012}%
  \BibitemOpen
  \bibfield  {author} {\bibinfo {author} {\bibfnamefont {L.~P.}\ \bibnamefont {Rokhinson}}, \bibinfo {author} {\bibfnamefont {X.}~\bibnamefont {Liu}},\ and\ \bibinfo {author} {\bibfnamefont {J.~K.}\ \bibnamefont {Furdyna}},\ }\bibfield  {title} {\bibinfo {title} {The fractional a.c. {{Josephson}} effect in a semiconductor-superconductor nanowire as a signature of {Majorana} particles},\ }\href {https://doi.org/10.1038/nphys2429} {\bibfield  {journal} {\bibinfo  {journal} {Nat. Phys.}\ }\textbf {\bibinfo {volume} {8}},\ \bibinfo {pages} {795} (\bibinfo {year} {2012})}\BibitemShut {NoStop}%
\bibitem [{\citenamefont {Wiedenmann}\ \emph {et~al.}(2016)\citenamefont {Wiedenmann}, \citenamefont {Bocquillon}, \citenamefont {Deacon}, \citenamefont {Hartinger}, \citenamefont {Herrmann}, \citenamefont {Klapwijk}, \citenamefont {Maier}, \citenamefont {Ames}, \citenamefont {Br{\"u}ne}, \citenamefont {Gould}, \citenamefont {Oiwa}, \citenamefont {Ishibashi}, \citenamefont {Tarucha}, \citenamefont {Buhmann},\ and\ \citenamefont {Molenkamp}}]{wiedenmann4pperiodicJosephsonSupercurrent2016}%
  \BibitemOpen
  \bibfield  {author} {\bibinfo {author} {\bibfnamefont {J.}~\bibnamefont {Wiedenmann}}, \bibinfo {author} {\bibfnamefont {E.}~\bibnamefont {Bocquillon}}, \bibinfo {author} {\bibfnamefont {R.~S.}\ \bibnamefont {Deacon}}, \bibinfo {author} {\bibfnamefont {S.}~\bibnamefont {Hartinger}}, \bibinfo {author} {\bibfnamefont {O.}~\bibnamefont {Herrmann}}, \bibinfo {author} {\bibfnamefont {T.~M.}\ \bibnamefont {Klapwijk}}, \bibinfo {author} {\bibfnamefont {L.}~\bibnamefont {Maier}}, \bibinfo {author} {\bibfnamefont {C.}~\bibnamefont {Ames}}, \bibinfo {author} {\bibfnamefont {C.}~\bibnamefont {Br{\"u}ne}}, \bibinfo {author} {\bibfnamefont {C.}~\bibnamefont {Gould}}, \bibinfo {author} {\bibfnamefont {A.}~\bibnamefont {Oiwa}}, \bibinfo {author} {\bibfnamefont {K.}~\bibnamefont {Ishibashi}}, \bibinfo {author} {\bibfnamefont {S.}~\bibnamefont {Tarucha}}, \bibinfo {author} {\bibfnamefont {H.}~\bibnamefont {Buhmann}},\ and\ \bibinfo {author} {\bibfnamefont {L.~W.}\ \bibnamefont {Molenkamp}},\ }\bibfield  {title} {\bibinfo
  {title} {4{$\pi$}-periodic {Josephson} supercurrent in {{HgTe-based}} topological {Josephson} junctions},\ }\href {https://doi.org/10.1038/ncomms10303} {\bibfield  {journal} {\bibinfo  {journal} {Nat. Commun.}\ }\textbf {\bibinfo {volume} {7}},\ \bibinfo {pages} {10303} (\bibinfo {year} {2016})}\BibitemShut {NoStop}%
\bibitem [{\citenamefont {Bocquillon}\ \emph {et~al.}(2017)\citenamefont {Bocquillon}, \citenamefont {Deacon}, \citenamefont {Wiedenmann}, \citenamefont {Leubner}, \citenamefont {Klapwijk}, \citenamefont {Br{\"u}ne}, \citenamefont {Ishibashi}, \citenamefont {Buhmann},\ and\ \citenamefont {Molenkamp}}]{bocquillonGaplessAndreevBound2017}%
  \BibitemOpen
  \bibfield  {author} {\bibinfo {author} {\bibfnamefont {E.}~\bibnamefont {Bocquillon}}, \bibinfo {author} {\bibfnamefont {R.~S.}\ \bibnamefont {Deacon}}, \bibinfo {author} {\bibfnamefont {J.}~\bibnamefont {Wiedenmann}}, \bibinfo {author} {\bibfnamefont {P.}~\bibnamefont {Leubner}}, \bibinfo {author} {\bibfnamefont {T.~M.}\ \bibnamefont {Klapwijk}}, \bibinfo {author} {\bibfnamefont {C.}~\bibnamefont {Br{\"u}ne}}, \bibinfo {author} {\bibfnamefont {K.}~\bibnamefont {Ishibashi}}, \bibinfo {author} {\bibfnamefont {H.}~\bibnamefont {Buhmann}},\ and\ \bibinfo {author} {\bibfnamefont {L.~W.}\ \bibnamefont {Molenkamp}},\ }\bibfield  {title} {\bibinfo {title} {Gapless {Andreev} bound states in the quantum spin {Hall} insulator {{HgTe}}},\ }\href {https://doi.org/10.1038/nnano.2016.159} {\bibfield  {journal} {\bibinfo  {journal} {Nat. Nanotech.}\ }\textbf {\bibinfo {volume} {12}},\ \bibinfo {pages} {137} (\bibinfo {year} {2017})}\BibitemShut {NoStop}%
\bibitem [{\citenamefont {Li}\ \emph {et~al.}(2018)\citenamefont {Li}, \citenamefont {{de Boer}}, \citenamefont {{de Ronde}}, \citenamefont {Ramankutty}, \citenamefont {{van Heumen}}, \citenamefont {Huang}, \citenamefont {{de Visser}}, \citenamefont {Golubov}, \citenamefont {Golden},\ and\ \citenamefont {Brinkman}}]{li4pperiodicAndreevBound2018}%
  \BibitemOpen
  \bibfield  {author} {\bibinfo {author} {\bibfnamefont {C.}~\bibnamefont {Li}}, \bibinfo {author} {\bibfnamefont {J.~C.}\ \bibnamefont {{de Boer}}}, \bibinfo {author} {\bibfnamefont {B.}~\bibnamefont {{de Ronde}}}, \bibinfo {author} {\bibfnamefont {S.~V.}\ \bibnamefont {Ramankutty}}, \bibinfo {author} {\bibfnamefont {E.}~\bibnamefont {{van Heumen}}}, \bibinfo {author} {\bibfnamefont {Y.}~\bibnamefont {Huang}}, \bibinfo {author} {\bibfnamefont {A.}~\bibnamefont {{de Visser}}}, \bibinfo {author} {\bibfnamefont {A.~A.}\ \bibnamefont {Golubov}}, \bibinfo {author} {\bibfnamefont {M.~S.}\ \bibnamefont {Golden}},\ and\ \bibinfo {author} {\bibfnamefont {A.}~\bibnamefont {Brinkman}},\ }\bibfield  {title} {\bibinfo {title} {4{$\pi$}-periodic {Andreev} bound states in a {Dirac} semimetal},\ }\href {https://doi.org/10.1038/s41563-018-0158-6} {\bibfield  {journal} {\bibinfo  {journal} {Nat. Mater.}\ }\textbf {\bibinfo {volume} {17}},\ \bibinfo {pages} {875} (\bibinfo {year} {2018})}\BibitemShut {NoStop}%
\bibitem [{\citenamefont {Dartiailh}\ \emph {et~al.}(2021)\citenamefont {Dartiailh}, \citenamefont {Cuozzo}, \citenamefont {Elfeky}, \citenamefont {Mayer}, \citenamefont {Yuan}, \citenamefont {Wickramasinghe}, \citenamefont {Rossi},\ and\ \citenamefont {Shabani}}]{dartiailhMissingShapiroSteps2021}%
  \BibitemOpen
  \bibfield  {author} {\bibinfo {author} {\bibfnamefont {M.~C.}\ \bibnamefont {Dartiailh}}, \bibinfo {author} {\bibfnamefont {J.~J.}\ \bibnamefont {Cuozzo}}, \bibinfo {author} {\bibfnamefont {B.~H.}\ \bibnamefont {Elfeky}}, \bibinfo {author} {\bibfnamefont {W.}~\bibnamefont {Mayer}}, \bibinfo {author} {\bibfnamefont {J.}~\bibnamefont {Yuan}}, \bibinfo {author} {\bibfnamefont {K.~S.}\ \bibnamefont {Wickramasinghe}}, \bibinfo {author} {\bibfnamefont {E.}~\bibnamefont {Rossi}},\ and\ \bibinfo {author} {\bibfnamefont {J.}~\bibnamefont {Shabani}},\ }\bibfield  {title} {\bibinfo {title} {Missing {{Shapiro}} steps in topologically trivial {{Josephson}} junction on {{InAs}} quantum well},\ }\href {https://doi.org/10.1038/s41467-020-20382-y} {\bibfield  {journal} {\bibinfo  {journal} {Nat Commun}\ }\textbf {\bibinfo {volume} {12}},\ \bibinfo {pages} {78} (\bibinfo {year} {2021})}\BibitemShut {NoStop}%
\bibitem [{\citenamefont {Elfeky}\ \emph {et~al.}(2023)\citenamefont {Elfeky}, \citenamefont {Cuozzo}, \citenamefont {Lotfizadeh}, \citenamefont {Schiela}, \citenamefont {Farzaneh}, \citenamefont {Strickland}, \citenamefont {Langone}, \citenamefont {Rossi},\ and\ \citenamefont {Shabani}}]{elfekyEvolution4pPeriodicSupercurrent2023}%
  \BibitemOpen
  \bibfield  {author} {\bibinfo {author} {\bibfnamefont {B.~H.}\ \bibnamefont {Elfeky}}, \bibinfo {author} {\bibfnamefont {J.~J.}\ \bibnamefont {Cuozzo}}, \bibinfo {author} {\bibfnamefont {N.}~\bibnamefont {Lotfizadeh}}, \bibinfo {author} {\bibfnamefont {W.~F.}\ \bibnamefont {Schiela}}, \bibinfo {author} {\bibfnamefont {S.~M.}\ \bibnamefont {Farzaneh}}, \bibinfo {author} {\bibfnamefont {W.~M.}\ \bibnamefont {Strickland}}, \bibinfo {author} {\bibfnamefont {D.}~\bibnamefont {Langone}}, \bibinfo {author} {\bibfnamefont {E.}~\bibnamefont {Rossi}},\ and\ \bibinfo {author} {\bibfnamefont {J.}~\bibnamefont {Shabani}},\ }\bibfield  {title} {\bibinfo {title} {Evolution of 4{$\pi$}-periodic supercurrent in the presence of an in-plane magnetic field},\ }\href {https://doi.org/10.1021/acsnano.2c10880} {\bibfield  {journal} {\bibinfo  {journal} {ACS Nano}\ }\textbf {\bibinfo {volume} {17}},\ \bibinfo {pages} {4650} (\bibinfo {year} {2023})}\BibitemShut {NoStop}%
\bibitem [{\citenamefont {Seoane~Souto}\ \emph {et~al.}(2024)\citenamefont {Seoane~Souto}, \citenamefont {Leijnse}, \citenamefont {Schrade}, \citenamefont {Valentini}, \citenamefont {Katsaros},\ and\ \citenamefont {Danon}}]{seoanesoutoTuningJosephsonDiode2024}%
  \BibitemOpen
  \bibfield  {author} {\bibinfo {author} {\bibfnamefont {R.}~\bibnamefont {Seoane~Souto}}, \bibinfo {author} {\bibfnamefont {M.}~\bibnamefont {Leijnse}}, \bibinfo {author} {\bibfnamefont {C.}~\bibnamefont {Schrade}}, \bibinfo {author} {\bibfnamefont {M.}~\bibnamefont {Valentini}}, \bibinfo {author} {\bibfnamefont {G.}~\bibnamefont {Katsaros}},\ and\ \bibinfo {author} {\bibfnamefont {J.}~\bibnamefont {Danon}},\ }\bibfield  {title} {\bibinfo {title} {Tuning the {{Josephson}} diode response with an ac current},\ }\href {https://doi.org/10.1103/PhysRevResearch.6.L022002} {\bibfield  {journal} {\bibinfo  {journal} {Phys. Rev. Res.}\ }\textbf {\bibinfo {volume} {6}},\ \bibinfo {pages} {L022002} (\bibinfo {year} {2024})}\BibitemShut {NoStop}%
\bibitem [{\citenamefont {Hansen}(2016)}]{hansenCMAEvolutionStrategy2016}%
  \BibitemOpen
  \bibfield  {author} {\bibinfo {author} {\bibfnamefont {N.}~\bibnamefont {Hansen}},\ }\href {https://doi.org/10.48550/ARXIV.1604.00772} {\bibinfo {title} {The {{CMA Evolution Strategy}}: {{A Tutorial}}}} (\bibinfo {year} {2016})\BibitemShut {NoStop}%
\bibitem [{Note1()}]{Note1}%
  \BibitemOpen
  \bibinfo {note} {We use three ``restarts'' of the of the CMA-ES algorithm (i.e., three independent consecutive fitting attempts) for each fit, which leads to convergence to the global minimum around 99 out of 100 times. For larger $m_\protect \text {max}$ or a higher desired success rate, it could be beneficial to increase the number of restarts.}\BibitemShut {Stop}%
\bibitem [{Note2()}]{Note2}%
  \BibitemOpen
  \bibinfo {note} {This of course assumes that the boundaries are not symmetric around $I_\protect \text {dc}=0$, i.e., $I^+_c(I_\protect \text {ac})=-I^-_c(I_\protect \text {ac})$ for all $I_\protect \text {ac}$, which happens when there is only one dominating harmonic or when $\gamma _2=n\pi $}\BibitemShut {NoStop}%
\bibitem [{Note3()}]{Note3}%
  \BibitemOpen
  \bibinfo {note} {In practice, we did a current-biased measurement and numerically reconstructed $V_\protect \text {dc}(I_\protect \text {dc})=\DOTSI \intop \ilimits@ _0^{I_\protect \text {dc}} \left ( dV/dI \right )_\protect \text {meas} dI$. Plotted as a function of $V_\protect \text {dc}$, Shapiro steps are easy to identify as horizontal resistance dips in the measured differential resistance map.}\BibitemShut {Stop}%
\bibitem [{Note4()}]{Note4}%
  \BibitemOpen
  \bibinfo {note} {$\omega \tau _\protect \text {J}=hf/2eI_cR\simeq 0.32$, for $I_c\simeq 150\protect \,\protect \mathrm {nA}$, $R=10\protect \,\Omega $, and $f = 230\protect \,\protect \mathrm {MHz}$, which is the case for half a flux quantum ($\varphi _{\protect \text {ext}} = 0.5$) threading the loop}\BibitemShut {NoStop}%
\bibitem [{Note5()}]{Note5}%
  \BibitemOpen
  \bibinfo {note} {We also tried even higher frequencies and noticed that for $f \gtrsim 5.5\protect \,$GHz the technique starts to fail. We believe that this could be related to Landau--Zener transitions induced by the driving, see the Appendix~\ref {appendix:D} for more details.}\BibitemShut {Stop}%
\bibitem [{Note6()}]{Note6}%
  \BibitemOpen
  \bibinfo {note} {Taylor expanding the Beenakker formula for ABS energy $E(\varphi )=\Delta \protect \sqrt {1-T\sin ^2(\phi /2)}$, for $T=0.7$, typical for SNS junctions, gives $E(\phi )\propto [\cos (\phi ) - 0.072\cos (2\phi )+0.011\cos (3\phi )+...\protect \,]$, with $I(\phi )=(2e/h)\partial _\phi E(\phi )$. Here, truncation to two harmonics is fairly accurate. Even for $T=0.95$, the CMA-ES fit can be implemented if errors of $\sim I_6/I_1=4.7\%$ are tolerable.}\BibitemShut {Stop}%
\bibitem [{\citenamefont {Wallraff}\ \emph {et~al.}(2003)\citenamefont {Wallraff}, \citenamefont {Lukashenko}, \citenamefont {Coqui}, \citenamefont {Kemp}, \citenamefont {Duty},\ and\ \citenamefont {Ustinov}}]{Switching_current_wallraff}%
  \BibitemOpen
  \bibfield  {author} {\bibinfo {author} {\bibfnamefont {A.}~\bibnamefont {Wallraff}}, \bibinfo {author} {\bibfnamefont {A.}~\bibnamefont {Lukashenko}}, \bibinfo {author} {\bibfnamefont {C.}~\bibnamefont {Coqui}}, \bibinfo {author} {\bibfnamefont {A.}~\bibnamefont {Kemp}}, \bibinfo {author} {\bibfnamefont {T.}~\bibnamefont {Duty}},\ and\ \bibinfo {author} {\bibfnamefont {A.~V.}\ \bibnamefont {Ustinov}},\ }\bibfield  {title} {\bibinfo {title} {Switching current measurements of large area {Josephson} tunnel junctions},\ }\href {https://doi.org/10.1063/1.1588752} {\bibfield  {journal} {\bibinfo  {journal} {Rev. Sci. Instr.}\ }\textbf {\bibinfo {volume} {74}},\ \bibinfo {pages} {3740} (\bibinfo {year} {2003})}\BibitemShut {NoStop}%
\bibitem [{\citenamefont {Gloos}\ and\ \citenamefont {Anders}(1999)}]{Gloos1999}%
  \BibitemOpen
  \bibfield  {author} {\bibinfo {author} {\bibfnamefont {K.}~\bibnamefont {Gloos}}\ and\ \bibinfo {author} {\bibfnamefont {F.}~\bibnamefont {Anders}},\ }\bibfield  {title} {\bibinfo {title} {Phase diffusion and suppression of the supercurrent by quantum-mechanical fluctuations of the josephson plasma at small superconducting point contacts},\ }\href {https://doi.org/10.1023/A:1021874709355} {\bibfield  {journal} {\bibinfo  {journal} {Journal of Low Temperature Physics}\ }\textbf {\bibinfo {volume} {116}},\ \bibinfo {pages} {21} (\bibinfo {year} {1999})}\BibitemShut {NoStop}%
\bibitem [{\citenamefont {Beenakker}\ and\ \citenamefont {van Houten}(1991)}]{beenakkerJosephsonCurrentSuperconducting1991}%
  \BibitemOpen
  \bibfield  {author} {\bibinfo {author} {\bibfnamefont {C.~W.~J.}\ \bibnamefont {Beenakker}}\ and\ \bibinfo {author} {\bibfnamefont {H.}~\bibnamefont {van Houten}},\ }\bibfield  {title} {\bibinfo {title} {Josephson current through a superconducting quantum point contact shorter than the coherence length},\ }\href {https://doi.org/10.1103/physrevlett.66.3056} {\bibfield  {journal} {\bibinfo  {journal} {Phys. Rev. Lett.}\ }\textbf {\bibinfo {volume} {66}},\ \bibinfo {pages} {3056} (\bibinfo {year} {1991})}\BibitemShut {NoStop}%
\bibitem [{\citenamefont {Averin}\ and\ \citenamefont {Bardas}(1995)}]{ac_Josephson_Effect_Averin}%
  \BibitemOpen
  \bibfield  {author} {\bibinfo {author} {\bibfnamefont {D.}~\bibnamefont {Averin}}\ and\ \bibinfo {author} {\bibfnamefont {A.}~\bibnamefont {Bardas}},\ }\bibfield  {title} {\bibinfo {title} {ac {Josephson} effect in a single quantum channel},\ }\href {https://doi.org/10.1103/PhysRevLett.75.1831} {\bibfield  {journal} {\bibinfo  {journal} {Phys. Rev. Lett.}\ }\textbf {\bibinfo {volume} {75}},\ \bibinfo {pages} {1831} (\bibinfo {year} {1995})}\BibitemShut {NoStop}%
\bibitem [{Note7()}]{Note7}%
  \BibitemOpen
  \bibinfo {note} {We use $I_0$ instead of $i_0$ for scaling $\sigma $ since we require it to scale with the randomly generated $I_m$ in Fig.~\ref {fig:noise1} for consistency.}\BibitemShut {Stop}%
\bibitem [{\citenamefont {Chrestin}\ \emph {et~al.}(1994)\citenamefont {Chrestin}, \citenamefont {Matsuyama},\ and\ \citenamefont {Merkt}}]{SNS_IcRn_Chrestin}%
  \BibitemOpen
  \bibfield  {author} {\bibinfo {author} {\bibfnamefont {A.}~\bibnamefont {Chrestin}}, \bibinfo {author} {\bibfnamefont {T.}~\bibnamefont {Matsuyama}},\ and\ \bibinfo {author} {\bibfnamefont {U.}~\bibnamefont {Merkt}},\ }\bibfield  {title} {\bibinfo {title} {Critical currents and supercurrent oscillations in {Josephson} field-effect transistors},\ }\href {https://doi.org/10.1103/PhysRevB.49.498} {\bibfield  {journal} {\bibinfo  {journal} {Phys. Rev. B}\ }\textbf {\bibinfo {volume} {49}},\ \bibinfo {pages} {498} (\bibinfo {year} {1994})}\BibitemShut {NoStop}%
\bibitem [{\citenamefont {Furusaki}\ \emph {et~al.}(1991)\citenamefont {Furusaki}, \citenamefont {Takayanagi},\ and\ \citenamefont {Tsukada}}]{furusakiTheoryQuantumConduction1991}%
  \BibitemOpen
  \bibfield  {author} {\bibinfo {author} {\bibfnamefont {A.}~\bibnamefont {Furusaki}}, \bibinfo {author} {\bibfnamefont {H.}~\bibnamefont {Takayanagi}},\ and\ \bibinfo {author} {\bibfnamefont {M.}~\bibnamefont {Tsukada}},\ }\bibfield  {title} {\bibinfo {title} {Theory of quantum conduction of supercurrent through a constriction},\ }\href {https://doi.org/10.1103/physrevlett.67.132} {\bibfield  {journal} {\bibinfo  {journal} {Phys. Rev. Lett.}\ }\textbf {\bibinfo {volume} {67}},\ \bibinfo {pages} {132} (\bibinfo {year} {1991})}\BibitemShut {NoStop}%
\bibitem [{Note8()}]{Note8}%
  \BibitemOpen
  \bibinfo {note} {A typical induced gap in shallow Ge quantum wells is of the order of $\Delta \simeq 150\protect \,\protect \mathrm {\mu eV}$ \cite {valentiniParityconservingCooperpairTransport2024}, such that $2\Delta \equiv 72~\protect \mathrm {GHz}$}\BibitemShut {NoStop}%
\end{thebibliography}
\end{document}